\documentclass[journal=jctcce,manuscript=article]{achemso}

\usepackage[version=3]{mhchem} % Formula subscripts using \ce{}

\author{Fabian M. Faulstich}
\affiliation{Department of Mathematics, University of California, Berkeley}
\email{f.m.faulstich@berkeley.edu}
\author{H\aa kon E. Kristiansen}
\affiliation{Hylleraas Centre for Quantum Molecular Sciences, Department of Chemistry, University of Oslo, Norway}
\author{Mihaly A. Csirik}
\affiliation{Hylleraas Centre for Quantum Molecular Sciences, Department of Chemistry, University of Oslo, Norway}
\author{Simen Kvaal}
\affiliation{Hylleraas Centre for Quantum Molecular Sciences, Department of Chemistry, University of Oslo, Norway}
\author{Thomas Bondo Pedersen}
\affiliation{Hylleraas Centre for Quantum Molecular Sciences, Department of Chemistry, University of Oslo, Norway}
\author{Andre Laestadius}
\affiliation{Department of Computer Science, Oslo Metropolitan University, Norway}
\alsoaffiliation[Second University]
{Hylleraas Centre for Quantum Molecular Sciences, Department of Chemistry, University of Oslo, Norway}

\title[An \textsf{achemso} demo]
  {The $S$-diagnostic---an a posteriori error assessment for single-reference coupled-cluster methods}

\abbreviations{IR,NMR,UV}

\usepackage{color, xcolor, colortbl}
\usepackage{graphicx}
\usepackage{geometry}
\usepackage{amsthm,amsmath,amssymb,amsfonts}
\usepackage{algorithm}
\usepackage{algorithmic}
\usepackage{dcolumn}% Align table columns on decimal point
\usepackage{epstopdf}
\usepackage{bm}
\usepackage{appendix}
\usepackage{multirow}
\usepackage{braket}
\usepackage[english]{babel}
\usepackage{hyperref}
\usepackage[capitalize]{cleveref}
\usepackage{xpatch}
\usepackage{tikz}
\usepackage{adjustbox}
\usepackage{xspace}
\usepackage{enumitem} 
\usepackage{comment}
\usepackage{pifont}
\usepackage{pgffor}
\usepackage{url}
\usepackage{hypcap}
\usepackage{dsfont}
\usepackage{algorithm}
\usepackage{algorithmic}
\usepackage{soul}
\usepackage{diagbox}
\usepackage{graphicx}
\usepackage{amsthm}
\usepackage{thmtools}
\usepackage{mathtools}
\usepackage{thm-restate}
\usepackage{amsmath}
\usepackage{amssymb}
\usepackage{comment}
\usepackage{placeins}
\usepackage{epstopdf}
\usetikzlibrary{calc,backgrounds}
\usepackage{booktabs}
\usepackage{rotating}

\usepackage{caption}
\usepackage{subcaption}

\definecolor{airforceblue}{rgb}{0.36, 0.54, 0.66}

\usepackage[normalem]{ulem}

\newcommand{\msvt}{\sigma(t)}
\newcommand{\msvz}{\sigma(z)}

\begin{document}

\begin{tocentry}

\includegraphics{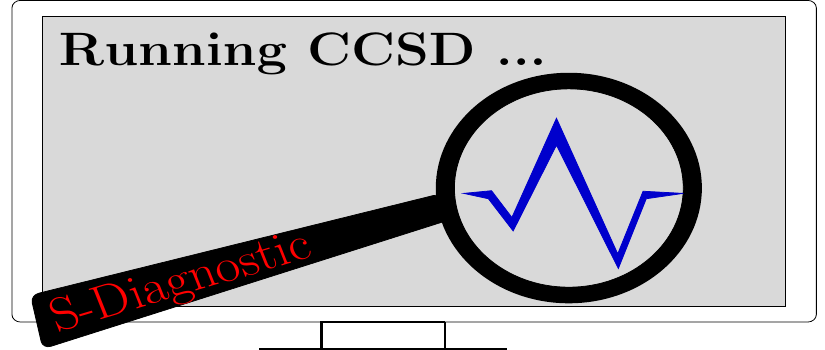}

\end{tocentry}

\begin{abstract}
We propose a novel {\it a posteriori} error assessment for the single-reference coupled-cluster (SRCC) method called the $S$-diagnostic. 
We provide a derivation of the $S$-diagnostic that is rooted in the mathematical analysis of different SRCC variants. 
We numerically scrutinized the $S$-diagnostic, testing its performance for (1) geometry optimizations, (2) electronic correlation simulations of systems with varying numerical difficulty, and (3) the square-planar copper complexes [CuCl$_4$]$^{2-}$, [Cu(NH$_3$)$_4$]$^{2+}$, and [Cu(H$_2$O)$_4$]$^{2+}$. 
Throughout the numerical investigations, the $S$-diagnostic is compared to other SRCC diagnostic procedures, that is, the $T_1$, $D_1$, and $D_2$ diagnostics as well as different indices of multi-determinantal and multi-reference character in coupled-cluster theory. 
Our numerical investigations show that the $S$-diagnostic outperforms the $T_1$, $D_1$, and $D_2$ diagnostics and is comparable to the indices of multi-determinantal and multi-reference character in coupled-cluster theory in their individual fields of applicability.
The experiments investigating the performance of the $S$-diagnostic for geometry optimizations using SRCC reveal that the $S$-diagnostic correlates well with different error measures at a high level of statistical relevance. 
The experiments investigating the performance of the $S$-diagnostic for electronic correlation simulations show that the $S$-diagnostic correctly predicts strong multi-reference regimes. 
The $S$-diagnostic moreover correctly detects the successful SRCC computations for [CuCl$_4$]$^{2-}$, [Cu(NH$_3$)$_4$]$^{2+}$, and [Cu(H$_2$O)$_4$]$^{2+}$, which have been known to be misdiagnosed by $T_1$ and $D_1$ diagnostics in the past. 
This shows that the $S$-diagnostic is a promising candidate for an \emph{a posteriori} diagnostic for SRCC calculations.
\end{abstract}

\section{Introduction}

While the underlying mathematical theory of the quantum many-body problem is, on a fundamental level, well described, the governing equation, namely, the {\it many-body Schr\"odinger equation}, remains numerically intractable for a large number of particles.
In fact, the many-body Schr\"odinger equation poses one of today's hardest numerical challenges, mainly
due to the exponential growth in computational complexity with the number of electrons.
Over the past century, numerous numerical approximation techniques of various levels of cost and accuracy have been developed in order to overcome this \emph{curse of dimensionality}.
Arguably, the most successful approaches are based on coupled-cluster (CC) theory~\cite{Bartlett2007},
which defines a cost-efficient hierarchy of increasingly accurate methods, including
the so-called \emph{gold standard} of quantum chemistry---the coupled-cluster singles-and-doubles with
perturbative triples (CCSD(T))~\cite{Raghavachari1989} model.

Despite the great success of CC theory, its reliability is not yet fully quantifiable.
More precisely, aside from a few heuristically derived results, there exists no universally reliable \emph{diagnostic} that indicates if the computational result is to be trusted.
This shortcoming is most apparent in the regime of transition metal compounds and molecular bond breaking/making processes, systems dominated by {\it strong nondynamic electron-correlation effects}, where several methods based on CC theory tend to fail along with all other numerically tractable approaches.

Therefore, {\it a posteriori} error diagnostics are urgently needed in the field. 
Until very recently, the diagnostic approaches available were limited to the so-called $T_1$ (also called $\tau_1$)~\cite{Lee1989,Lee1989b}, $D_1$, and $D_2$ diagnostic~\cite{Janssen1998,Nielsen1999}.
Despite clear numerical evidence that diagnostics based on the single excitation amplitudes, such as the  $T_1$ and $D_1$ diagnostics, do not provide reliable indicators~\cite{giner2018interplay}, they are commonly used due to the lack of alternatives. 
Recently, an alternative set of multi-reference indices was introduced which provided a number of  {\it a posteriori}  diagnostic tools~\cite{bartlett2020index} christened the indices of multi-determinantal and multi-reference character in coupled-cluster theory. These tools are highly descriptive and able to determine different molecular scenarios in which CC theory may fail. 

We provide an alternative error diagnostic that is based on assumptions employed in the mathematical analysis CC theory.
%The mathematical results referred to are within the realm of \emph{nonlinear functional analysis}.
More precisely, our diagnostic is derived from the mathematical analysis of CC theory that provides sufficient conditions for a locally unique and quasi-optimal solution to the CC working equations. 
%The said conditions are also believed to be necessary for a CC solution to be well-behaved. \SK{There are systems with small HOMO-LUMO gap where CC seems to do just fine.}
Central to our derivation is the \emph{strong monotonicity property}, as introduced by Schneider~\cite{schneider2009analysis}, which is eponymous for our {\it S-diagnostic}. 
Compared to the recently suggested nine indices that describe the multi-determinantal and multi-reference character in coupled-cluster
theory~\cite{bartlett2020index}, the $S$-diagnostic is a diagnostic technique that can be applied to multi-determinantal and multi-reference scenarios alike.
%\MCS{In what sense general? Do you mean better?}
%\FF{No, I meant that we propose one diagnostic for all situations, whereas Bartlett suggest nine indices, all applicable to difference situations}
%\TBP{Maybe spell this out explicitly in the text?}
We complement our theoretical derivation of the $S$-diagnostic with numerical simulations scrutinizing its validity for different geometry optimizations, and electronic correlation computations for systems of varying numerical difficulty for single reference coupled-cluster methods.

The rest of the article is structured as follows. We begin with a brief review of CC theory, followed by a short summary of the mathematical results derived in previous works which lay the mathematical foundation for the proposed $S$-diagnostics. Then, we derive the main result, i.e., the $S$-diagnostic which is subsequently numerically scrutinized.

\section{Theory}

\subsection{Brief overview of coupled-cluster theory}
In CC theory the wave function is parametrized by the exponential $\vert \psi \rangle = e^{\hat T } \vert \phi_0 \rangle$. 
Here, $\vert \phi_0 \rangle$ is the reference determinant defining the occupied spin orbitals, and $\hat T = \sum_\mu t_\mu \hat X_\mu= \sum_k \hat T_k$ is a cluster operator, where $\hat T_k$  excites $k=1,\dots, N$ electrons---$k$ is the excitation rank of a given $\hat T_k$---from the occupied spin orbitals into the virtual spin-orbitals. 
All possible excited determinants can be expressed as $\vert\mu \rangle = \hat X_\mu \vert \phi_0 \rangle$ for some multi-index $\mu$ labeling occupied and virtual spin-orbitals. 
The governing equations determining amplitudes $(t_\mu)$, and therewith also the CC energy $\mathcal{E}_{\rm CC}(t)$, are given by $f_\mathrm{CC}(t)=0$, where
\begin{equation}
\label{eq:cceqs_poly}
\left\lbrace
\begin{aligned}
\mathcal{E}_{\rm CC}(t) &= \langle \phi_0 \vert e^{-\hat T}\hat He^{\hat T} \vert \phi_0 \rangle\\
(f_\mathrm{CC}(t))_\mu&= \langle \mu \vert e^{-\hat T}\hat He^{\hat T} \vert \phi_0 \rangle .
\end{aligned}
\right.
\end{equation}
%
%Note that in the case of untruncated cluster operators, solving the CC equations is equivalent to finding the solution of the Schr\"odinger equation, i.e., $\hat H \psi_* = E_0 \psi_*$.
%Based on Arponen's \emph{bivariational principle}~\cite{Arponen1982,Arponen1983}, 
More compactly,~\cref{eq:cceqs_poly} can be expressed using the CC Lagrangian~\cite{Helgaker1988,kvaal2022three}
\begin{equation}
    \mathcal L(t,z) = \mathcal{E}_\mathrm{CC}(t) + \sum_\mu z_\mu (f_\mathrm{CC}(t))_\mu
    = \langle  \phi_0 \vert (\hat I + \hat Z^\dagger) e^{-\hat T}\hat He^{\hat T} \vert \phi_0 \rangle  ,
%    \mathcal L(t,z) 
%    =  \langle \phi_0 \vert e^{-\hat T}\hat He^{\hat T} \vert \phi_0 \rangle + \sum_\mu z_\mu \langle \mu \vert e^{-\hat T}\hat He^{\hat T} \vert \phi_0 \rangle
%    = \langle (\hat I + \hat Z) \phi_0 \vert e^{-\hat T}\hat He^{\hat T} \vert \phi_0 \rangle  ,
\end{equation}
where $(z_\mu)$ are the Lagrange multipliers which are the {\it dual variables} corresponding to $(t_\mu)$. 
In the extended CC theory~\cite{Arponen1983,Arponen1987,Arponen1987b} (ECC), which will be used to introduce additional information to our $S$-diagnostic, the Lagrangian is replaced with the more general energy expression
%the so-called bivariate quotient can be expressed by 
\begin{equation}
    \mathcal E_\mathrm{ECC}(t,\lambda)
    = \langle \phi_0 \vert e^{\hat{\Lambda}^\dagger} e^{-\hat T} \hat H e^{\hat T} \vert \phi_0 \rangle .
\end{equation}
Consequently, through the substitution $e^{\hat{\Lambda}} = \hat I + \hat Z$, we have $\mathcal E_\mathrm{ECC}(t,\lambda) =  \mathcal L(t,z)$. 
The stationarity condition can then be formulated as $F_\mathrm{ECC}=0$, where
\begin{equation}
\label{eq:flipped_gradient}
F_\mathrm{ECC}=(\partial_\Lambda \mathcal E_\mathrm{ECC}, \partial_T\mathcal E_\mathrm{ECC})
\end{equation}
is the so-called flipped gradient~\cite{laestadius2018analysis}.
The partial derivatives with respect to the amplitudes in Eq.~\eqref{eq:flipped_gradient} are given by
\begin{equation}
\label{eq:Feq}
\begin{aligned}
    \partial_{\lambda_\mu} \mathcal E_\mathrm{ECC}	 
    &= \langle  \mu \vert e^{ \hat \Lambda^\dagger} e^{-\hat T} \hat H e^{\hat T} \vert \phi_0\rangle, \\
	\partial_{t_\mu}\mathcal E_\mathrm{ECC} 
	&= \langle \phi_0 \vert e^{\hat \Lambda^\dagger}[ e^{-\hat T} \hat H e^{\hat T},\hat X_\mu] \vert
	\phi_0 \rangle. 
\end{aligned}
\end{equation}

Since the number of determinants, and therewith the size of the system's governing equations, suffer in general from the \emph{curse of dimensionality} (i.e., it grows exponentially fast with the number of electrons), restrictions are necessary to ensure the system's numerical tractability.
In practice this is achieved by restricting excitations to excited determinants that correspond to a preselected index set---this is referred to as \emph{truncation}.
Such excitation hierarchies are commonly denoted as singles (S), doubles (D), etc.
We emphasize that the CC working equations, as a system of polynomial equations, typically have a large number of roots, and the corresponding landscape of said roots is highly non-trivial~\cite{faulstich2022coupled}.
Consequently, different limit processes have to be considered separately and carefully studied.
More precisely, the convergence of the CC roots with respect to the basis set discretization, i.e., convergence towards the complete basis set limit, is a fundamentally different limit process from the convergence with respect to the coupled-cluster truncations.
Hence, it is important to note that the convergence of the numerical root finding procedure for the truncated standard (or extended) CC equations does not by itself imply convergence of the roots to the corresponding exact roots.
In other words, whether the discrete roots converge to the exact roots cannot simply be assumed to be true in general. 
%\TBP{You might want to consider a more precise description. The JCTC readership is mostly chemists and some will likely not have ever considered the fact that the CC equations could have more than one solution and that the solutions do not necessarily correspond to what is wanted.}

Before proceeding further with the derivation of the $S$-diagnostic, we wish to provide the reader with a more precise description of the underlying mathematical conventions in coupled-cluster theory. 
We first emphasize the distinction between the cluster amplitudes and the corresponding wave function. Although related, these objects live in different spaces which we shall elaborate on subsequently. 
First, the wave function object $|\psi \rangle = e^{\hat T} |\phi_0\rangle$ lives in the $N$-particle Hilbert space of square-integrable functions, i.e., $L^2 = \{ \psi : \int |\psi|^2 < +\infty \}$, with finite kinetic energy.\footnote{
Mathematically, assuming finite kinetic energy is important for the well-posedness of the Schr\"odinger equation. In a ``weak'' formulation this is given by (here for simplicity leaving out spin degrees of freedom)
\[
\int_{\mathbb R^{3N}} |\nabla \psi(\mathbf r_1, \dots,\mathbf r_N )|^2 \mathrm{d} \mathbf r_1 \dots \mathrm{d} \mathbf r_N < +\infty.
\]
In the mathematical literature this can be summarized by $\psi \in H^1$ (Sobolev space)~\cite{aubin2011applied}. This extra constraint of finite kinetic energy is moreover important for the ``continuous'' (i.e., infinite dimensional) formulation of coupled-cluster~\cite{rohwedder2013continuous}. 
}
We remind the reader of the notation for the $L^2$-inner product $\langle \psi'|\psi\rangle $, and its induced norm $\Vert \psi \Vert^2_{L^2} = \langle \psi |  \psi\rangle$.
Second, operators that act on the wave function, e.g., the Hamiltonian or excitation operators. In this case, we can introduce a norm expression for the operator inherited from the function space it is defined on. For example, let $O$ be an operator defined on $L^2$ then we define the $L^2$ operator norm 
\begin{equation}
\Vert O \Vert_{L^2}
=
\sup \{ \Vert O\Psi \Vert_{L^2} ~:~  \Vert \Psi \Vert_{L^2} = 1~\text{and}~ \Psi \in L^2\}.
\end{equation}
Note that this reduces to the conventional matrix norm in the finite dimensional case.
Third, the CC amplitudes $(t_\mu)$ live in the Hilbert space of finite square summable sequences denoted the $\ell^2$-space. This space is equipped with the $\ell^2$-inner product~\cite{aubin2011applied}, i.e., let $x=(x_\mu)$ and $y=(y_\mu)$ be two finite sequences, the $\ell^2$-inner product is defined as 
$$ \langle x, y \rangle_{\ell^2} = \sum_\mu x_\mu y_\mu,$$ which induces the norm $\Vert x \Vert^2_{\ell^2} = \langle x, x\rangle_{\ell^2}$.
Henceforth, we shall denote the full amplitude space by $\mathcal{V}$, and the truncated amplitude space, e.g., the space only containing single and double amplitudes, by $\mathcal{V}^{(d)}$; note that we use ``$d$'' in this section to distinguish objects that are subject to imposed truncations.
We moreover follow the mathematically convenient convention that uses a generic constant $C$, independent of the main variables under consideration,  for the different estimations performed subsequently.

Having laid down the basic definitions, we now recall a result that gives insight into the root convergence of CC theory which can be established using a basic existence result of nonlinear analysis~\cite{schneider2009analysis,rohwedder2013continuous,rohwedder2013error,laestadius2018analysis,Laestadius2019}. 
To state this result, we need two more definitions. 

First, \emph{local strong monotonicity}. 
Let $t,t',t_*$ be cluster amplitudes with $\hat T$, $\hat T'$ and $\hat T_*$ denoting the corresponding cluster operators. 
Set
\begin{equation}  
\label{eq:delta-def}
\Delta(t,t') 
= 
%\sum_\mu (f_\mathrm{CC}(t) - f_\mathrm{CC}(t'))_\mu (t - t')_\mu
%=
\langle f_\mathrm{CC}(t) - f_\mathrm{CC}(t'), t - t' \rangle_{\ell^2},
\end{equation}
and furthermore $\Delta \hat T = \hat T- \hat T'$. 
Then the CC function $f_\mathrm{CC}$ is said to be locally strongly monotone at $t_*$ if for some $r>0$, $\gamma>0$ and all $t,t'$ within the distance $r$ of $t_*$
\begin{equation}  
\label{eq:Start1}
\Delta(t,t') \geq \gamma \Vert t-t' \Vert_{\ell^2}^2.
\end{equation}

Second, \emph{local Lipschitz continuity}.  
The function $f_\mathrm{CC}$ is said to be locally Lipschitz continuous at $t_*$ with Lipschitz constant $L > 0$ if 
\begin{equation}
    \Vert f_\mathrm{CC}(t) - f_\mathrm{CC}(t') \Vert_{\ell^2} \leq L \Vert t - t' \Vert_{\ell^2}
\end{equation} 
for any $t,t'$ in a ball around $t_*$. Note that in the finite-dimensional case, $f_\mathrm{CC}$ is indeed locally Lipschitz since it is continuously differentiable. 

With these definitions at hand, we can recall the following result~\cite{rohwedder2013error,schneider2009analysis}:\\
\textit{
Let $f_\mathrm{CC}(t_*)=0$ and assume that $f_\mathrm{CC}$ is locally strongly monotone with constant $\gamma>0$ at $t_*$. 
Furthermore, let $\mathcal{V}^{(d)} \subset \mathcal{V}$ be a truncated amplitude space with $P_d$ being the orthogonal projector onto $\mathcal{V}^{(d)}$ and $f_d$ a  discretization of $f_\mathrm{CC}$, i.e., $f_d = P_d f_\mathrm{CC}$. 
Then, the following holds:
\begin{enumerate}
\item \label{bullet1} $t_*$ is locally unique, i.e., $| \psi_* \rangle= e^{T_*} | \phi_0\rangle$ is the only solution within a sufficiently small ball.  
\item \label{bullet2} There exists a sufficiently large $d_0$, such that for any $d>d_0$, there exists $t_{*}^{(d)}\in\mathcal{V}^{(d)}$ such that $f_d(t_{*}^{(d)})=0$. 
This root is unique in a ball centered at $t_*$ (for some radius $r$) and we have quasi-optimality of the discrete solution $t_{*}^{(d)}$ i.e.
\begin{equation} 
\label{eq:CCerror-est1}
\|t_{*}^{(d)} - t_*\|_{\ell^2} 
\leq 
\frac{L}{\gamma} \mathrm{dist}(t_*, \mathcal{V}^{(d)}), 
\end{equation}
where $\mathrm{dist}(v,\mathcal{V}^{(d)})$ is the distance from $v$ to $\mathcal{V}^{(d)}$ measured using the norm of $\mathcal{V}$, and $L$ is the Lipschitz constant of $f_\mathrm{CC}$ at $t_*$.
\item 
For $d>d_0$, the discrete equations $f_d(t_*^{(d)}) = 0$ have locally unique solutions, and in addition to the error estimate~\eqref{eq:CCerror-est1}, we have the quadratic energy error bound
\begin{equation}  
\label{eq:error-est2}
|\mathcal{E}_\mathrm{CC}(t_{*}^{(d)}) - E_0| 
\leq C_1 \Vert t_* - t_{*}^{(d)} \Vert_{\ell^2} ^2 + C_2  \Vert t_* - t_{*}^{(d)} \Vert_{\ell^2}  \Vert z_* - z_{*}^{(d)} \Vert_{\ell^2} , 
\end{equation}
where $E_0$ is the ground state energy and $z_*$ and $z_{*}^{(d)}$ are the Lagrange multiplier of the exact and truncated equations, respectively. 
The constants $C_1,C_2>0$ arise in general from particular continuity considerations~\cite{rohwedder2013error,rohwedder2013continuous} which shall not be further characterized here.  
%\TBP{$E_0$ has not been defined and $C_1$, $C_2$ should be described.}
\end{enumerate}
}

We emphasize that the result in Ref.~\citenum{rohwedder2013continuous} is more elaborate since it is concerned with an infinite dimensional amplitude space. Here, we implicitly assume a finite-dimensional amplitude space which allows us to present the result in the simpler but equivalent $\ell^2$-topology.  
This result ensures that the CC method is convergent as the truncated cluster amplitude space $\mathcal{V}^{(d)}$ approaches the untruncated limit and that the energy converges quadratically.
Note also that the above results hold for conventional single-reference CC theory but can be formulated for the extended CC theory as well with some slight modifications (see Ref.~\citenum{laestadius2018analysis}).

\subsection{Strong Monotonicity Property}

The local strong monotonicity at a root of the CC equations is the mathematical basis of what we deem as a reliable solution obtained from a truncated CC calculation since this implies a unique solution of $f_d=0$ for sufficiently good approximate $\mathcal V^{(d)}$ as well as a quadratic convergence in the energy. Moreover, it follows that the Jacobian of both $f_\mathrm{CC}$ and $f_d$ are non-degenerate at such a solution.
In order to derive the $S$-diagnostic, we start with a brief review of the proof presented in the literature~\cite{Laestadius2019,rohwedder2013continuous,laestadius2018analysis} while  making some slight improvements. 
%We emphasize that in the subsequent derivation, we refrain from specifying the the norm topologies used, that is, either the Sobolev norm or the $L^2$-norm. 
%For more details on the respective topologies during the subsequent derivation, we refer the reader to Ref.~\citenum{Laestadius2019}; the final result will be formulated in the $L^2$-topology. 
We subsequently establish Eq.~\eqref{eq:Start1} up to second order in $\Vert t -t'\Vert_{\ell^2}$ under certain assumptions.
%\TBP{Again, define the norm (if it is intended to be different than above where the $\ell^2$ is not used).}
To that end, we define
\begin{equation}
\Delta_2(t_*;t,t') 
= 
\langle \Delta \hat T \phi_0| e^{-\hat T_*} (\hat H-E_0) e^{\hat T_*} |\Delta \hat T \phi_0 \rangle .
\end{equation} 

Now, suppose that $f_\mathrm{CC}(t_*) =0$, then by Taylor expansion we find 
\begin{equation} 
\label{eq:arg2-3}
\Delta(t,t')
= 
\Delta_2(t_*;t,t') + \mathcal O((\Delta t)^3).
\end{equation}
For the proof, we refer the reader to Ref.~\citenum{rohwedder2013error}.
We emphasize that the core idea of the proof is a Taylor expansion of $e^{\hat T}$ and $e^{\hat T'}$ around $\hat T_*$, which does not require $t_*$ itself to be small, rather, the assumption is that we are within a certain neighborhood of $t_*$. 

By Eq.~\eqref{eq:arg2-3}, if $\Delta_2(t_*;t,t') \geq \gamma' \Vert t-t' \Vert_{\ell^2}^2$ with $\gamma'>0$ for $t$, $t'$ within distance $r'$ from $t_*$, then it is possible to find $r>0$ such that Eq.~\eqref{eq:Start1} is true for $\gamma\in (0,\gamma']$ for $t$, $t'$ at distance at most $r \leq r'$) from $t_*$. 
Consequently, we wish to establish
\begin{equation}
\label{eq:Start2}
\Delta_2(t_*;t,t') \geq \gamma' \Vert t-t' \Vert_{\ell^2}^2
\end{equation}
for some $\gamma'=\gamma'(t_*)>0$.

%We state some assumptions on the systems we consider our analysis applicable to.  
We subsequently assume that 
%the Hamiltonian $\hat{H}$ is self-adjoint \MCS{Maybe we dont need to explicity state this?} 
%(\HEK{In practice, when computing excited states/the spectral gap within EOM-CC theory, the cluster operators are truncated and the EOM-CC Hamiltonian is non-Hermitian. Does this have to be considered at all?})
the ground state of $\hat{H}$ exists and is non-degenerate, and that $\hat{H}$ admits a spectral gap $\gamma_*>0$ between the ground-state energy $E_0$ and the rest of the spectrum of $\hat H$, i.e.,
\begin{equation} 
\label{eq:spec-gap}
\gamma_* 
= 
\inf \left\{ \frac{\langle \psi | \hat H - E_0 |\psi  \rangle}{\langle\psi |\psi\rangle} : |\psi\rangle  \perp | \psi_*\rangle  \right\} > 0 .
\end{equation}
Moreover, we assume that the reference $|\phi_0\rangle$ is such that it is not orthogonal to the ground-state wave function. 
%[Maybe enough to just say that we are interested in molecular Hamiltonians.]
With these assumptions, we can establish an improved version of Lemma~11 in Ref.\citenum{laestadius2018analysis} and Lemma~3.5 in Ref.~\citenum{rohwedder2013error}:  
If $t_*$ solves $f_\mathrm{CC}(t_*)=0$ then for $| \psi\rangle  \perp |\phi_0\rangle  $ 
\begin{equation} 
\label{eq:PP}
\langle  \psi |\hat H-E_0 | \psi \rangle 
\geq 
\gamma_*^\mathrm{eff} \Vert \psi \Vert_{L^2}^2,
\end{equation}
where 
\begin{equation}
\label{eq:effctiveGap}
\gamma_*^\mathrm{eff} 
= 
\frac{\gamma_*}{\Vert e^{T_*} \phi_0   \Vert_{L^2}^2} .
\end{equation}
For the sake of clarity, we here display the used $L^2$-norm.
%\TBP{Define the $L^2$ norm.}
Equation~\ref{eq:PP} can be obtained as follows: Let $\mathcal{P}_*$ be the projection onto the solution $| \psi_* \rangle$, then
\begin{equation}\label{eq:est1}
\begin{aligned}
\langle  \psi | (\hat H-E_0) \psi \rangle
&= \langle  \psi - \mathcal{P}_*(\psi)| \hat H-E_0 |\psi - \mathcal{P}_*(\psi) \rangle\\
& \geq \gamma_* \Vert \psi - \mathcal{P}_*(\psi)\Vert_{L^2}^2 \\
& = \Vert\psi \Vert_{L^2}^2 -2 \mathrm{Re} \langle \psi | \mathcal{P}_*(\psi) \rangle +   \Vert \mathcal{P}_*(\psi) \Vert_{L^2}^2 \\
&=  \Vert\psi \Vert_{L^2}^2   - \frac{\vert \langle \psi | \psi_* \rangle\vert^2}
{\Vert \psi_*\Vert_{L^2}^2} \\
& = \Vert\psi \Vert_{L^2}^2   - \frac{\vert \langle \psi | (e^{T_*} - I)\phi_0 \rangle\vert^2}
{\Vert \psi_*\Vert_{L^2}^2}.
\end{aligned}
\end{equation}
We next note that
\begin{equation*}
\begin{aligned}
\frac{\vert \langle \psi | (e^{T_*} - I)\phi_0 \rangle\vert^2}{\Vert \psi_*\Vert_{L^2}^2} 
\leq 
\Vert \psi\Vert_{L^2}^2 \frac{\Vert (e^{T_*} - I)\phi_0 \Vert_{L^2}^2}{\Vert \psi_*\Vert_{L^2}^2}  
=
 \Vert \psi\Vert_{L^2}^2 \left( 1 - \frac{1}{\Vert \psi_*\Vert_{L^2}^2}\right),
\end{aligned}
\end{equation*}
which inserted in Eq.~\eqref{eq:est1} yields the desired result.

With the inequality~\eqref{eq:PP} at hand, we can establish the inequality
\begin{equation}
\begin{aligned}
\label{eq:Delta2cc}
\Delta_2(t_*;t,t')
&=
\langle \Delta \hat T \phi_0| e^{-\hat T_*} (\hat H-E_0) e^{\hat T_*} |\Delta \hat T \phi_0 \rangle\\
&\geq  
\gamma_*^\mathrm{eff}   
\Vert \Delta \hat T \phi_0 \Vert_{L^2}^2 - C \mathcal G_\mathrm{CC}(T_*) \Vert \Delta \hat T \phi_0 \Vert_{H^1}^2, 
\end{aligned}
\end{equation}
where $C$ is a constant that depends on the Hamiltonian $\hat H$ and 
\begin{equation} 
\label{eq:gCC}
\mathcal G_\mathrm{CC}(T_*) 
=  
\Vert e^{\hat T_*}-I \Vert_{L^2} + \Vert e^{-\hat T_*^\dagger} -I \Vert_{L^2} \Vert e^{\hat T_*} \Vert_{L^2} .
\end{equation}
%and $\Vert \cdot \Vert_{H^1}$, $\Vert \cdot \Vert_{\mathcal B}$ are the $H^1$-norm (that also measures the kinetic energy) and the operator norm on $H^1$, respectively. 
Equation~\eqref{eq:Delta2cc} follows from the definition of $\Delta_2$ and that 
\begin{align*}
    \Delta_2 & = \langle \Delta \hat T \phi_0| \hat H -E_0 | \Delta \hat T \phi_0 \rangle +
        \langle \Delta \hat T \phi_0 | \hat H -E_0 |  (e^{\hat T_*} -I) \Delta \hat T \phi_0    \rangle  \\
    &+ \langle (e^{-\hat T_*^\dagger} -I) \Delta  \hat T \phi_0| \hat H - E_0 |  e^{\hat T_*} \Delta \hat T \phi_0 \rangle ,
\end{align*}    
then, using that $\hat H$ is a bounded operator in the energy norm and the estimate in Eq.~\eqref{eq:PP}, we obtain the desired result in Eq.~\eqref{eq:Delta2cc}.
%\TBP{The previous sentence needs rewriting}
% We emphasize that this in general only holds in the more restrictive $H^1$ topology, i.e.,  the $H^1$ Sobolev norm that also measures the kinetic energy~\cite{aubin2011applied,Laestadius2019}.%\TBP{citation needed here?}
% Note that with some further assumptions on $\hat H$, namely a G\aa rding  inequality~\cite{renardy2006introduction,yserentant2010regularity}, Eq.~\eqref{eq:Delta2cc} can be turned into
% \begin{align*}
%     \Delta_2(t_*;t,t') \geq \eta (\gamma_*^\mathrm{eff} - C \mathcal G_\mathrm{CC}(T_*)) \Vert \Delta \hat T \phi_0  \Vert_{H^1}^2, \quad \eta >0,
% \end{align*}
% %\TBP{What is $\eta$?}
% which establishes strong monotonicity in the $H^1$ topology.
% \AL{This is not true, and also not needed}

\section{The $S$-Diagnostic}

Given the reformulation of the strong monotonicity property in~\cref{eq:Delta2cc}, we consider a computation to be successful if the results fulfill~\cref{eq:Delta2cc}. 
In order to derive an {\it a posterioi} diagnostic, we reformulate this inequality in a way that yields a function that indicates a reliable computation. To ensure the tractability of the said function we introduce the following approximations, which will yield diagnostic functions of different flavors, later referred to as $S_1$, $S_2$, and $S_3$, respectively. 

\paragraph{Approximation (i)} 
A first-order Taylor approximation of $e^{\hat T_*}$ and the trivial operator norm inequality~\footnote{\[ \Vert \hat T_*  \phi_0 \Vert_{L^2} \leq \Vert \hat T_*\Vert_{L^2} \Vert \phi_0 \Vert_{L^2} = \Vert \hat T_*\Vert_{L^2}\]} yields
\begin{equation}
\label{eq:approximation1}
\Vert e^{\hat T_*} \phi_0 \Vert_{L^2}^2  \approx 1 + \Vert \hat T_* \Vert_{L^2}^2.
\end{equation}

\paragraph{Approximation (ii)} For $\mathcal G_\mathrm{CC}$ we use (i) and make the approximation (linearization) 
\begin{equation}
\label{eq:approximation3}
\mathcal G_\mathrm{CC}(T) \approx  2 \Vert \hat T \Vert_{L^2} . % \leq C \Vert t \Vert_{\ell^2}.
\end{equation}

\paragraph{Approximation (iii)} As outlined in Ref.~\citenum{Laestadius2019}, we can moreover estimate
\begin{equation}
(1 + \Vert \hat Z_* \Vert_{L^2}^2)^{1/2} \approx ( 1 + \Vert \hat T_* \Vert_{L^2}^2)^{-1/2}.
\end{equation}
%\MCS{Make it clear that THIS is the assumption.}
This approximation follows by equating the bra and ket wave functions (in the bivariational formulation)  $e^{-\hat T_*^\dagger}(\hat I + \hat Z_*) | \phi_0\rangle = \Vert e^{\hat T_*}  \phi_0 \Vert_{L^2}^{-2} e^{\hat T_*} | \phi_0\rangle$ with $e^{\hat \Lambda_*} = \hat I + \hat Z_*$ and approximating
\begin{equation}
\label{eq:approximation4}
e^{-\hat T_*^\dagger}(\hat I + \hat Z_*) | \phi_0 \rangle  
\approx 
(\hat I + \hat Z_*) | \phi_0\rangle .
\end{equation}
With these approximations at hand, we can derive three variants of the $S$-diagnostic that we shall investigate subsequently.

\subsection{The $S_1$-diagnostic}
Starting from Eq.~\eqref{eq:Delta2cc}, we first note that we are considering the finite-dimensional case, and therefore there exists a constant $C>0$ such that 
\begin{equation} 
\Delta_2(t_*;t,t') 
\geq  
\left[ \gamma_*^\mathrm{eff} - C \mathcal G_\mathrm{CC}(\hat T_*) \right] \Vert \Delta \hat T \phi_0 \Vert_{L^2}^2
\end{equation}
holds.
Next, we employ Approximation (ii) in the definition of $\mathcal G_\mathrm{CC}(\hat T_*)$, and combine Approximation (i) with the definition of $\gamma_*^\mathrm{eff}$ in Eq.~\eqref{eq:effctiveGap}, i.e., 
\begin{equation}
\gamma_*^\mathrm{eff}
\approx
\frac{\gamma_*}{1 + \Vert \hat T_*  \Vert_{L^2 }^2}.
\end{equation}
This yields
\begin{equation}
\gamma_*^\mathrm{eff}  - C \mathcal G_\mathrm{CC}(\hat T_*)  
\approx  
\frac{\gamma_*}{1 + \Vert \hat T_*  \Vert_{L^2 }^2} -2 C \Vert \hat T_* \Vert_{L^2}  .
\end{equation}
Requiring that this expression is positive, we 
%can rearrange it to obtain $(1 +  \Vert \hat T_* \Vert_{L^2}^2) \Vert \hat T_* \Vert_{L^2}  < \frac{\gamma_*}{2C}$, or, equivalently,
obtain the success condition
\begin{equation}
\label{eq:FirstDiag}
\frac{1}{2} > \frac{C}{\gamma_*} (1+ \Vert \hat T_* \Vert_{L^2}^2)\Vert \hat T_* \Vert_{L^2}.
\end{equation}

% \FF{Note that I changes this, previously it said  
% \begin{align}
% (1 +  \Vert \hat T_* \Vert_\mathcal{B}^2) \Vert \hat T_* \Vert_\mathcal{B}   
% < 2C\gamma_* ,
% \end{align}
% which I don't see...
% }
%let depend on the number of electrons and optimized numerically in Section~\ref{sec:N-dep}. 

\subsection{The $S_2$-diagnostic}
By applying Approximation (iii) 
%\TBP{The list of approximations above only has 3 entries, not 4\ldots}
to Eq.~\eqref{eq:FirstDiag}, we obtain a success condition that involves the Lagrange multipliers, namely, 
\begin{equation}
\label{S2_cond}
\frac{1}{2} > \frac{C}{\gamma_*} \frac{ \Vert \hat T_* \Vert_{L^2}^2}{(1+ \Vert \hat Z_* \Vert_{L^2}^2)}.
\end{equation}
%\MCS{Why is it beneficial to introduce an approximation here? In the S1 version you could readily compute everything. Does this yield a more stringent condition?}

% \FF{Note that I changes this, previously it said
% \begin{align*}
%     \frac{\Vert \hat T_* \Vert_\mathcal{B}}{1 + \Vert \hat Z_* \Vert_\mathcal{B}^2}   
%   < 2C \gamma_*  .
% \end{align*}
% }

\subsection{The $S_3$-diagnostic}

% \SK{It is not clear why we go to ECC theory at all. How can this help, when we are doing standard CC? I think reviewers will not understand this. Can we avoid introducing ECC at all? This will make the paper clearer, and give us a higher impact in the end.} 
% \AL{The argument is for std CC, but uses ECC. In this way was both ``bra'' and ``ket'' are given a more even treatment. An extra sentence has been added. }

To obtain a diagnostic that includes the Lagrangian multipliers without making use of Approximation (iii), we shall follow the argument on strong monotonicity of the extended CC function $F_\mathrm{ECC}$ defined above.
Note that although we use the extended CC formalism in this section (i.e., where the Lagrange multipliers are treated as a second set of cluster amplitudes), the derived diagnostic is for the conventional single reference CC method.
Subsequently, we assume that truncations of $\hat T$ and $\hat \Lambda$ are 
at the same rank, i.e., the truncated scheme follows as described above for $\mathcal V^{(d)}$ but takes the \emph{double} form $\mathcal V^{(d)} \times \mathcal V^{(d)}$ and with $P_d$ being the orthogonal projector onto $\mathcal{V}_d\times\mathcal{V}_d$.  
Note that this aligns with practical implementations of the CC Lagrangian.
For brevity, let $\hat U=(\hat T,\hat \Lambda)$, $\hat U_* = (\hat T_*,\hat \Lambda_*)$ and $\hat U_*^{(d)} = (\hat T_*^{(d)}, \hat \Lambda_*^{(d)})$ and 
furthermore, set $F_d$ to be the Galerkin discretization of $F_\mathrm{ECC}$, i.e., $F_d(\hat U^{(d)}) = P_d F_\mathrm{ECC}(\hat U^{(d)})$.

In Ref.~\citenum{laestadius2018analysis} strong monotonicity of $F_\mathrm{ECC}$ was established under certain assumptions, and 
recently generalized to a class of extended CC theories~\cite{KvaalLaestadiusBodenstein2020}. 
We, therefore, refer the reader to these references for the full proof, here we shall only address those parts relevant to our diagnostics. 

Similarly to the CC case, local strong monotonicity of $F_\mathrm{ECC}$ holds if
\begin{equation}
\Delta^\mathrm{ECC} 
:= 
\langle F_\mathrm{ECC}(u) - F_\mathrm{ECC}(u'), u-u' \rangle \geq \gamma \Vert u - u' \Vert^2
\end{equation}
for some positive constant $\gamma$. 
Note that we here extended the notation such that $u$ carries both the primal-, and dual variables. 
Furthermore, we let $\Delta^\mathrm{ECC}$ up to second order in $\Vert u-u'\Vert$ be denoted
$\Delta_2^\mathrm{ECC}$ and similarly to Eq.~\eqref{eq:Delta2cc} we have 
\begin{equation} 
\label{eq:monoECC}
\Delta_2^\mathrm{ECC}(u_*;u,u') 
\geq   
\gamma_*^\mathrm{eff} \Vert \Delta \hat U \phi_0 \Vert_{L^2}^2 - C \mathcal G_\mathrm{ECC}(\hat U_*) \Vert \Delta \hat U \phi_0 \Vert_{H^1}^2,
\end{equation}
where 
\begin{align*}
\mathcal G_{\mathrm{ECC}}(\hat U) 
&=
\mathcal G_{\mathrm{ECC}}(\hat T,\hat \Lambda)  \\
&= \Vert e^{-\hat T^\dagger} e^{\hat \Lambda} \Vert_{{L^2}} \Vert  e^{\hat T} -I  \Vert_{{L^2}}    +  \Vert e^{-\hat T^\dagger}e^{\hat \Lambda} -I \Vert_{{L^2}}  + K \Vert \phi_0 \Vert_{H^1} \Vert e^{-\hat T^\dagger} \Vert_{{L^2}} 
\Vert e^{\hat T} \Vert_{{L^2}} \Vert e^{\hat \Lambda} - I \Vert_{{L^2}} .
\end{align*}
for some positive constant $K$

Starting from Eq.~\eqref{eq:monoECC}, we note again that since we are considering finite-dimensional Hilbert spaces, there exists a constant $C>0$ such that 
\begin{equation}
\Delta_2^\mathrm{ECC} (u_*;u,u') 
\geq  
\left[ \gamma_*^\mathrm{eff}  - C \mathcal G_\mathrm{ECC}(\hat U_*)  \right] \Vert \Delta \hat U \phi_0 \Vert_{L^2}^2 . 
\end{equation}
We next employ a variation of Approximation (iii): 
For $\mathcal G_\mathrm{ECC}$ we make the substitution $e^{\hat \Lambda} = \hat I + \hat Z$ and approximate %\SK{I feel this step requires at least some explanation...} \AL{Taylor}
with a low-order Taylor expansion 
\begin{equation}
\tilde{\mathcal{G}}_\mathrm{ECC}(\hat T,\hat Z) 
:= 
\mathcal G_\mathrm{ECC}(\hat T,\hat \Lambda(\hat Z)) 
\approx 
C ( \Vert \hat T \Vert_{{L^2}} + \Vert \hat Z \Vert_{{L^2}} ) .
\end{equation}
Hence, we arrive at the approximation (and we remind the reader that $C$ is used as a generic constant)
% there exists \SK{Is this strictly true, or assuming something on the accuracy of the approximations?} a constant $C>0$ such that \AL{(we next approximate... )}
\begin{equation}
\gamma_*^\mathrm{eff} - C \mathcal G_\mathrm{ECC}(\hat U_*)  
\approx  
\frac{\gamma_*}{1 + \Vert \hat T_*  \Vert_{L^2}^2} - C (\Vert \hat T_* \Vert_{L^2} + \Vert \hat Z_* \Vert_{L^2} ) .
\end{equation}
Requiring that this expression is positive, we find the condition
\begin{equation}
\label{S3_cond}
1
>
\frac{C}{\gamma_*}
\left(
(1 +  \Vert \hat T_* \Vert_{L^2}^2)( \Vert \hat T_* \Vert_{{L^2}}  +  \Vert \hat Z_*\Vert_{{L^2}})
\right)
\approx
\frac{C}{\gamma_*}
\left(
(1 +  \Vert \hat T_* \Vert_{{L^2}}^2) \Vert \hat T_* \Vert_{{L^2}}  
   +  \frac{\Vert \hat Z_*\Vert_{{L^2}}}{1 +\Vert \hat Z_*\Vert_{{L^2}} }
\right).
\end{equation}
%where $C$ is a constant to be optimized numerically. 
% \FF{I changed this since I could not rederive it, before it said
% \begin{align*}
%   (1 +  \Vert \hat T_* \Vert_\mathcal{B}^2)( \Vert \hat T_* \Vert_{\mathcal B}  +  \Vert \hat Z_*\Vert_{\mathcal B}) 
%   \approx
%   (1 +  \Vert \hat T_* \Vert_{\mathcal B}^2) \Vert \hat T_* \Vert_{\mathcal B}  
%   +  \frac{\Vert \hat Z_*\Vert_{\mathcal B}}{1 +\Vert \hat Z_*\Vert_{\mathcal B} }
%   < C \gamma_* ,
% \end{align*}}

\subsection{Approximation of operator norms using singular values}
%
%[Used refs. here are Beran~\cite{Beran2004} and Hackbusch.]

The above-derived success conditions~\cref{eq:FirstDiag,S2_cond,S3_cond} can be directly implemented, however, the quantities involved will depend on the system size. This can be illustrated by simply placing copies of a molecular system at a distance such that they are at least numerically non-interacting. In that case, the reliability of the overall CC calculation is determined by the CC calculations of a single copy, yet, the operator norm of the cluster operator $\Vert \hat T \Vert_{L^2}$ will scale with the system's size. 

To remedy this serious difficulty, we consider an alternative interpretation of the cluster operators~\cite{Beran2004}: The CCSD method yields a set of single amplitudes $(t_i^a)$ forming a matrix in $\mathbb{R}^{n_{\rm occ} \times n_{\rm virt}}$ and a set of double amplitudes $(t_{ij}^{ab})$ forming a fourth-order tensor in $\mathbb{R}^{n_{\rm occ} \times n_{\rm occ} \times n_{\rm virt} \times n_{\rm virt}}$. 
As outlined in Ref.~\citenum{Beran2004}, in order to capture the pair correlation we reshape the fourth-order tensor that describes the double amplitudes as a matrix in $\mathbb{R}^{n_{\rm occ}^2 \times n_{\rm virt}^2}$, an operation that is also known as ``matricization''. 
In order to include pair correlations captured by the single amplitudes, we can moreover extend $(t_{ij}^{ab})$ to also include products of single amplitudes which yields $M_T \in \mathbb R^{n_{\rm occ}^2 \times n_{\rm virt}^2}$ with matrix elements
\begin{equation}
[M_T]_{ij,ab} = t_{ij}^{ab}+ (t_i^at_j^b-t_i^bt_j^a).
\end{equation}
The singular value decomposition then yields 
\begin{equation}
M_T = U_T \Sigma_T V_T^\top,
\end{equation}
where $U_T, V_T$ are real orthogonal matrix and $\Sigma_T$ is diagonal. We will subsequently use the spectral norm, i.e., the largest singular value, here denoted as $ \sigma(M_T)$ to approximate the operator norm, i.e., 
\begin{equation}
\Vert \hat T \Vert_{L^2} \approx  \sigma(M_T) =: \msvt
\end{equation}
and similarly for the dual variable $z$. 
Incorporating this into the success conditions~\cref{eq:FirstDiag,S2_cond,S3_cond} yields the $S$-diagnostic functions used in this article
\begin{subequations}
\begin{align}
& S_1(t) := \frac {1} {\gamma_*} (1 +  \msvt^2) \msvt ,    \label{eq:MainS-diag-1}\\
& S_2(t,z) := \frac {1} {\gamma_*} \frac{ \msvt }{1 +  \msvz^2}
\label{eq:MainS-Diag-1},\\
&S_3(t,z) 
:= \frac {1} {\gamma_*} \Big[ (1 +  \msvt^2) \msvt  +   \frac{\msvz }{1 + \msvz^2}   \Big].
\label{eq:MainS-Diag-3} 
\end{align}
\end{subequations}

For computed cluster amplitudes $(t)$ and Lagrange multipliers $(z)$, the above functions will yield an $S$-diagnostic value. In the following numerical investigations, we will first investigate the statistical correlation between the computed $S$-diagnostic value and different measures of error. Second, we will investigate a quantitative bound for the $S$-diagnostic value beyond which the computations may not be reliable and further benchmark computations with more profound error classifications are advised.

\section{Numerical simulations}

%\begin{sidewaystable}
%\input{table}
%\end{sidewaystable}

In this section, we numerically scrutinize the proposed $S$-diagnostic procedures derived in the previous sections. 
All simulations are performed using the Python-based Simulations of Chemistry Framework (PySCF)~\cite{sun2018pyscf,sun2020recent,sun2015libcint}. 
First, we perform geometry optimizations on a medium-sized set of molecules comprising all molecules that were investigated in Refs.~\citenum{Lee1989,Nielsen1999,Janssen1998} to test the $T_1$, $D_1$, and $D_2$ diagnostic, respectively. 
With this data at hand, we can propose an initial set of values, beyond which our diagnostic suggests interpreting the computational results with caution and if possible benchmarking with additional methods that allow for a more profound error classification.
Second, we target small model systems whose multi-reference character can be controlled by simple geometric changes.
Third, we numerically investigate transition metal complexes that have been shown to be misdiagnosed by the $T_1$ and $D_1$ diagnostics~\cite{giner2018interplay}.   

\subsection{Correlation in Geometry Optimization}
\label{sec:GeomOpt}

In order to quantify the correlation between the $S$-diagnostics and the error of the CC method, we numerically investigate the Spearman correlation~\cite{myers2013research} between the error of {\it in silico} geometry optimizations and the corresponding value of the $S$-diagnostics. 
We perform geometry optimizations for 34 small to medium-sized molecules that were previously studied in relation to CC error classifications~\cite{Lee1989,Janssen1998,Nielsen1999}, see~\cref{tab:Molecules}.

\begin{table}[]
    \centering
    \begin{tabular}{ccccccc}
H$_2$N$_2$ & HOF & C$_2$H$_2$ & ClOH & H$_2$S & O$_3$ & FNO\\
ClNO & C$_2$ & C$_3$ & CO & HNO & HNC & HOF \\
Cl$_2$O & P$_2$ & N$_2$H$_2$ & HCN & CH$_2$NH & N$_2$ & C$_2$H$_4$ \\
F$_2$ & HOCl & Cl$_2$ & HF & CH$_4$ & H$_2$O & SiH$_4$ \\
NH$_3$ & HCl & CO$_2$ & BeO & H$_2$CO & CH$_2$ 
    \end{tabular}
    \caption{Molecules which are used in the geometry optimization presented here.}
    \label{tab:Molecules}
\end{table}

The calculations are performed using the CC method with singles and doubles (CCSD) using the cc-pVDZ basis set provided by PySCF; the geometry optimization is performed using the interface to PyBerny~\cite{pyberny}.
The numerically obtained results are compared with experimentally measured geometries of the considered systems in their gas phases extracted from the \emph{Computational Chemistry Comparison and Benchmark Data Base} (CCCBDB)~\cite{johnsoncomputational}. Since the computed atomic positions cannot be directly compared, we introduce the bond-length matrix that describes the pairwise distance between the atoms in the molecular compound. This bond-length matrix can be directly compared with the bond-length matrix provided by CCCBDB if we label and order the atoms of the corresponding system accordingly. We investigate the correlation between the $S$-diagnostics and three possible error characterizations obtained from the absolute difference of the bond-length matrices denoted $D^{\rm (diff)}$:
\begin{itemize}
    \item[i)] The maximal absolute error ($\Delta r_{\rm abs}^{\rm (max)}$): the maximal absolute deviation of the numerically obtained bond-length matrix to the experimentally obtained bond-length matrix, i.e., 
    \[
    \Delta r_{\rm abs}^{\rm (max)}= \max_{i,j}D^{\rm (diff)}_{ij}
    \]
    \item[ii)] The averaged absolute error ($\Delta r_{\rm abs}^{\rm (ave)}$): the averaged absolute deviation of the numerically obtained bond-length matrix to the experimentally obtained bond-length matrix, i.e., 
    \[
    \Delta r_{\rm abs}^{\rm (ave)} = \frac{\sum_{i,j} D^{\rm (diff)}_{i,j}}{N_{\rm atoms}}
    \]
    \item[iii)] The averaged relative error ($\Delta r_{\rm rel}^{\rm (ave)}$): the averaged relative deviation of the numerically obtained bond-length matrix to the experimentally obtained bond-length matrix, i.e.,
    \[
    \Delta r_{\rm rel}^{\rm (ave)} = \frac{\sum_{i,j} D^{\rm (diff)}_{i,j}}{N_{\rm atoms}\;\max_{i,j}D^{\rm (diff)}_{ij}}
    \]
\end{itemize}

Computing the Spearman correlation between the errors listed above and the proposed $S$-diagnostics, we find that all suggested $S$-diagnostics correlate well with all the error measures suggested, i.e., we consistently find correlations of $r_{\rm sp} > 0.5$ with $p < 0.0008$, see~\cref{tab:Spearman_corr_CC}. 
The largest correlation is observed between the maximal absolute error ($\Delta r_{\rm abs}^{\rm (max)}$) and $S_{2}$ and $S_3$ where we find a correlation of $r_{\rm sp}= 0.58476$ with $p=0.00018$. 
For comparison, we compute the Spearman correlation for the previously suggested $T_1$, $D_1$, and $D_2$ diagnostic in~\cref{tab:Spearman_corr_CC}. 
We find that $T_1$, and $D_1$, are uncorrelated to all the errors that we investigate here, i.e., $r_{\rm sp}<0.3$ with $p>0.1$. 
The $D_2$ diagnostic~\cite{Nielsen1999} shows a correlation with the averaged absolute error ($\Delta r_{\rm abs}^{\rm (ave)}$) and the averaged relative error ($\Delta r_{\rm rel}^{\rm (ave)}$), where we find a correlation of $r_{\rm sp}= 0.36886$ with $p=0.026847$ and $r_{\rm sp}=0.35496$ with $p= 0.033646$, respectively. 
We moreover compare the $S$-diagnostics with the recently suggested indices of multi-determinantal and multi-reference character in CC theory~\cite{bartlett2020index}. 
We find that similar to the $S$-diagnostics, the EEN index~\cite{bartlett2020index} correlates well with the maximal absolute error ($\Delta r_{\rm abs}^{\rm (max)}$); 
we observe a correlation of $r_{\rm sp}= 0.53572$ with $p=0.000759$.

Directly comparing the Spearman correlation of the $S$-diagnostics with the $T_1$, $D_1$, and $D_2$ diagnostic, we see that the $S$-diagnostics have a significantly higher correlation than the heuristically motivated diagnostics $T_1$, $D_1$ and $D_2$ diagnostics while exhibiting a higher level of stochastic significance. 
Comparing the Spearman correlation of the $S$-diagnostics with the indices of multi-determinantal and multi-reference character in CC theory, we find that the $S$-diagnostic and EEN show similar correlation with the maximal absolute error ($\Delta r_{\rm abs}^{\rm (max)}$) with a comparable level of stochastic significance. 

\begin{table}
    \centering
    \begin{tabular}{c|ccc}
         &  $\Delta r_{\rm abs}^{\rm (max)}$ & $\Delta r_{\rm abs}^{\rm (ave)}$ & $\Delta r_{\rm rel}^{\rm (ave)}$\\
         \hline
        $S_1$  & (0.57910, 0.000215) & (0.57761, 0.000225) & (0.53668, 0.000740)\\
        $S_2$ & (0.58476, 0.000180) & (0.58584, 0.000174) & (0.54543, 0.000581) \\
        $S_3$ & (0.58476, 0.000180) & (0.58584, 0.000174) & (0.54543, 0.000581)\\
        \hline
        $T_1$ & (0.03025, 0.863034) & (0.00489, 0.977416) & (0.02265, 0.895674) \\
        $D_1$ & (0.27675, 0.107522) & (-0.00541, 0.975040) & (-0.02034, 0.906294) \\
        $D_2$ & (0.16974, 0.329625) & (0.36886, 0.026847) & (0.35496, 0.033646) \\
        \hline
        EEN & (0.53572, 0.000759) & (0.42059, 0.010643) & (0.33694, 0.044488) \\
    \end{tabular}
    \caption{Spearman correlation between the $S$-diagnostic computed form CCSD amplitudes and different errors in geometry optimization. The pair-entries show the rank correlation and the corresponding $p$-value, i.e., ($r_{\rm sp}$, $p$).}
    \label{tab:Spearman_corr_CC}
\end{table}

% For comparison we also list the Spearman correlation between the errors in geometry optimization and the existing CC diagnostics

% \begin{table}[]
%     \centering
%     \begin{tabular}{c|ccc}
%     &  $\Delta r_{\rm abs}^{\rm (max)}$ & $\Delta r_{\rm abs}^{\rm (ave)}$ & $\Delta r_{\rm rel}^{\rm (ave)}$\\
%          \hline
%     $T_1$ & (0.03025, 0.863034) & (0.00489, 0.977416) & (0.02265, 0.895674) \\
%     $D_1$ & (0.27675, 0.107522) & (-0.00541, 0.975040) & (-0.02034, 0.906294) \\
%     $D_2$ & (0.16974, 0.329625) & (0.36886, 0.026847) & (0.35496, 0.033646) \\
%     \hline
%     EEN & (0.53572, 0.000759) & (0.42059, 0.010643) & (0.33694, 0.044488) \\
%     \end{tabular}
%     \caption{Spearman correlation between the existing CC diagnostics computed for CCSD}
%     \label{tab:other_diags}
% \end{table}

\begin{figure}
\centering
\begin{subfigure}[b]{0.48\textwidth}
    \centering
    \includegraphics[width=\textwidth]{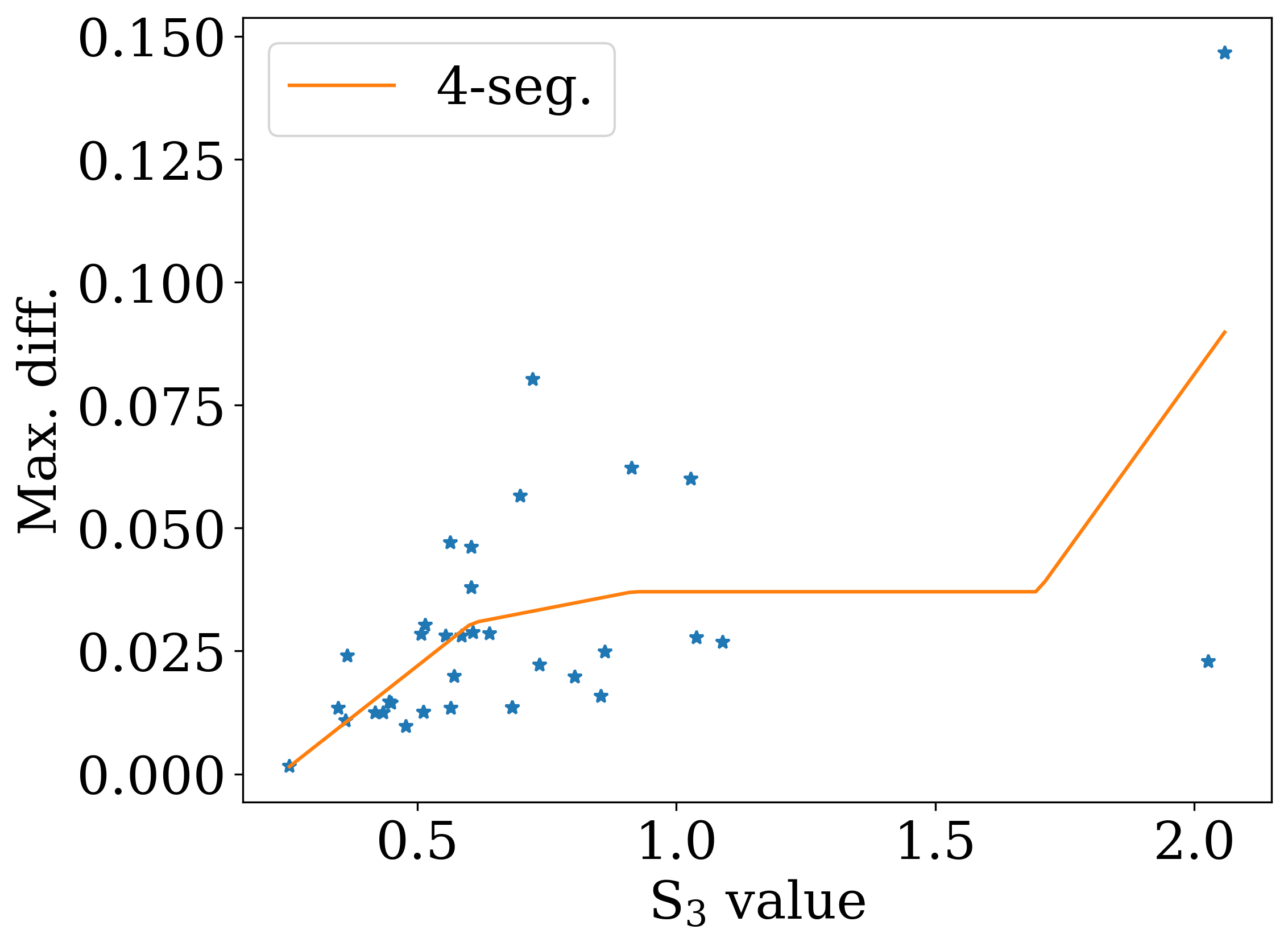}
    \caption{}
    \label{fig:smp_function}
\end{subfigure}
\hfill
\begin{subfigure}[b]{0.48\textwidth}
    \centering
    \includegraphics[width=\textwidth]{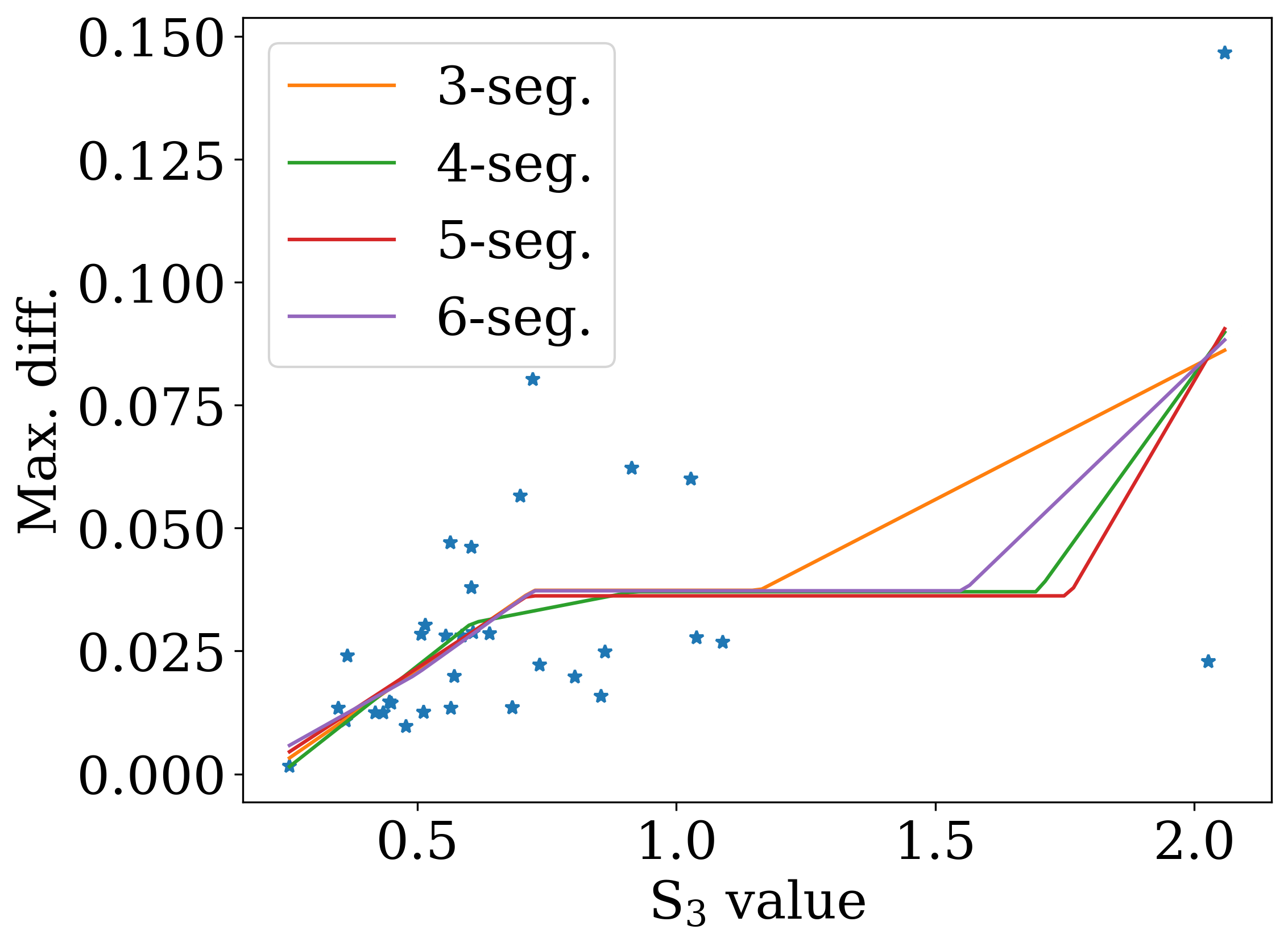}
    \caption{}
    \label{fig:smp_functions}
\end{subfigure}
\caption{\label{fig:smp}
The maximal error in geometry optimization as a function of the $S_2$ value.
(\subref{fig:smp_function}) 
The orange line corresponds to a piecewise linear fit to the data using four segments for the piecewise linear function.
(\subref{fig:smp_functions}) Piecewise linear fits to the data with a varying number of segments.
}
\end{figure}

In order to obtain an approximate trusted region suggested by the $S$-diagnostics, we require a descriptive function that maps the value obtained from the $S$-diagnostic to the error in geometry. 
Since the Spearman correlation describes a monotone relation between the quantities, we may not assume that this relation is linear.
Unfortunately, the Spearman correlation does not indicate the type of relation that connects the two measured quantities. 
We, therefore, perform a piecewise linear fit to the data obtained in this simulation, see~\cref{fig:smp}.
We here allow for four segments which are optimized to reach the best approximation by means of a piecewise linear and monotone function. 
We emphasize that larger numbers of segments yield similar approximations, see~\cref{fig:smp_functions}.
Performing this piecewise linear fit, we observe that the function is constant on some segments.
Based on the data distribution, we conclude that this constant behavior is artificial and caused by the test set not being sufficiently versatile.
In particular, no quantitative conclusions can be drawn from the piecewise linear fit function for values $S_3 > 1$. Therefore, from the geometry optimizations performed here, we can merely conjecture to raise a concern about the validity of CC calculations performed for values of the $S$-diagnostics $v_{\rm crit}^{(3)} \geq 1$. Based on the piecewise linear fit, $S_3=1$ corresponds to an error larger than $0.035\;a_0$. A larger statistical investigation with a larger variety of molecules and basis set discretizations is delegated to future works.
We emphasize that this first estimation of $v_{\rm crit}$ is particularly pessimistic since the data set is not versatile enough to give a precise estimation of $v_{\rm crit}$. Indeed, in the subsequently performed simulations, we show a more refined estimation of $v_{\rm crit}$ that reveals $v_{\rm crit}^{(2)}= 1.9$ and $v_{\rm crit}^{(3)}= 1.8$, for $S_2$, and $S_3$, respectively.
%
% The data obtained by our simulations yields a plateau of the piecewise linear fit function in the region for the values obtained from the $S$-diagnostics that are between 1.0 and 2.0. 
% Based on the data distribution, we conclude that this is an artificial plateau caused by the test set not being sufficiently versatile. \SK{I find this analysis a little strange. Looking at the plots, there is no data below $S_3 \approx 1$, except for two points that must be considered outliers here. Then how can we justify the piecewise linear fit? Moreover, we basically draw the conclusion that $S < 1$ is okay, based on data points that are all okay?? It may be that I do not understand our analysis, but then I feel it should be properly explained.} 
% A larger statistical investigation with a larger variety of molecules and basis set discretizations is delegated to future works.
% We therefore suggest that for values of the $S$-diagnostics larger than $v_{\rm crit} \geq 1$ to raise concern about the validity of the performed CC calculations. 
% Based on the piecewise linear fit, the values of the $S$-diagnostics of one corresponds to an error larger than $0.035\;a_0$.    
%\FF{This needs to be adjusted in case we want to advertise a different SMP variant.}

Aside from CC-based simulations, we can also perform MP2 simulations, and use the obtained doubles amplitudes to compute the $S$-diagnostics. 
We find that the proposed $S$-diagnostics correlate similarly well with MP2 based calculations as it does for CCSD, see~\cref{tab:Spearman_corr_MP}

\begin{table}
    \centering
    \begin{tabular}{c|c|c|c}
         &  $\Delta r_{\rm abs}^{\rm (max)}$ & $\Delta r_{\rm abs}^{\rm (ave)}$ & $\Delta r_{\rm rel}^{\rm (ave)}$\\
         \hline
        $S_1$ & (0.55992, 0.000384) & (0.54569, 0.000577) & (0.49781, 0.002006)\\
       $S_2$ & (0.56687, 0.000313) & (0.54801, 0.000541) & (0.49858, 0.001968)\\
       $S_3$ & (0.55992, 0.000384) & (0.54569, 0.000577) & (0.49781, 0.002006)\\
    \end{tabular}
    \caption{Spearman correlation between $S$-diagnostics computed from MP2 doubles amplitudes and different errors in geometry optimization.}
    \label{tab:Spearman_corr_MP}
\end{table}

\subsection{Model Systems}
%\TBP{Again, another name than "transition state" is needed}

In this section we investigate the use of the proposed $S$-diagnostics for four model systems whose multi-reference character can be controlled by simple geometric change: (1) twisting ethylene, (2) the C$_{2\text{v}}$ insertion pathway for BeH$_2$ (Be $\cdots$ H$_2$)~\cite{purvis1983c2v}, (3) the H$_4$ model (transition from square to linear geometry)~\cite{jankowski1980applicability} (4) the H$_4$ model (symmetrically disturbed on a circle); the computations are performed in cc-pVTZ basis.  

\subsubsection{Twisting ethylene}
\begin{figure}
    \centering
    \includegraphics[width = 0.7\textwidth]{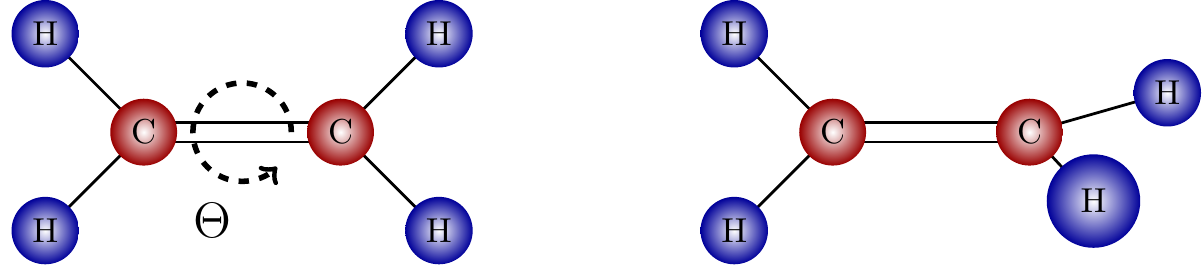}
    \caption{Depiction of the ethylene (C$_2$H$_4$) model with twist angle $\Theta$.}
    \label{figc2h4_mod}
\end{figure}

We begin by numerically investigating the proposed $S$-diagnostics for ethylene twisted around the carbon--carbon bond, see~\cref{figc2h4_mod}.

\begin{figure}
\centering
\begin{subfigure}[b]{0.48\textwidth}
    \centering
    \includegraphics[width=\textwidth]{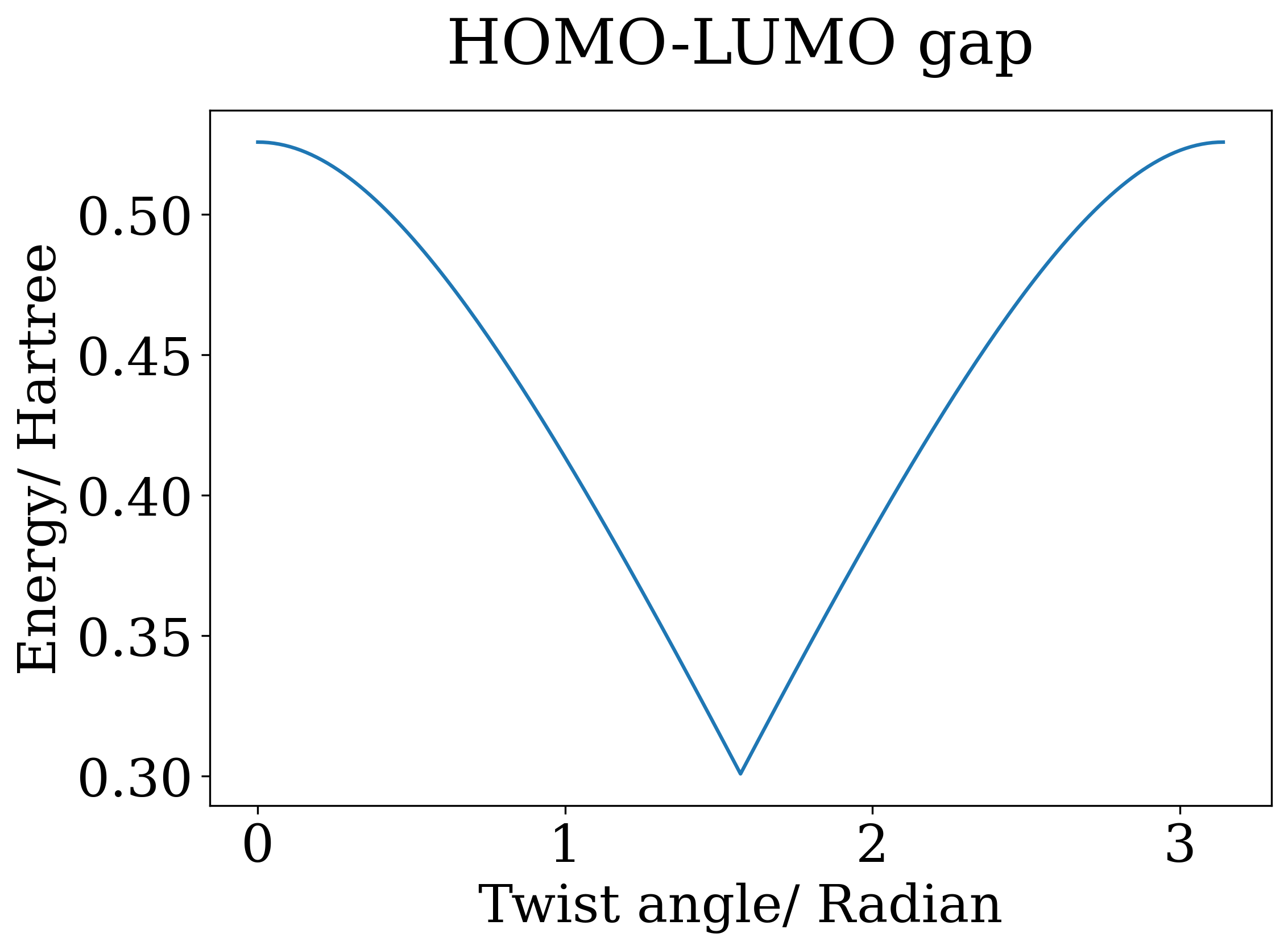}
    \caption{}
    \label{fig:homo_lumo_c2h4}
\end{subfigure}
\hfill
\begin{subfigure}[b]{0.48\textwidth}
    \centering
    \includegraphics[width=\textwidth]{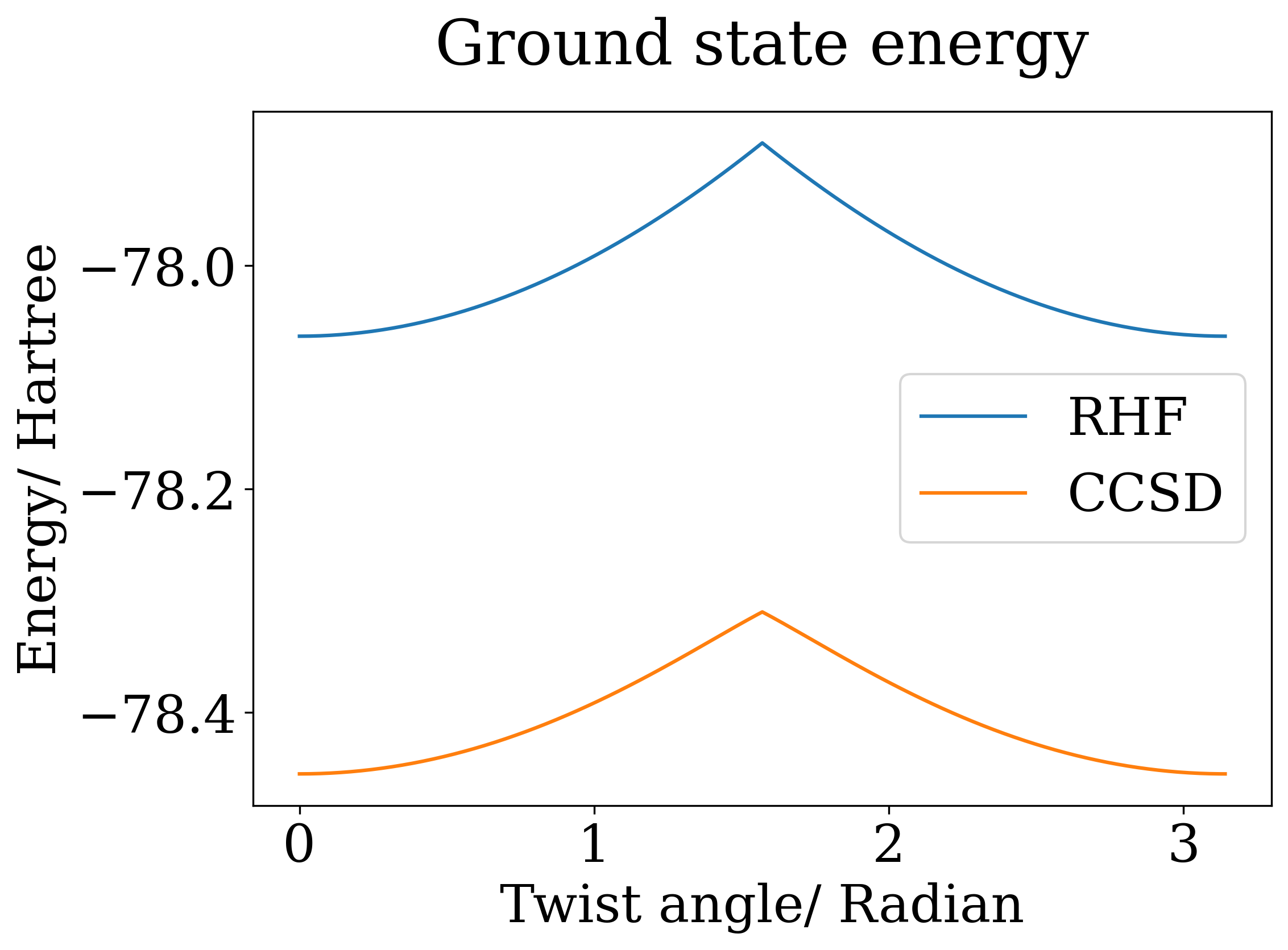}
    \caption{}
    \label{fig:energies_c2h4}
\end{subfigure}
\caption{\label{fig:c2h4}
(\subref{fig:homo_lumo_c2h4}) 
HOMO-LUMO gap of C$_2$H$_4$ as a function of the twist angle 
(\subref{fig:energies_c2h4}) RHF and RCCSD energies of C$_2$H$_4$ as a function of the twist angle
}
\end{figure}

At a twist angle of 90°, this system shows a strong multi-reference character. 
This can be seen as follows: At the equilibrium geometry, i.e., in a planar geometry, the two carbon $p$ orbitals are perpendicular to the molecular plane form bonding $\pi$ and anti-bonding $\pi^*$ orbitals. 
In this geometry, the ground state doubly occupies the $\pi$-orbital. 
As we twist around the carbon--carbon bond, the overlap between the two $p$ orbitals decreases and becomes zero at 90°. 
Therefore, at 90° the $\pi$ and $\pi^*$ orbitals become degenerate and the $\pi$-bond is broken.
This (quasi) degeneracy can also be observed numerically by computing the HOMO-LUMO gap as a function of the twist angle, see~\cref{fig:homo_lumo_c2h4}.
Computing the corresponding ground state energy as a function of the twist angle, we observe the characteristic energy cusp at exactly 90°, see~\cref{fig:energies_c2h4}. 

\begin{figure}
\centering
\begin{subfigure}[b]{0.48\textwidth}
    \centering
    \includegraphics[width=\textwidth]{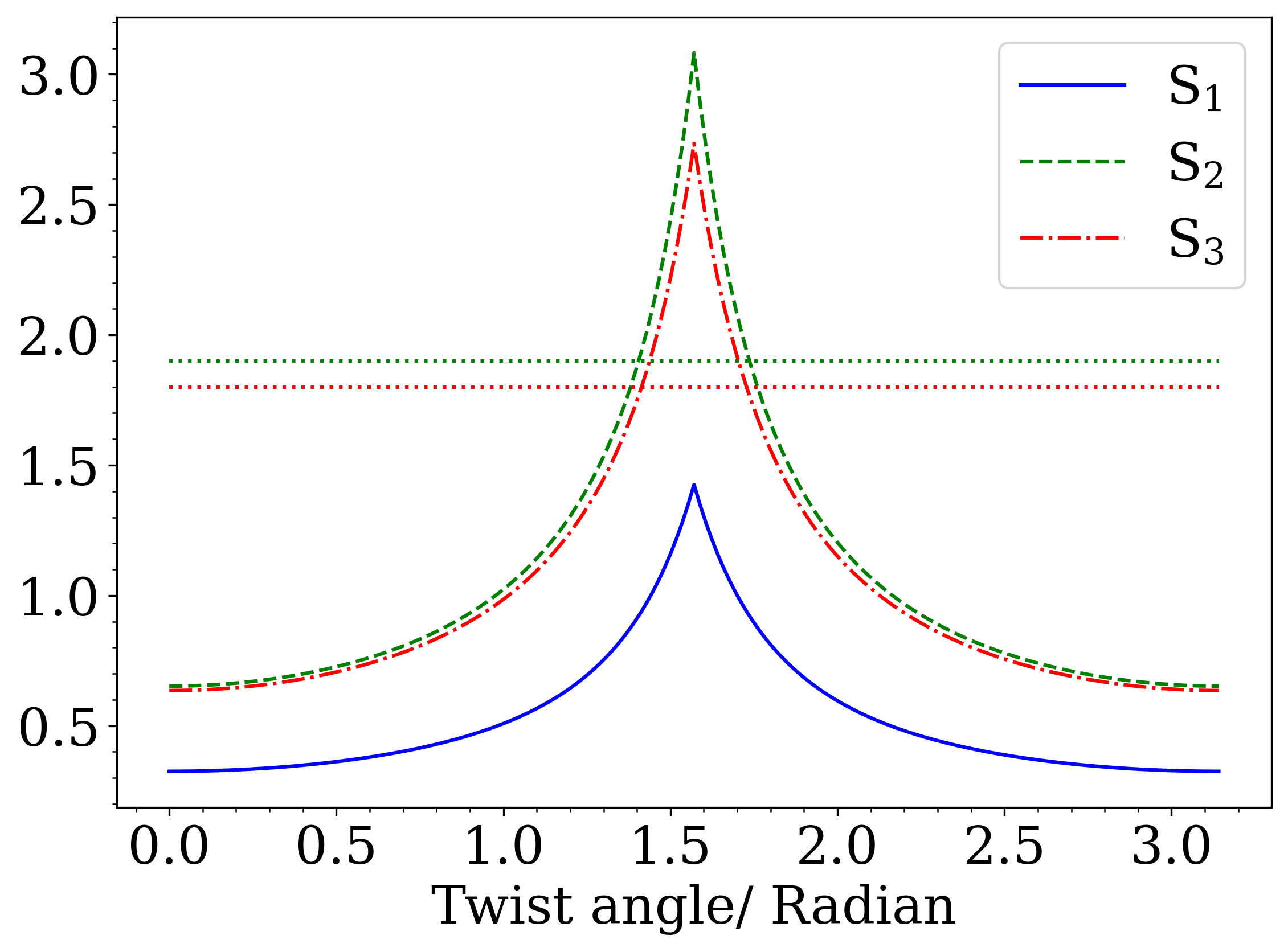}
    \caption{}
    \label{fig:smp_c2h4}
\end{subfigure}
\hfill
\begin{subfigure}[b]{0.48\textwidth}
    \centering
    \includegraphics[width=\textwidth]{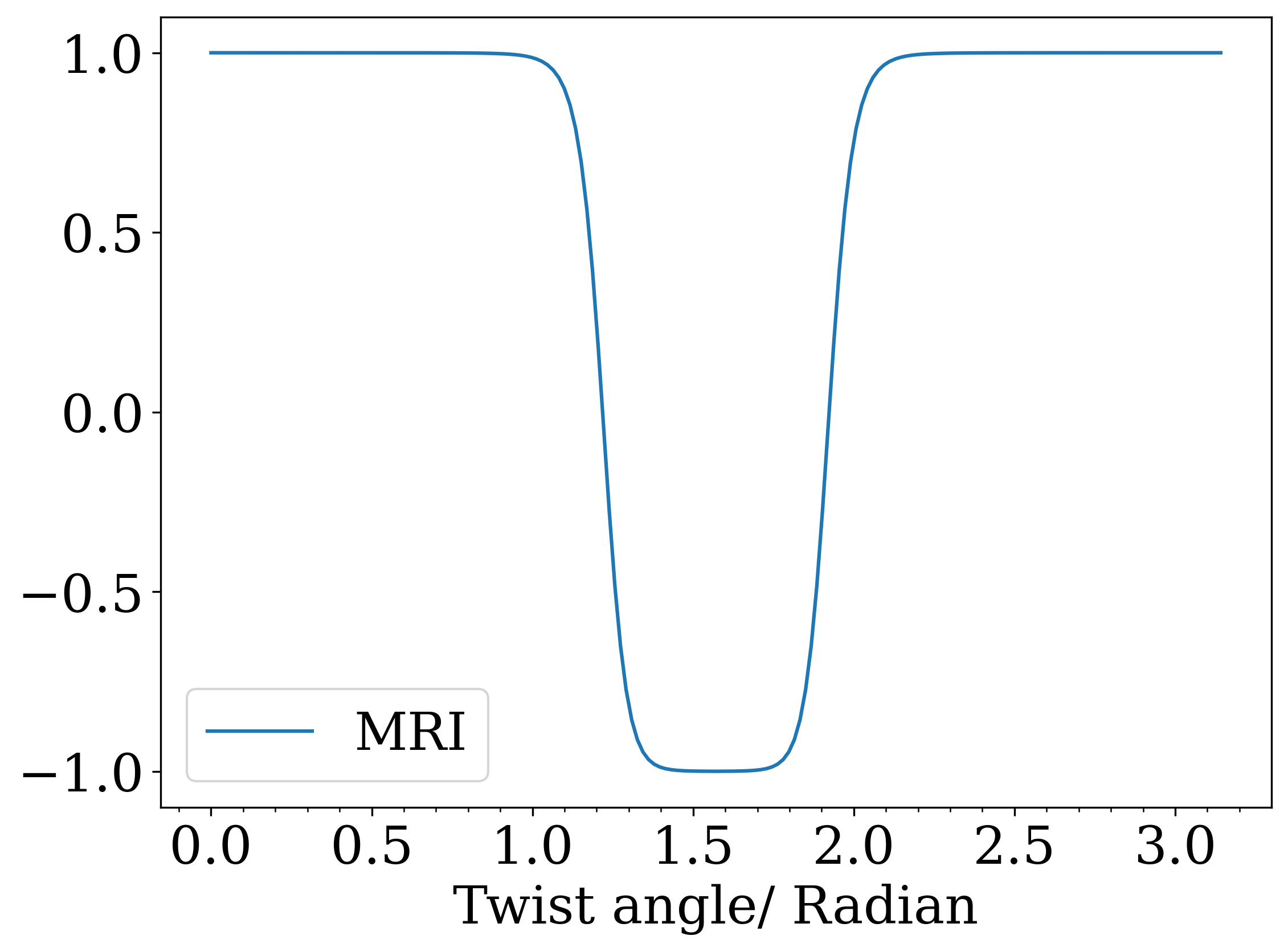}
    \caption{}
    \label{fig:mri_c2h4}
\end{subfigure}
\caption{\label{fig:diag_c2h4}
(\subref{fig:smp_c2h4}) 
The proposed $S$-diagnostics of C$_2$H$_4$ as a function of the twist angle, the dotted green and red horizontal lines correspond to $v_{\rm crit}^{(2)}= 1.9$ and $v_{\rm crit}^{(3)}= 1.8$, respectively.
(\subref{fig:mri_c2h4}) The
previously suggested MRI of C$_2$H$_4$ as a function of the twist angle
%\TBP{The left panel (a) needs more ticks on the y-axis! What are the horizontal lines?}
}
\end{figure}

Due to the quasi degeneracy around 90°, we compare the $S$-diagnostic with the MRI index suggested in Ref.~\citenum{bartlett2020index}. 
We clearly see the indication of the quasi degeneracy in the MRI index, see~\cref{fig:mri_c2h4}. 
The $S$-diagnostic also indicates the problematic region around 90°. 
By numerical comparison, we find that a cut-off value of $v_{\rm crit}^{(2)}= 1.9$ and $v_{\rm crit}^{(3)}= 1.8$ for $S_2$ and $S_3$, respectively, indicates the same region of quasi degeneracy as the MRI index.
%\TBP{How were the cutoff values determined? Plus, the superscripts 1/2 and 1 seem awkward}

\subsubsection{C$_{2v}$ insertion pathway for BeH$_2$}

\begin{figure}
    \centering
    \includegraphics[width = 0.6\textwidth]{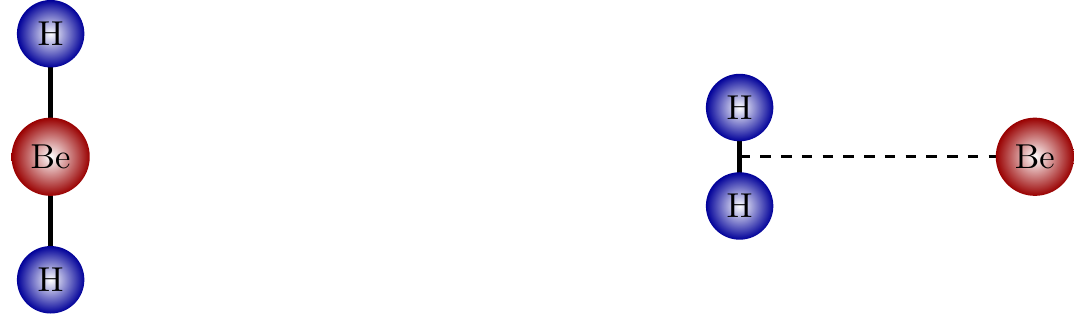}
    \caption{Depiction of the C$_{2v}$ insertion pathway for BeH$_2$.}
    \label{fig:BeH2_mod}
\end{figure}

Next we shall investigate the C$_{2v}$ insertion pathway for BeH$_2$ (Be $\cdots$ H$_2$)~\cite{purvis1983c2v}.
The model represents an insertion of the Be atom into the H$_2$ molecule.
The transformation coordinate connects the non-interacting subsystems (Be + H$_2$) with the linear equilibrium state (H-Be-H), see~\cref{fig:BeH2_mod}

\begin{figure}
\centering
\begin{subfigure}[b]{0.48\textwidth}
    \centering
    \includegraphics[width=\textwidth]{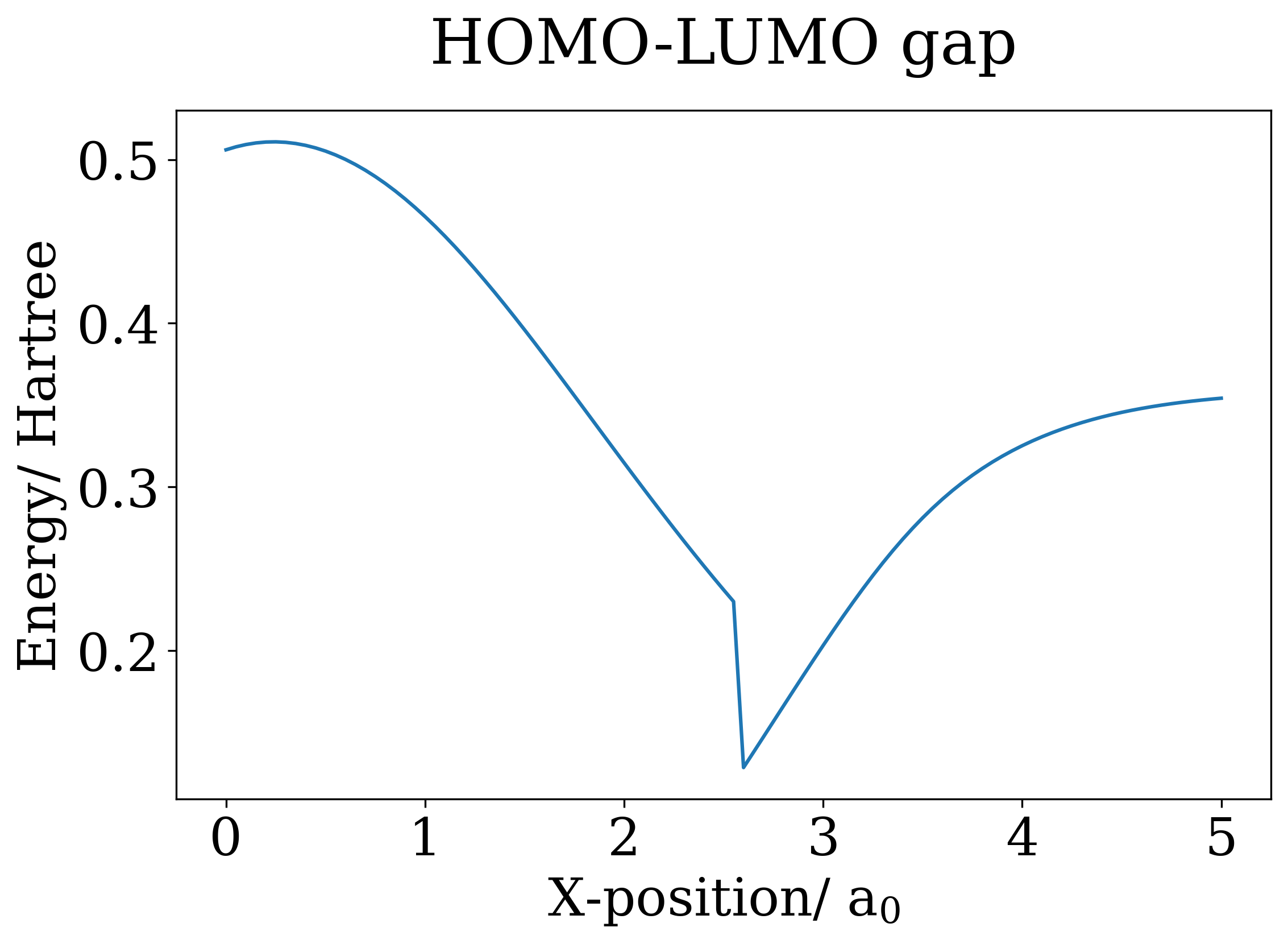}
    \caption{}
    \label{fig:homo_lumo_beh2}
\end{subfigure}
\hfill
\begin{subfigure}[b]{0.48\textwidth}
    \centering
    \includegraphics[width=\textwidth]{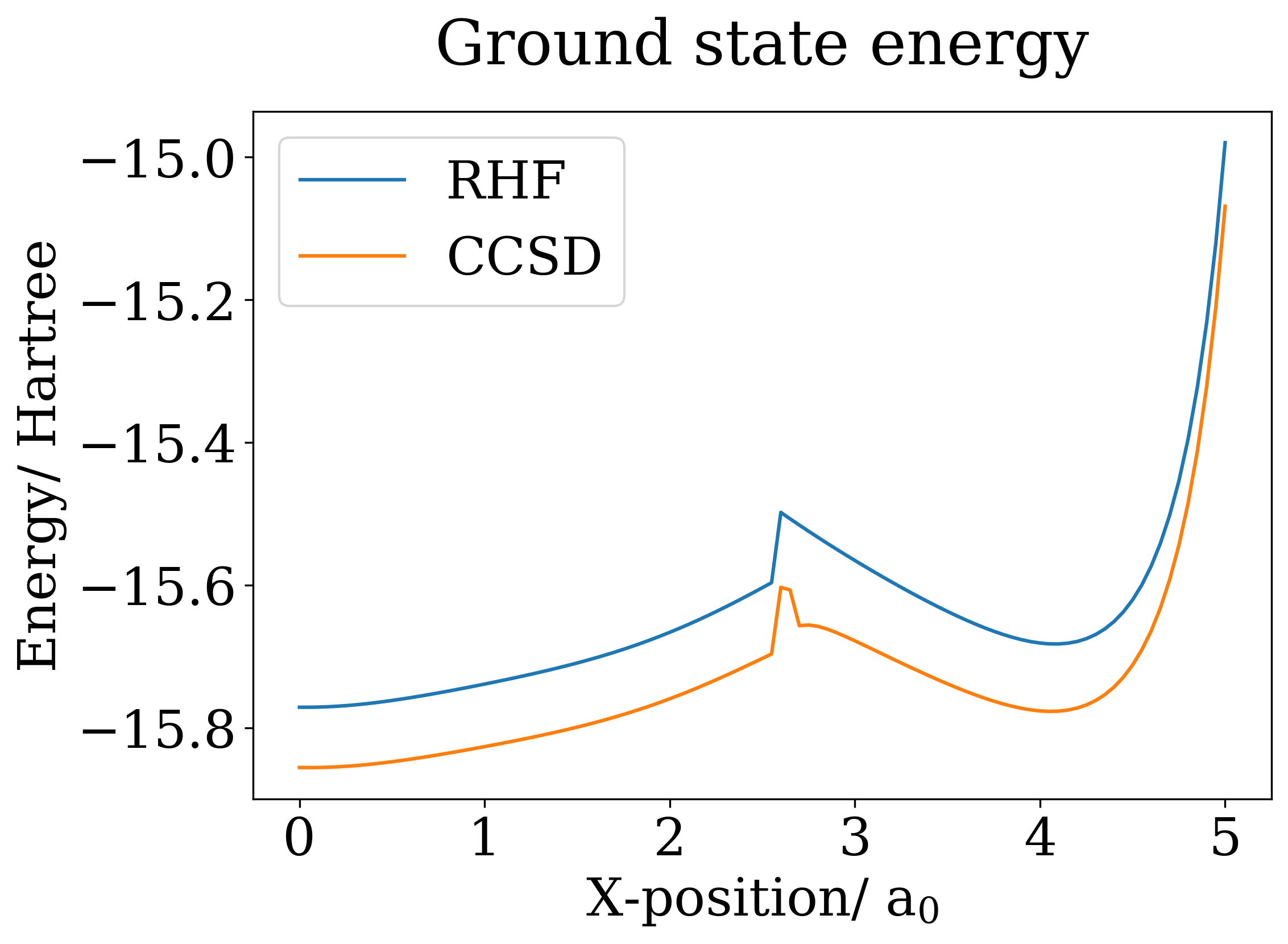}
    \caption{}
    \label{fig:energies_beh2}
\end{subfigure}
\caption{\label{fig:beh2}
(\subref{fig:homo_lumo_beh2}) 
HOMO-LUMO gap as a function of the $X$-position
(\subref{fig:energies_beh2}) RHF and RCCSD energies  as a function of the $X$-position.
%\TBP{What is the "$X$-position"?}
}
\end{figure}

We here follow the insertion pathway outlined in Ref.~\citenum{purvis1983c2v} and denote the position of the beryllium atom by X-position, where X-position equal to zero corresponds to the linear equilibrium state and X-position equal to five corresponds to the non-interacting subsystems.
The transition state of this chemical transformation has a pronounced multi-reference character.
Another distinguishing feature of this model system is a change in the character of the dominating determinant in the wave function along the potential energy surface. 
There are two leading determinants in the wave function, each of which dominates in a certain region of the potential energy surface while both are quasi-degenerate around the transition-state geometry. 
This leads yields to discontinuities as can be seen in~\cref{fig:energies_beh2,fig:homo_lumo_beh2}

\begin{figure}
\centering
\begin{subfigure}[b]{0.48\textwidth}
    \centering
    \includegraphics[width=\textwidth]{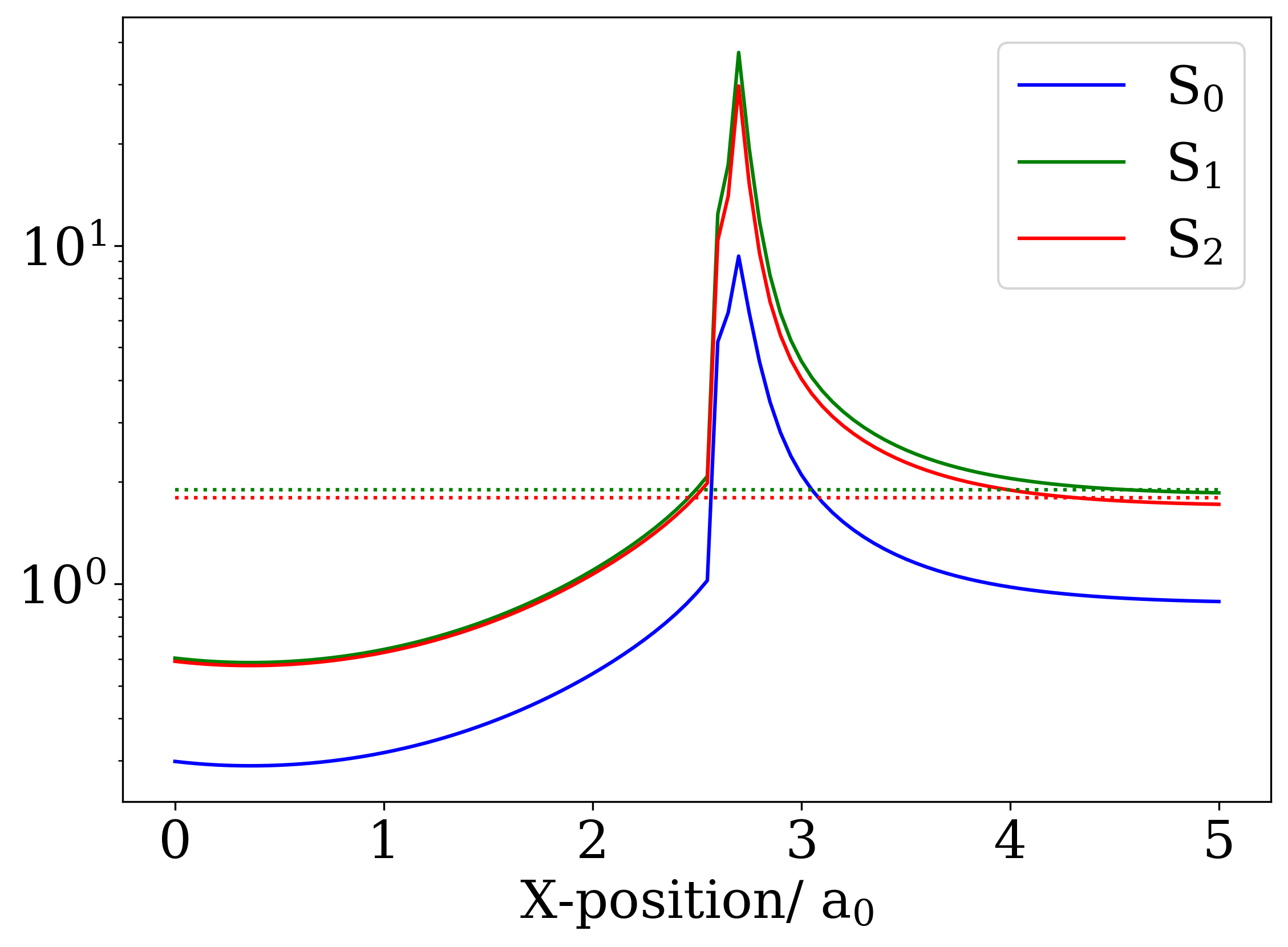}
    \caption{}
    \label{fig:smp_beh2}
\end{subfigure}
\hfill
\begin{subfigure}[b]{0.48\textwidth}
    \centering
    \includegraphics[width=\textwidth]{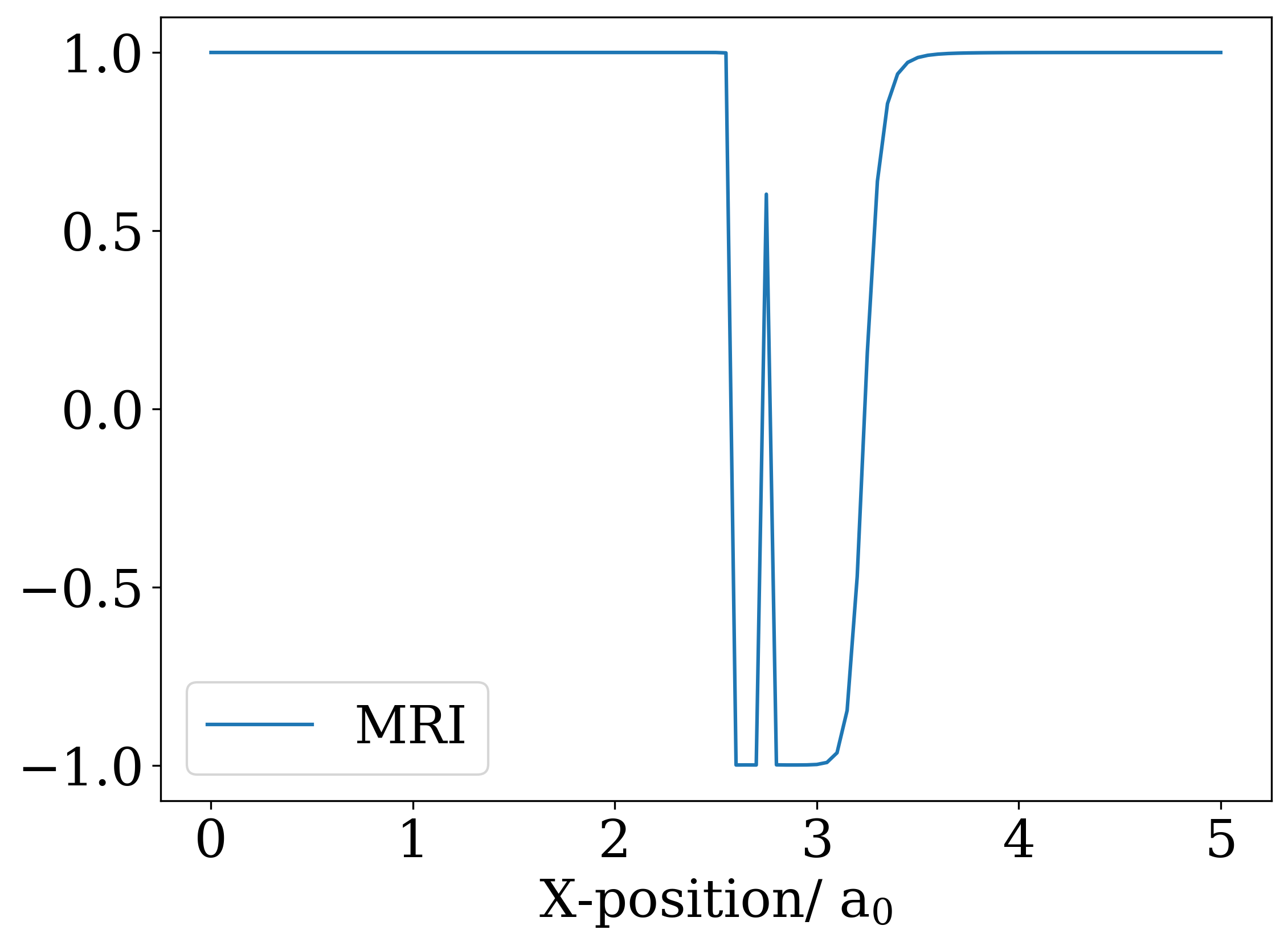}
    \caption{}
    \label{fig:mri_beh2}
\end{subfigure}
\caption{\label{fig:diag_beh2}
(\subref{fig:smp_beh2}) 
shows the $S$-diagnostics, the dotted green, and red horizontal lines correspond to $v_{\rm crit}^{(2)}= 1.9$ and $v_{\rm crit}^{(3)}= 1.8$, respectively.
(\subref{fig:mri_beh2}) shows the
previously suggested MRI
}
\end{figure}

Due to the quasi-degeneracy that appears along the transition path, we again compare the proposed $S$-diagnostics with the MRI index suggested in Ref.~\citenum{bartlett2020index}. 
We clearly see the indication of the quasi degeneracy in the MRI index, see~\cref{fig:mri_beh2}.
The region indicated by MRI$<-0.99$ corresponds to $x\in [2.6, 3.05]$.
The $S$-diagnostic also indicates a region where the CC computations are potentially unreliable. 
It is worth mentioning that choosing the critical values similar to the previous example, i.e., $v_{\rm crit}^{(2)}= 1.9$ and $v_{\rm crit}^{(3)}= 1.8$, the predicted region corresponds to $x\in [2.5, 4.5]$ and $x\in [2.5, 4.25]$, respectively.
In order to reproduce the same region of quasi-degeneracy as indicated by the MRI index, the critical values have to be adjusted to $v_{\rm crit}^{(2)}= 3.8$ and $v_{\rm crit}^{(3)}= 3.5$, respectively.

%\SK{Why don't we compare CC with FCI to see if the diagnostics are actually correct?}

\subsubsection{H$_4$ model (transition from square to linear geometry)}

\begin{figure}
    \centering
    \includegraphics[width = \textwidth]{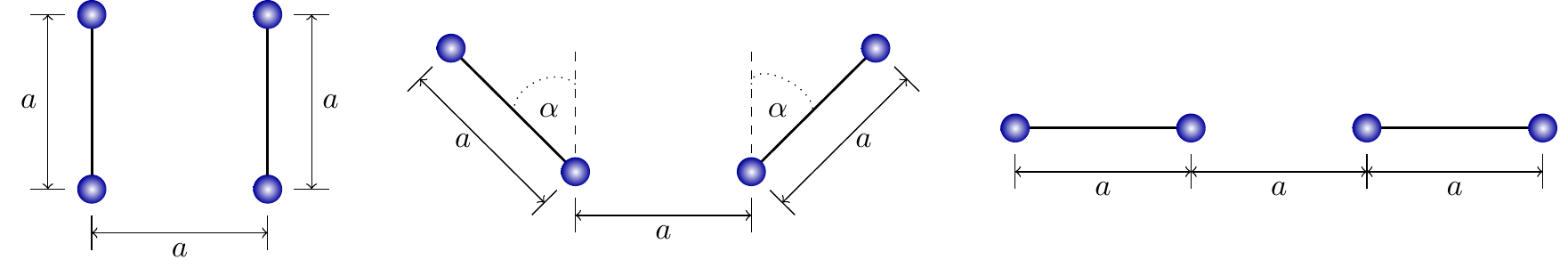}
    \caption{Depiction of the H$_4$ model undergoing the transition from a square geometry to linear geometry model by the angle $\alpha$.}
    \label{fig:H4_model}
\end{figure}

Next, we shall investigate the proposed $S$-diagnostics applied to the H$_4$ model. The H$_4$ model is a standard transition model that allows steering the quasi-degeneracy using a single parameter, namely, the transition angle $\alpha$ where $\alpha = 0$ corresponds to a square geometry and $\alpha =\pi/2$ corresponds to a linear geometry. 
Following Ref.~\cite{jankowski1980applicability}, we set a = 2.0 (a.u.), see~\cref{fig:H4_model}.

\begin{figure}
\centering
\begin{subfigure}[b]{0.48\textwidth}
    \centering
    \includegraphics[width=\textwidth]{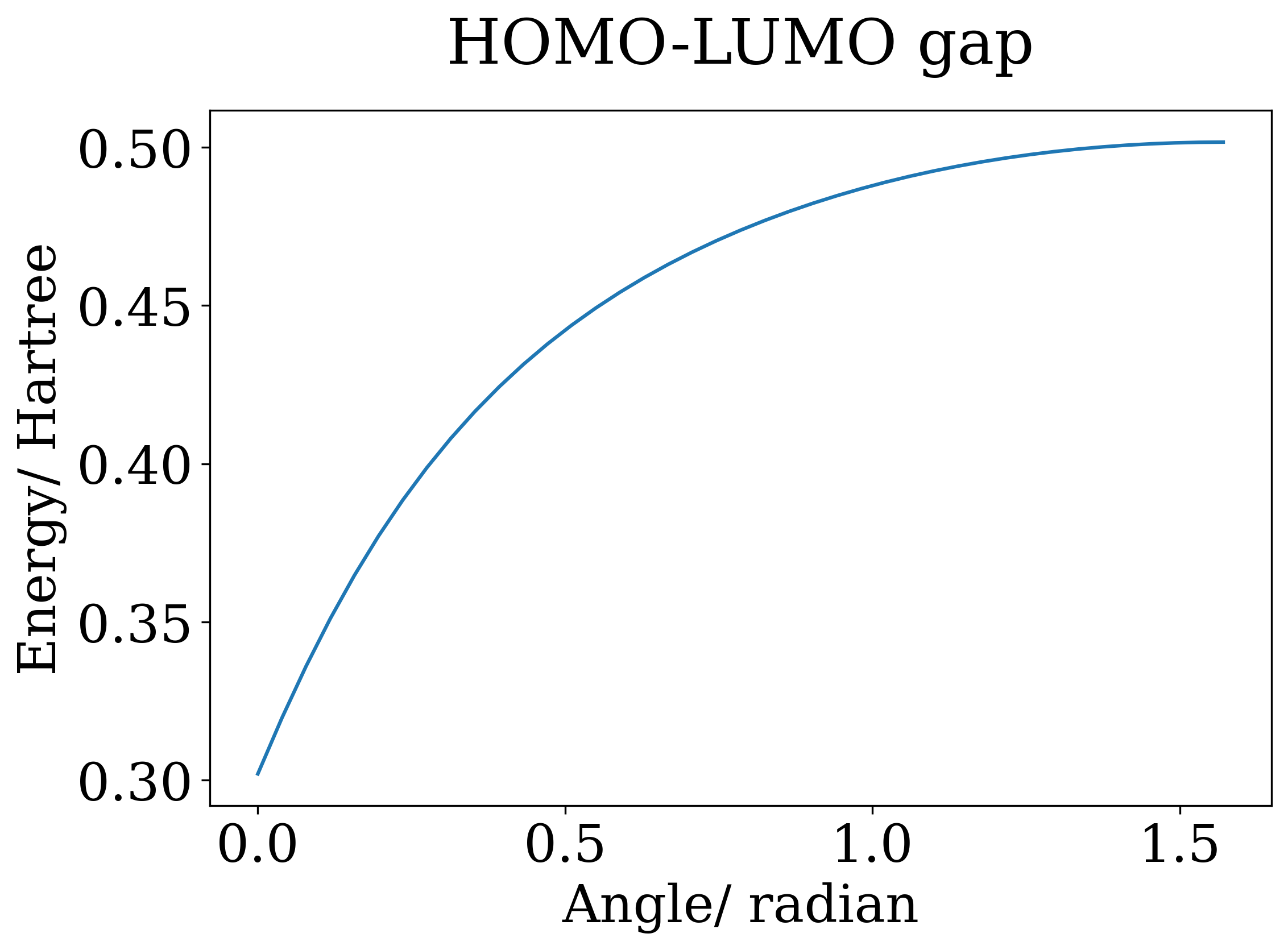}
    \caption{}
    \label{fig:homo_lumo_h4}
\end{subfigure}
\hfill
\begin{subfigure}[b]{0.48\textwidth}
    \centering
    \includegraphics[width=\textwidth]{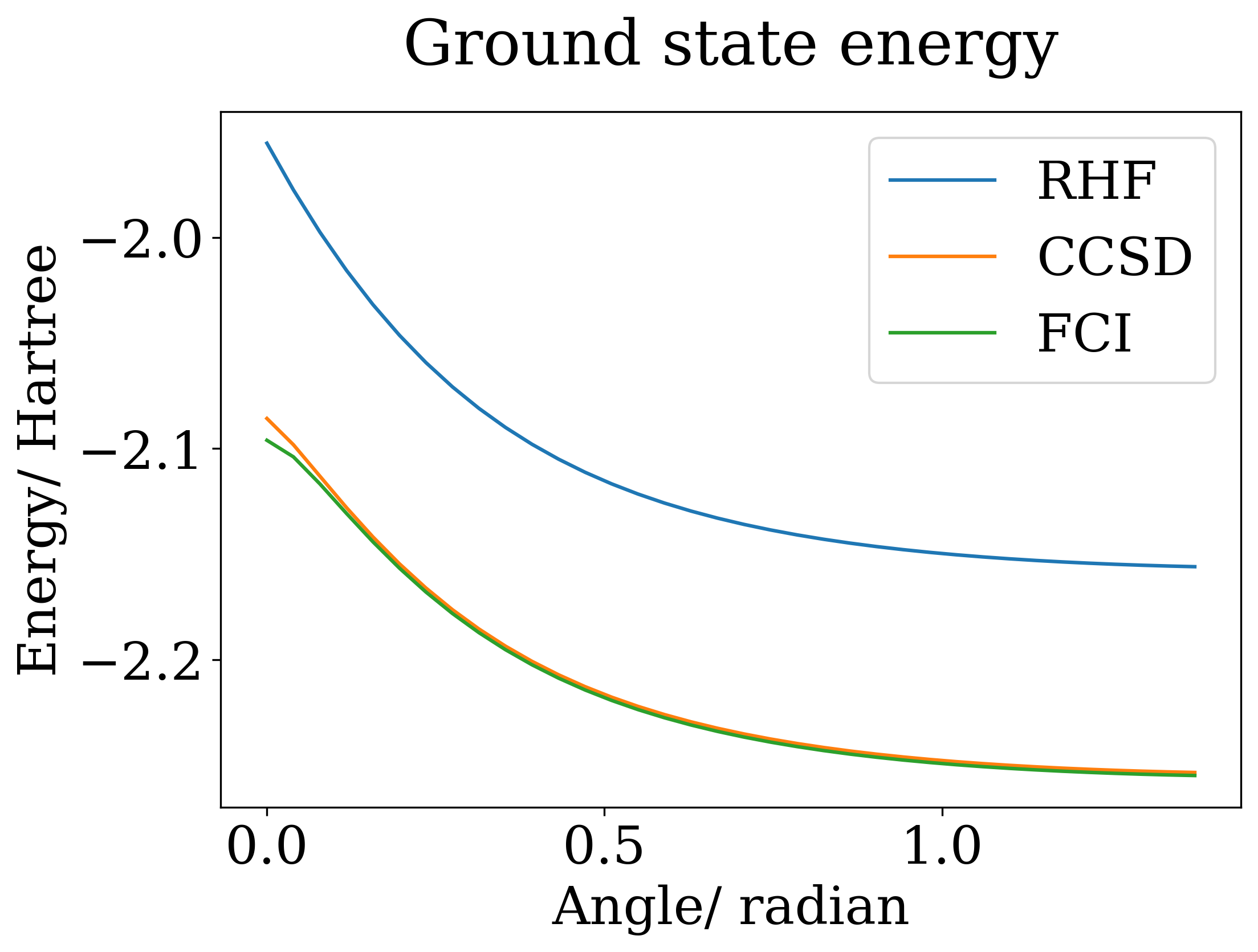}
    \caption{}
    \label{fig:energies_h4}
\end{subfigure}
\caption{\label{fig:h4}
(\subref{fig:homo_lumo_h4}) 
HOMO-LUMO gap of H$_4$ as a function of the transition angle 
(\subref{fig:energies_h4}) RHF, CCSD and FCI energies of H$_4$ as a function of the transition angle
}
\end{figure}

We see that as the transition angle $\alpha$ tends to zero, the HOMO-LUMO gap closes and the system shows signs of (quasi-) degeneracy, see~\cref{fig:homo_lumo_h4}

\begin{figure}
\centering
\begin{subfigure}[b]{0.48\textwidth}
    \centering
    \includegraphics[width=\textwidth]{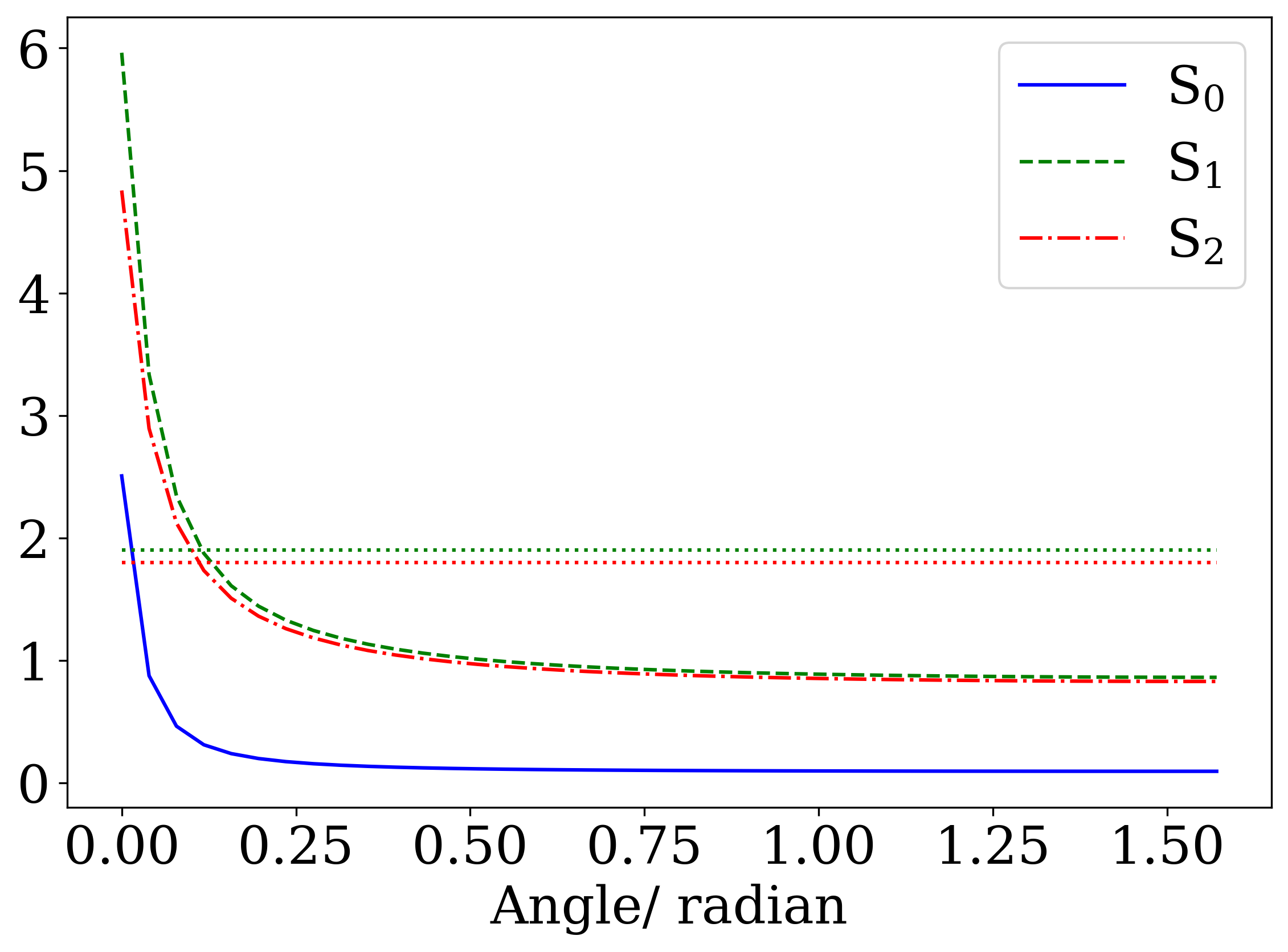}
    \caption{}
    \label{fig:smp_h4}
\end{subfigure}
\hfill
\begin{subfigure}[b]{0.48\textwidth}
    \centering
    \includegraphics[width=\textwidth]{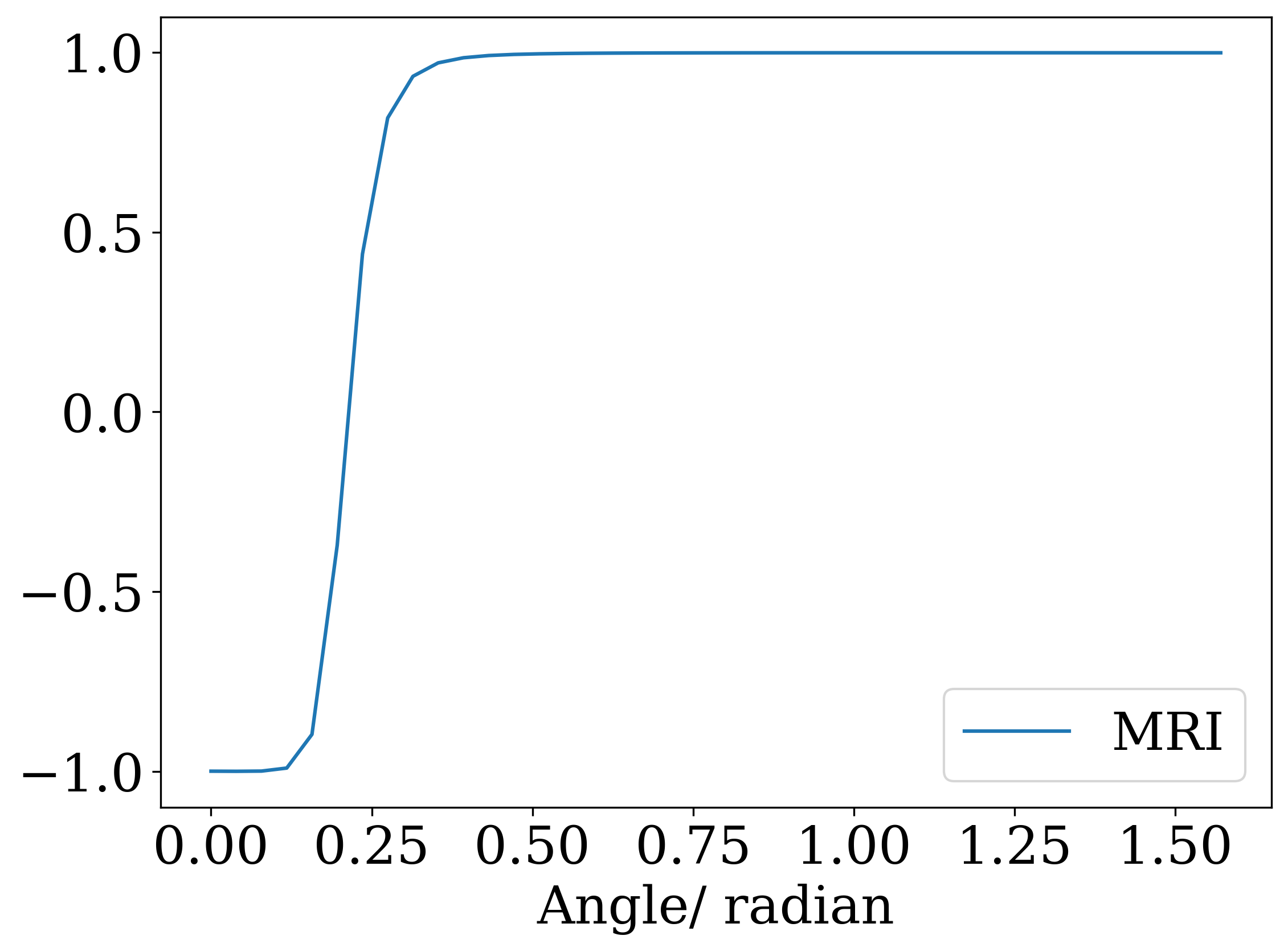}
    \caption{}
    \label{fig:mri_h4}
\end{subfigure}
\caption{\label{fig:diag_h4}
(\subref{fig:smp_h4}) The $S$-diagnostics of H$_4$ as a function of the transition angle, the dotted green, and red horizontal lines correspond to $v_{\rm crit}^{(2)}= 1.9$ and $v_{\rm crit}^{(3)}= 1.8$, respectively.
(\subref{fig:mri_h4}) The
previously suggested MRI of H$_4$ as a function of the transition angle.
}
\end{figure}

Due to the quasi degeneracy near $\alpha=0$, we again compare the proposed $S$-diagnostics with the MRI index.
We clearly see the indication of the quasi degeneracy in the MRI index, see~\cref{fig:mri_h4}. 
The $S$-diagnostic also indicates the problematic region near zero transition angle. 
A cut-off value of $v_{\rm crit}^{(2)}= 1.9$ and $v_{\rm crit}^{(3)}= 1.8$ results in $S_2$ and $S_3$, respectively, indicating the same region of quasi degeneracy as the MRI index.

\begin{figure}
    \includegraphics[width=0.5\textwidth]{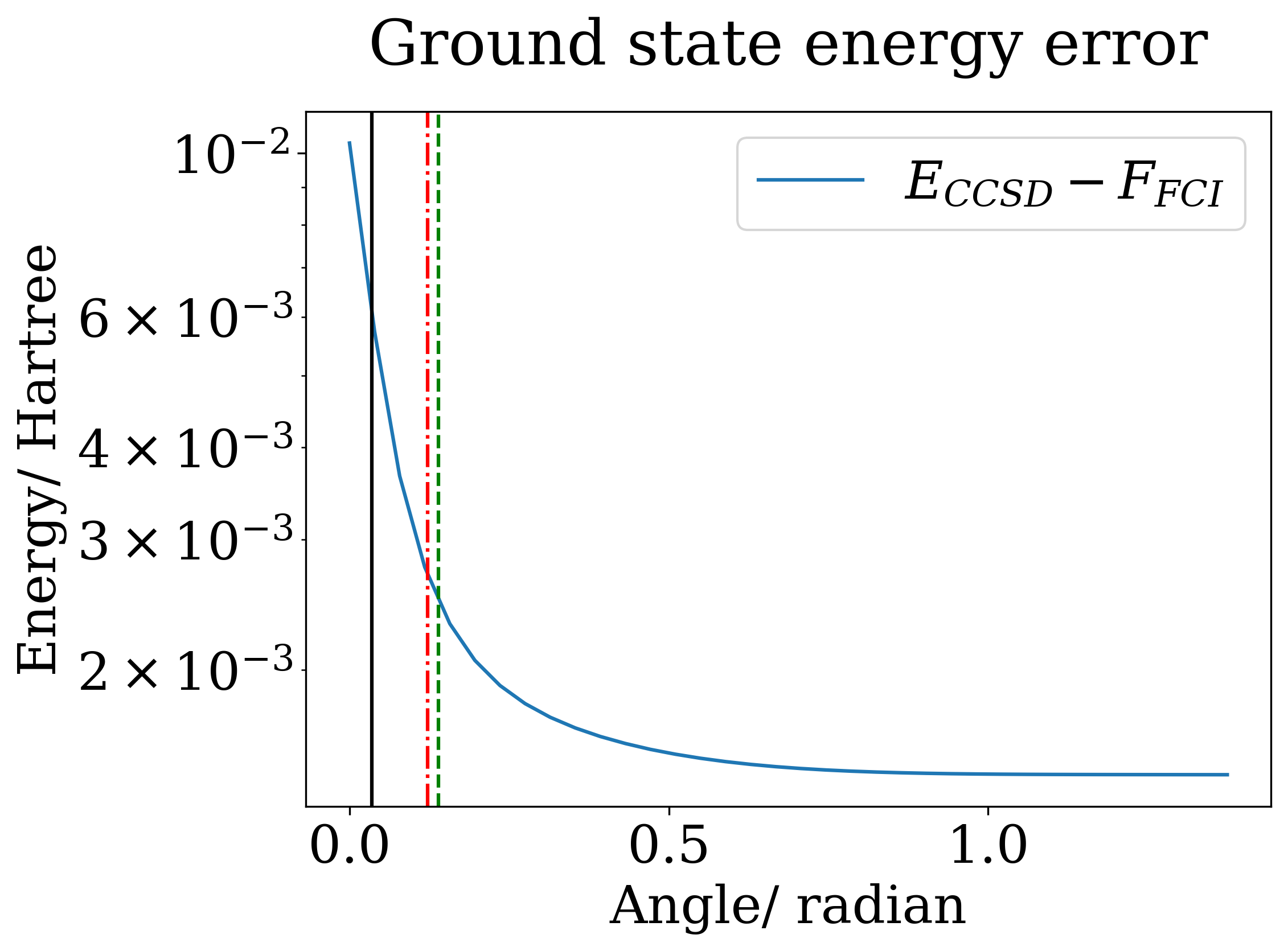}
    \caption{}
    \label{fig:fci_ccsd_energy_loglin_diff_2}
\caption{\label{fig:h4_fci_energies_2}
The energy error of CCSD compared to the FCI reference energy using semi-log scales. The area left of the vertical solid (black), dashed (green), and dotted-dashed (red) lines correspond to the regions where the MRI, $S_2$, and $S_3$ diagnostic indicate a potential failure of CCSD, respectively.
}
\end{figure}

For this small model Hamiltonian, it is moreover feasible to perform computations at the FCI level of theory, see~\cref{fig:h4_fci_energies_2}. This comparison yields a quantitative comparison of error and $S$-diagnostic.

\subsubsection{H$_4$ model (symmetrically disturbed on a circle)}

\begin{figure}
    \centering
    \includegraphics[width = 0.6\textwidth]{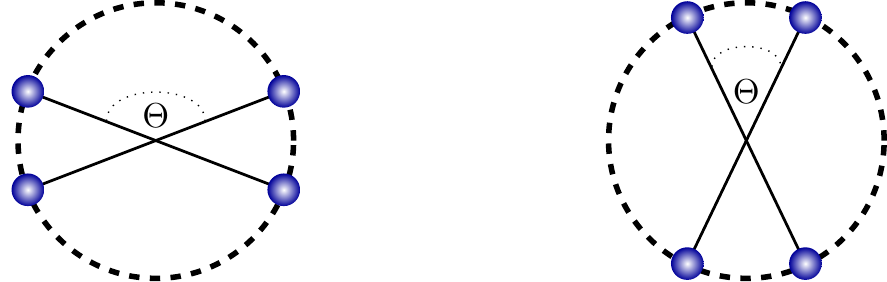}
    \caption{Depiction of the H$_4$ model undergoing a symmetric disturbance on a circle modeled by the angle  $\Theta$.}
    \label{fig:H4_circ_mod}
\end{figure}

Another variant of the H$_4$ model that is commonly employed to evaluate CC methods consists of four hydrogen atoms symmetrically distributed on a circle of radius $R = 1.738$~\AA~\cite{van2000benchmark}. 
For small or large angles, the system resembles two H$_2$ molecules that are reasonably well separated, but as the angle passes through $90°$, the four atoms form a square yielding a degenerate ground state.
The exact energy is smooth as a function of the angle, but at the RHF level, we observe a cusp at $90°$, similar to the rotation of the carbon-carbon bond in ethylene.
We follow the system's geometry configuration outlined in Ref.~\cite{bulik2015can}, see~\cref{fig:H4_circ_mod}.

\begin{figure}
\centering
\begin{subfigure}[b]{0.48\textwidth}
    \centering
    \includegraphics[width=\textwidth]{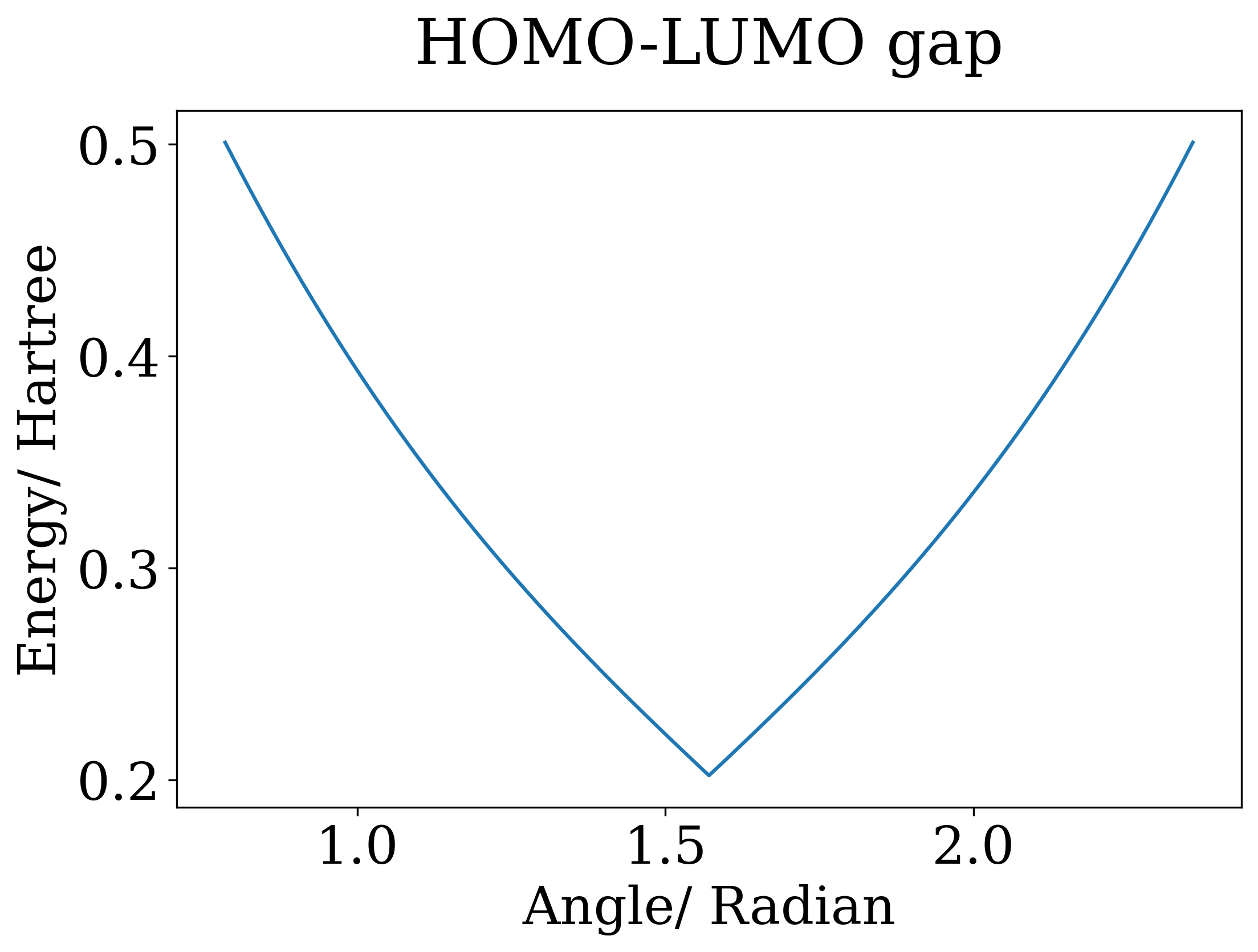}
    \caption{}
    \label{fig:homo_lumo_h4_scus}
\end{subfigure}
\hfill
\begin{subfigure}[b]{0.48\textwidth}
    \centering
    \includegraphics[width=\textwidth]{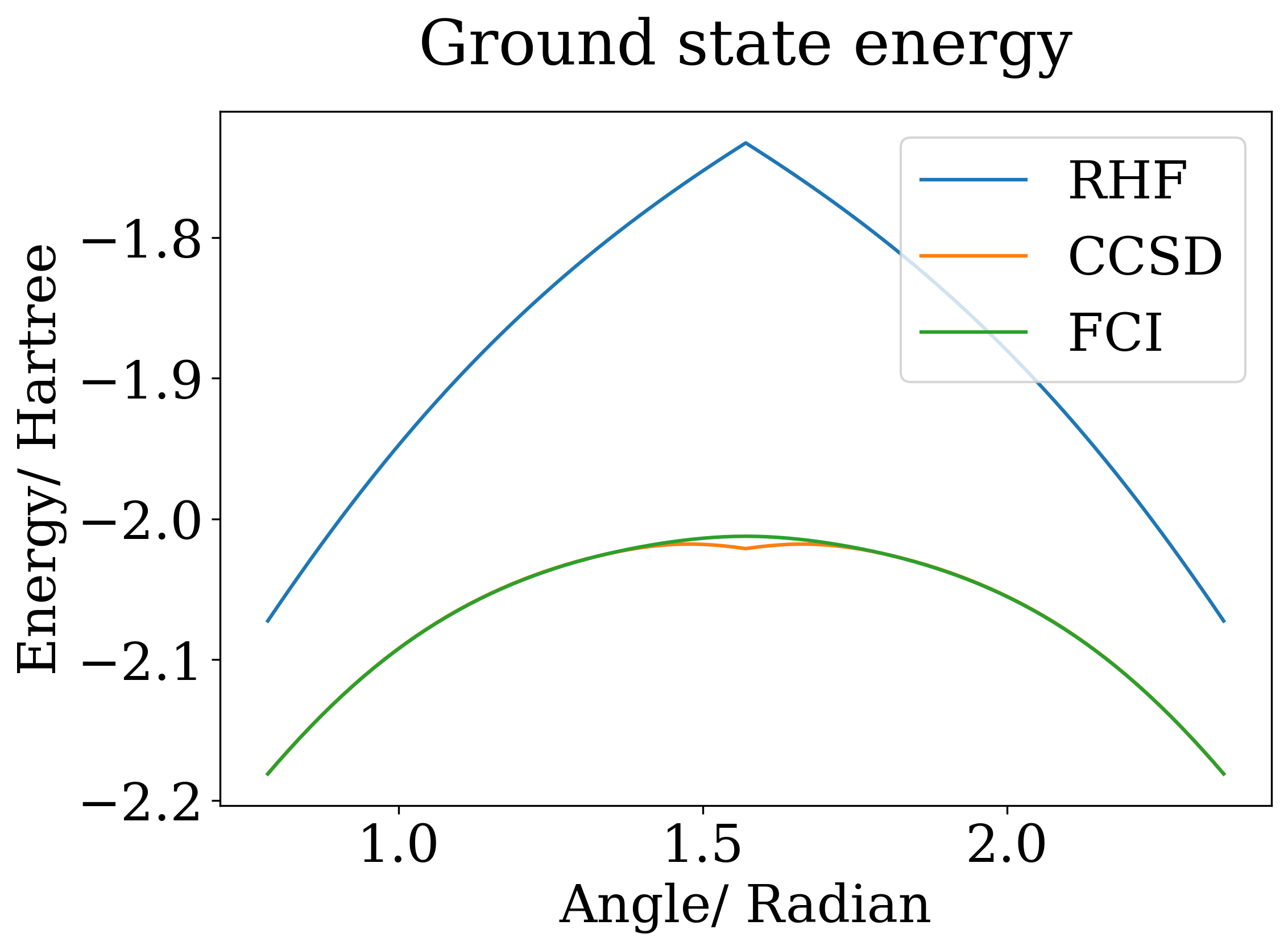}
    \caption{}
    \label{fig:energies_h4_scus}
\end{subfigure}\\
%\hfill
% \begin{subfigure}[b]{0.48\textwidth}
%     \centering
%     \includegraphics[width=\textwidth]{Graphics/fci_ccsd_abs_diff_semilogy_h4_scus_ccpvtz.png}
%     \caption{}
%     \label{fig:fci_ccsd_energy_diff}
% \end{subfigure}
\caption{\label{fig:h4_scus}
(\subref{fig:homo_lumo_h4_scus}) 
HOMO-LUMO gap of H$_4$ as a function of the transition angle 
(\subref{fig:energies_h4}) RHF, RCCSD energies of H$_4$ as a function of the transition angle.
%and FCI energy. Note that in the region 75-105 angle/radian the RCCSD energy is lower than the FCI ground-state energy, which indicates the variational collapse of the RCCSD energy in this region.
}
\end{figure}

We see that as the transition angle $\Theta$ tends to $\pi / 2$ radians (90°), the HOMO-LUMO gap closes and the system shows signs of (quasi) degeneracy, see~\cref{fig:homo_lumo_h4_scus}

\begin{figure}
\centering
\begin{subfigure}[b]{0.48\textwidth}
    \centering
    \includegraphics[width=\textwidth]{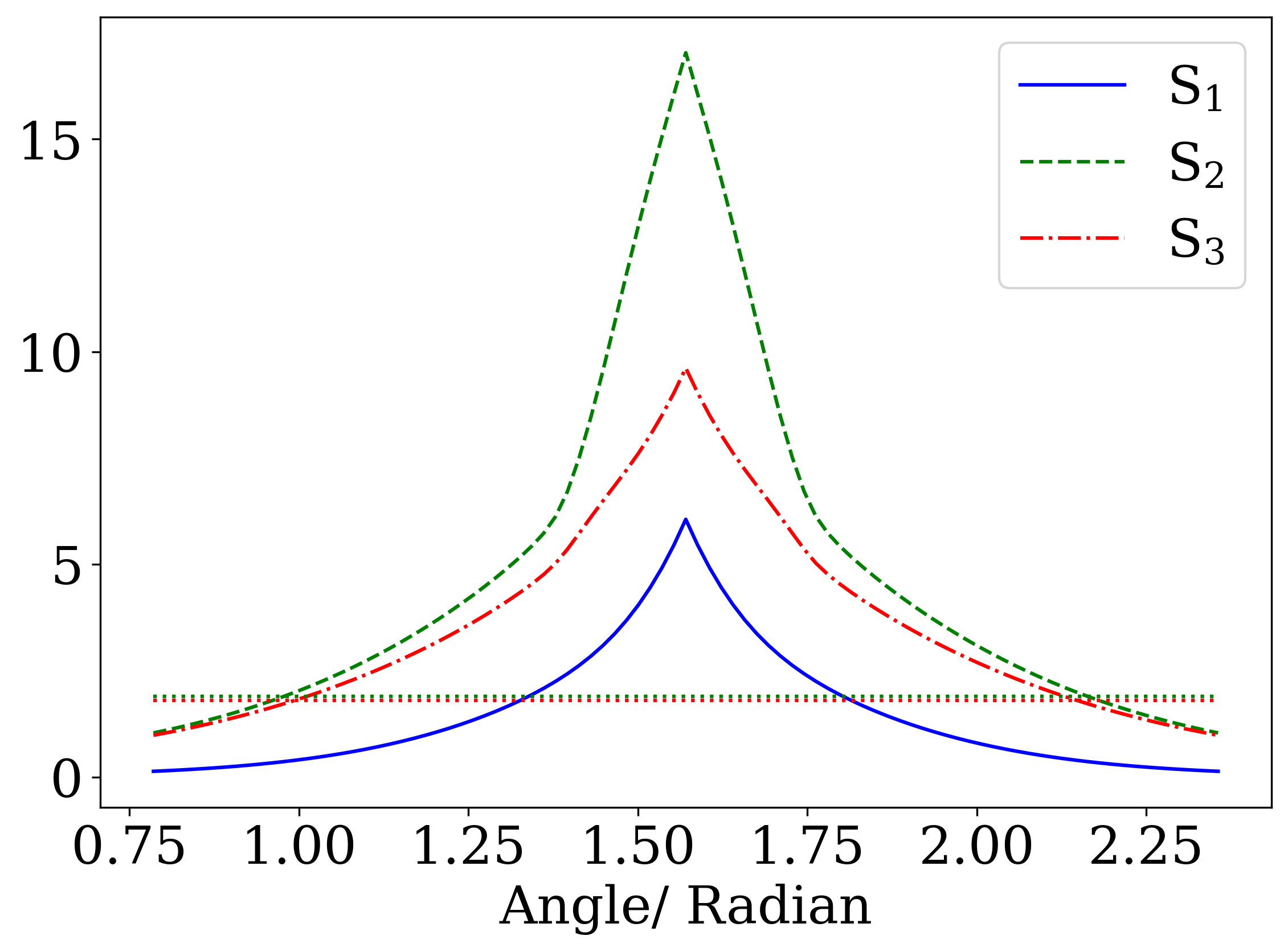}
    \caption{}
    \label{fig:smp_h4_scus}
\end{subfigure}
\hfill
\begin{subfigure}[b]{0.48\textwidth}
    \centering
    \includegraphics[width=\textwidth]{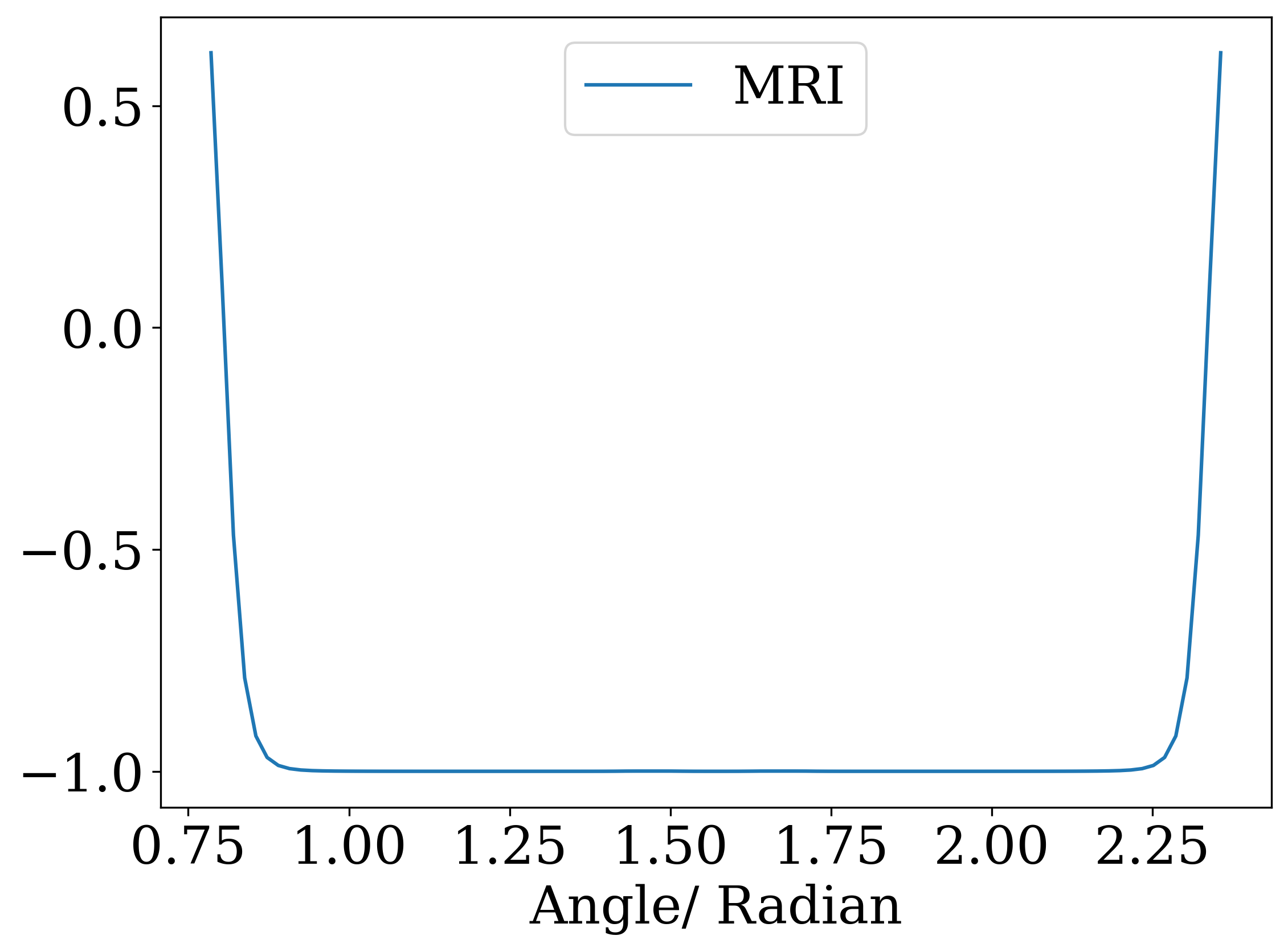}
    \caption{}
    \label{fig:mri_h4_scus}
\end{subfigure}
\caption{\label{fig:diag_h4_scus}
(\subref{fig:smp_h4_scus}) The $S$-diagnostics of H$_4$ as a function of the transition angle, the dotted green, and red horizontal lines correspond to $v_{\rm crit}^{(2)}= 1.9$ and $v_{\rm crit}^{(3)}= 1.8$, respectively.
(\subref{fig:mri_h4_scus}) The
previously suggested MRI of H$_4$ as a function of the transition angle.
}
\end{figure}

Due to the quasi degeneracy near $\Theta=\pi / 2$ (90°), we again compare the proposed $S$-diagnostics with the MRI index.
We clearly see the indication of the quasi degeneracy in the MRI index, see~\cref{fig:mri_h4_scus}. 
The $S$-diagnostic also indicates the problematic region near zero transition angle. 
A cut-off value of $v_{\rm crit}^{(2)}= 1.9$ and $v_{\rm crit}^{(3)}= 1.8$ results in $S_2$ and $S_3$, respectively, indicating the same region of quasi degeneracy as the MRI index.

\begin{figure}
\centering
\begin{subfigure}[b]{0.48\textwidth}
    \centering
    \includegraphics[width=\textwidth]{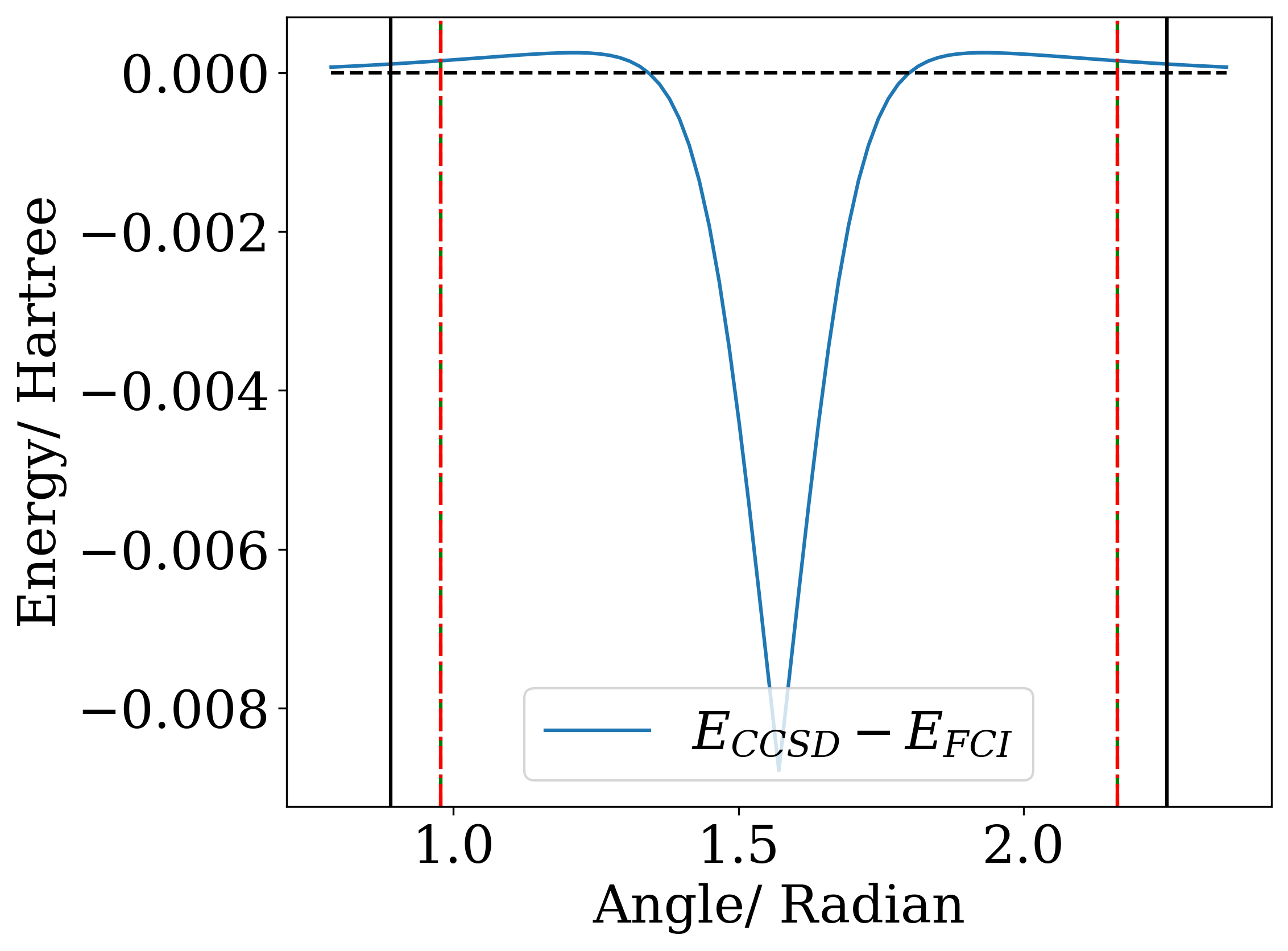}
    \caption{}
    \label{fig:fci_ccsd_energy_diff}
\end{subfigure}
\hfill
\begin{subfigure}[b]{0.48\textwidth}
    \centering
    \includegraphics[width=\textwidth]{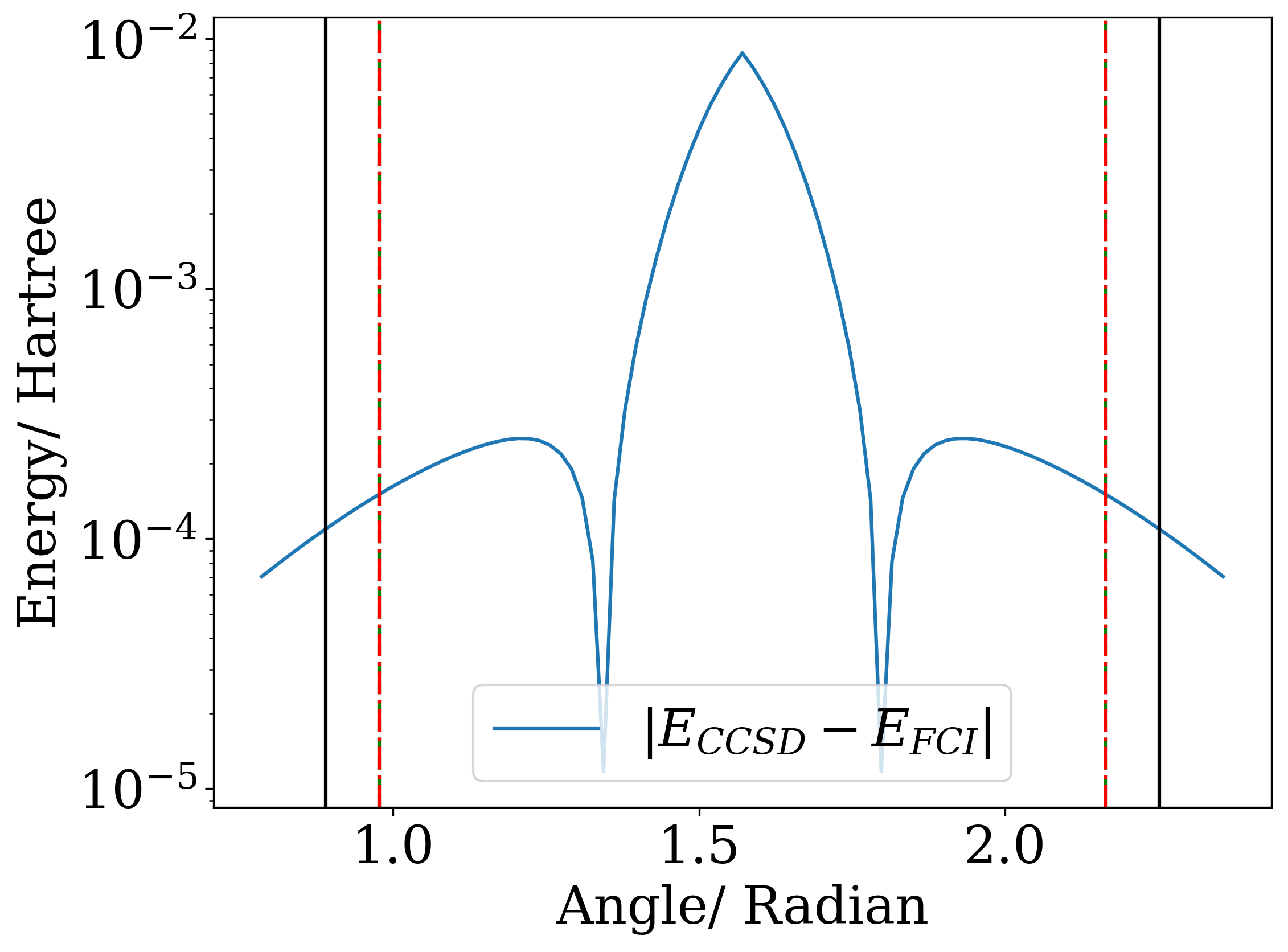}
    \caption{}
    \label{fig:fci_ccsd_energy_loglin_diff}
\end{subfigure}
\caption{\label{fig:h4_fci_energies}
(\subref{fig:fci_ccsd_energy_diff}) The energy error of CCSD compared to the FCI reference energy. Note that in the region of 1.3-1.8 radians the CCSD energy is lower than the FCI reference energy, which indicates the variational collapse of the CCSD energy in this region.
(\subref{fig:fci_ccsd_energy_loglin_diff}) The absolute value of the energy error of CCSD compared to the FCI reference energy using semi-log scales. The area between the vertical solid (black), dashed (green), and dotted-dashed (red) lines correspond to the regions where the MRI, $S_2$, and $S_3$ diagnostic indicate a potential failure of CCSD, respectively.
}
\end{figure}

For this small model Hamiltonian, it is moreover feasible to perform computations at the FCI level of theory, see~\cref{fig:h4_fci_energies}. This comparison reveals the variational collapse of the CCSD energy, see~\cref{fig:fci_ccsd_energy_diff}, and moreover yields a quantitative comparison of error and $S$-diagnostic.
The trusted region suggested by the $S$-diagnostic corresponds to a CCSD energy error smaller than $2\cdot 10^{-4}$ a.u.~which is below the chemical accuracy threshold.

Since the simulations performed in the previous section suggest that the previously used $T_1$, $D_1$, and $D_2$ diagnostics are uncorrelated, or merely weakly correlated, we do not report their performance here. The computations showing the performance of the $T_1$, $D_1$, and $D_2$ diagnostics can be found in the Appendix, see~\cref{fig:prev_diag_c2h4,fig:prev_diag_beh2,fig:prev_diag_h4,fig:prev_diag_h2_scus}

\subsection{Transition metal complexes}

In this section we investigate three square-planar copper complexes [CuCl$_4$]$^{2-}$, [Cu(NH$_3$)$_4$]$^{2+}$, and [Cu(H$_2$O)$_4$]$^{2+}$. 
Transition metal complexes are in general considered to be strongly correlated systems and complete active space self-consistent field (CASSCF) theory is commonly applied, with multi-reference perturbation or truncated CI corrections for dynamic correlation. 
However, as shown in Ref.~\citenum{giner2018interplay}, the single reference CC method performs very well despite the large $D_1$ diagnostic value. 
We use these systems to scrutinize the proposed $S$-diagnostics for larger systems that are known to be misleadingly diagnosed by the $D_1$ diagnostics.

Similar to Ref.~\citenum{giner2018interplay}, we perform the simulation of [CuCl$_4$]$^{2-}$, [Cu(NH$_3$)$_4$]$^{2+}$, and [Cu(H$_2$O)$_4$]$^{2+}$ in 6-31G basis using UHF and ROHF as reference states. 
Also, He, Ne, and Ar cores were frozen in the nitrogen, chlorine, and copper atoms, respectively, resulting in 41 electrons in 50, 66, and 74 orbitals for the [CuCl$_4$]$^{2-}$, [Cu(H$_2$O)$_4$]$^{2+}$, and [Cu(NH$_3$)$_4$]$^{2+}$ molecules, respectively. 
We list the ground state energies obtained at the mean-field level of theory and the corresponding CCSD results in~\cref{tab:energies_transition_metals}; we moreover list the HOMO-LUMO gap which enters in the $S$-diagnostics.  

\begin{table}[]
    \centering
    \begin{tabular}{c|ccc|ccc}
         & UHF & $\gamma_{UHF}$ & UCCSD & RHOF & $\gamma_{ROHF}$ & UCCSD \\
          \hline
~ [CuCl$_4$]$^{2-}$       & -3476.764 & 0.453 & -3477.119 & -3476.763 & 0.146 & -3477.119\\
~ [Cu(NH$_3$)$_4$]$^{2+}$ & -1862.977 & 0.564 & -1863.663 & -1862.976 & 0.351 & -1863.663\\
~ [Cu(H$_2$O)$_4$]$^{2+}$ & -1942.225 & 0.677 & -1942.914 & -1942.224 & 0.340 & -1942.914
    \end{tabular}
    \caption{Energies values and HOMO-LUMO gap obtained with UHF, ROHF, and UCCSD calculations given the reference state from UHF and ROHF, respectively. 
    %\SK{I think there are errors in the header of the table.} 
    }
    \label{tab:energies_transition_metals}
\end{table}

The results in~\cref{tab:energies_transition_metals} show that UHF and ROHF calculations predict similar energy values. 
Moreover, using the UHF, or ROHF reference state results in similar CCSD energy values. 
It is worth noticing that ROHF yields a generally smaller HOMO-LUMO gap. 
Since the performed CCSD calculations differ in their reference, we can compute the $S$-diagnostics for both sets of calculations.  
The results obtained from a UHF and ROHF reference are listed in~\cref{tab:TransitionMetals_1} and in~\cref{tab:TransitionMetals_2}, respectively.

% \begin{table}[]
%     \centering
%     \begin{tabular}{c|ccc|ccc|ccc}
%         & $S_1$ & $S_{2}$ & $S_3$ & w$_1$ & w$_2$ & w$_3$ & $T_1$ &  $D_1$ & $D_2$ \\
%         \hline
%         ~ [CuCl$_4$]$^{2-}$ &  0.208 & 0.409 & 0.406 & 0.831 & 0.030 & 0.139 & 0.019 & 0.158 & 0.110\\
%         ~ [Cu(NH$_3$)$_4$]$^{2+}$ &  0.203& 0.403 & 0.398& 0.735 & 0.025 & 0.241& 0.014 & 0.130 & 0.121\\
%         ~ [Cu(H$_2$O)$_4$]$^{2+}$ & 0.155 & 0.308 & 0.305 & 0.786 & 0.072 & 0.116& 0.011 & 0.072 & 0.116  
%     \end{tabular}
%     \caption{$S$-diagnostics obtained for the three square-planar copper complexes [CuCl$_4$]$^{2-}$, [Cu(NH$_3$)$_4$]$^{2+}$, and [Cu(H$_2$O)$_4$]$^{2+}$ in spin unrestricted formulation with UHF reference.}
%     \label{tab:TransitionMetals_1}
% \end{table}

\begin{table}[]
    \centering
    \begin{tabular}{c|ccc|ccc}
        & $S_1$ & $S_{2}$ & $S_3$ & $T_1$ &  $D_1$ & $D_2$ \\
        \hline
        ~ [CuCl$_4$]$^{2-}$ &  0.208 & 0.409 & 0.406 & 0.019 & 0.158 & 0.110\\
        ~ [Cu(NH$_3$)$_4$]$^{2+}$ &  0.203& 0.403 & 0.398 & 0.014 & 0.130 & 0.121\\
        ~ [Cu(H$_2$O)$_4$]$^{2+}$ & 0.155 & 0.308 & 0.305 & 0.011 & 0.072 & 0.116  
    \end{tabular}
    \caption{$S$-diagnostics obtained for the three square-planar copper complexes [CuCl$_4$]$^{2-}$, [Cu(NH$_3$)$_4$]$^{2+}$, and [Cu(H$_2$O)$_4$]$^{2+}$ in spin unrestricted formulation with UHF reference.}
    \label{tab:TransitionMetals_1}
\end{table}

We see that all $S$-diagnostic variants suggest that the CCSD calculations were successful, and do not require additional numerical confirmation.
This is opposed to the $D_1$ diagnostics, which aligns with the results reported in Ref.~\citenum{giner2018interplay}. 
% We moreover compute CCSD configurational weights, i.e., the overlap of the CCSD solution with the reference Slater determinant (w$_1$), the sum of the overlaps of the CCSD solution with singly excited Slater determinants (w$_2$) and the sum of the overlaps of the CCSD solution with doubly excited Slater determinants (w$_3$).
% \FF{@Hakon, please add an interpretation of the w$_1$, w$_2$ and w$_3$ results here.}

% \begin{table}[]
%     \centering
%     \begin{tabular}{c|ccc|ccc|ccc}
%         & $S_0$ & $S_1$ & $S_2$ & w$_1$ & w$_2$ & w$_3$ & $T_1$ &  $D_1$ & $D_2$ \\
%         \hline
%         ~ [CuCl$_4$]$^{2-}$ &  0.645 & 1.285 & 1.27 & 0.827 & 0.033 & 0.140 & 0.020 & 0.167 & 0.110\\
%         ~ [Cu(NH$_3$)$_4$]$^{2+}$ & 0.326 & 0.646 & 0.638 & 0.731 & 0.028 & 0.241 & 0.015 & 0.139 & 0.121 \\
%         ~ [Cu(H$_2$O)$_4$]$^{2+}$ & 0.309 & 0.614 & 0.607 & 0.785 & 0.013 & 0.202 & 0.011 & 0.077 & 0.116
%     \end{tabular}
%     \caption{$S$-diagnostics obtained for the three square-planar copper complexes [CuCl$_4$]$^{2-}$, [Cu(NH$_3$)$_4$]$^{2+}$, and [Cu(H$_2$O)$_4$]$^{2+}$ in spin unrestricted formulation with ROHF reference.}
%     \label{tab:TransitionMetals_2}
% \end{table}

\begin{table}[]
    \centering
    \begin{tabular}{c|ccc|ccc}
        & $S_0$ & $S_1$ & $S_2$ & $T_1$ &  $D_1$ & $D_2$ \\
        \hline
        ~ [CuCl$_4$]$^{2-}$ &  0.645 & 1.285 & 1.27 & 0.020 & 0.167 & 0.110\\
        ~ [Cu(NH$_3$)$_4$]$^{2+}$ & 0.326 & 0.646 & 0.638 & 0.015 & 0.139 & 0.121 \\
        ~ [Cu(H$_2$O)$_4$]$^{2+}$ & 0.309 & 0.614 & 0.607 & 0.011 & 0.077 & 0.116
    \end{tabular}
    \caption{$S$-diagnostics obtained for the three square-planar copper complexes [CuCl$_4$]$^{2-}$, [Cu(NH$_3$)$_4$]$^{2+}$, and [Cu(H$_2$O)$_4$]$^{2+}$ in spin unrestricted formulation with ROHF reference.}
    \label{tab:TransitionMetals_2}
\end{table}

Similar to the results in~\cref{tab:TransitionMetals_1}, we see that all variants of the $S$-diagnostic suggest that the CCSD calculations were successful.
However, it is worth noticing that the $S$-diagnostic values have increased compared to the values reported in~\cref{tab:TransitionMetals_1}.

\section{Conclusion}

In this article, we proposed three {\it a posteriori} diagnostics for single-reference CC calculations which we called $S$-diagnostics, due to their origin in the strong monotonicity analysis.
Contrary to previously suggested CC diagnostics, the $S$-diagnostics are motivated by mathematical principles that have been used to analyze CC methods of different flavors in the past~\cite{schneider2009analysis,rohwedder2013continuous,rohwedder2013error,laestadius2018analysis,faulstich2019analysis}. 

We performed a set of geometry optimizations for small to medium-sized molecules in order to reveal the correlation between the $S$-diagnostics and the error in geometry from CCSD calculations. 
The test set comprised all molecules that were used in previous articles concerning CC diagnostics~\cite{Lee1989,Lee1989b,Janssen1998,Nielsen1999}. 
Our investigations revealed that the $S$-diagnostics correlate well and with large statistical relevance with different errors in geometry. 
This yields a first estimate of the critical values for the $S$-diagnostics beyond which the computational results should be confirmed using further and more careful numerical investigations.  
The observed correlation between the $S$-diagnostics and the different errors in geometry are comparable to the recently suggested EEN index~\cite{bartlett2020index}.
A heuristic test revealed that the $S$-diagnostics also correlate well and with large statistical relevance with the error in geometry at the MP2 level of theory. 
This suggests that the $S$-diagnostics can also be used as an {\it a posteriori} diagnostic for MP2 calculations. 
Our numerical simulations moreover showed that diagnostics based on single excitation cluster amplitudes, i.e., $D_1$ and $T_1$, are uncorrelated to errors in geometry optimization.

Following we investigated the $S$-diagnostics for transition state models that undergo a transition from a region in which CC calculations are reliable to a regime where the CC calculations require further numerical investigations---in this case, due to (quasi-) degeneracy of the ground state. 
The $S$-diagnostic detects the corresponding regions of (quasi-) degeneracy well. 
In fact, its performance is comparable to the recently suggested MRI indicator---an {\it a posteriori} indicator for multi-reference character~\cite{bartlett2020index}.  

The last set of numerical simulations targeted transition metal complexes which have recently been carefully benchmarked~\cite{giner2018interplay}. 
The previously performed benchmark calculations revealed that diagnostics based on single excitation amplitudes severely misdiagnose the performance of CCSD for these transition metal complexes. 
Our computations confirm this, and moreover, show that the $S$-diagnostic correctly confirms the accuracy of the CCSD results outlined in~Ref.~\citenum{giner2018interplay}.

These carefully performed numerical investigations suggest that the $S$-diagnostic is a promising candidate for an {\it a posteriori} diagnostic for single-reference CC and MP2 calculations. 
To further confirm this, benchmarks on a larger set of molecules will be performed in the future.
Moreover, since the mathematical analysis of the single-reference CC method generalizes to periodic systems as well, we believe that the $S$-diagnostic can moreover be applied to simulations of solids at the CC and MP2 level of theory.   

Throughout our numerical investigations, we observe a subpar performance of the $T_1$ and $D_1$ diagnostics. This suggests that those diagnostics should once and for all be removed as {\it a posteriori} diagnostic tools for single-reference CC calculations. 

\section*{Acknowledgement}
This work was partially supported by the Air Force Office of Scientific Research under the award
number FA9550-18-1-0095 and by the Simons Targeted Grants in Mathematics and Physical
Sciences on Moir\'e Materials Magic (F.M.F.), by the Peder Sather
Grant Program (A.L., M.A.C., F.M.F.,),
and by the Research Council of Norway (A.L., M.A.C.) through Project No. 287906 (CCerror) and its Centres of Excellence scheme (Hylleraas Centre) Project No.\ 262695.
Some of the calculations were performed on resources provided by
Sigma2 - the National Infrastructure for High Performance Computing and
Data Storage in Norway (Project No.\ NN4654K).
We also want to thank Prof.~Lin Lin, Prof.~Trygve Helgaker, Prof.~Anna Krylov, Dr.~Pavel Pokhilko, Dr.~Tanner P. Culpitt, Dr.~Laurens Peters, and Dr. Tilmann Bodenstein for fruitful discussions.
\bibliography{lib}

\newpage
\section{Appendix}

\subsection{Correlation in Geometry Optimization}

Since we can correlate three $S$-diagnostic variants with three error measures, we can in principle perform the piecewise linear fit that is presented in Section on {\it Correlation in Geometry Optimization} for nine different scenarios. 
We here present the piecewise linear fits which were not addressed in the above article.

\subsubsection{S$_1$-diagnostic Correlations}

\begin{figure}
\centering
\begin{subfigure}[b]{0.48\textwidth}
    \centering
    \includegraphics[width=\textwidth]{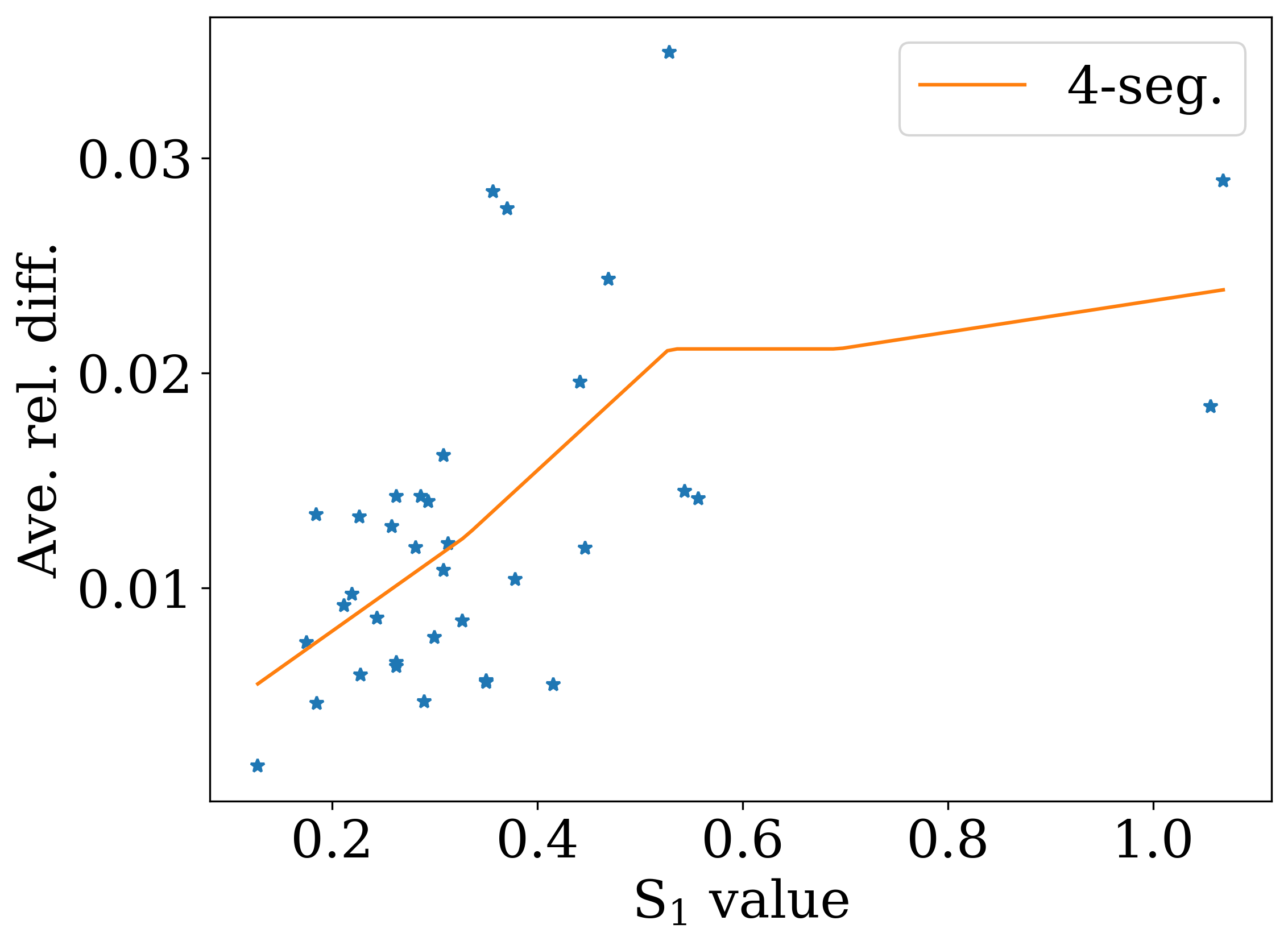}
    \caption{}
    \label{fig:smp_function_0_0}
\end{subfigure}
\hfill
\begin{subfigure}[b]{0.48\textwidth}
    \centering
    \includegraphics[width=\textwidth]{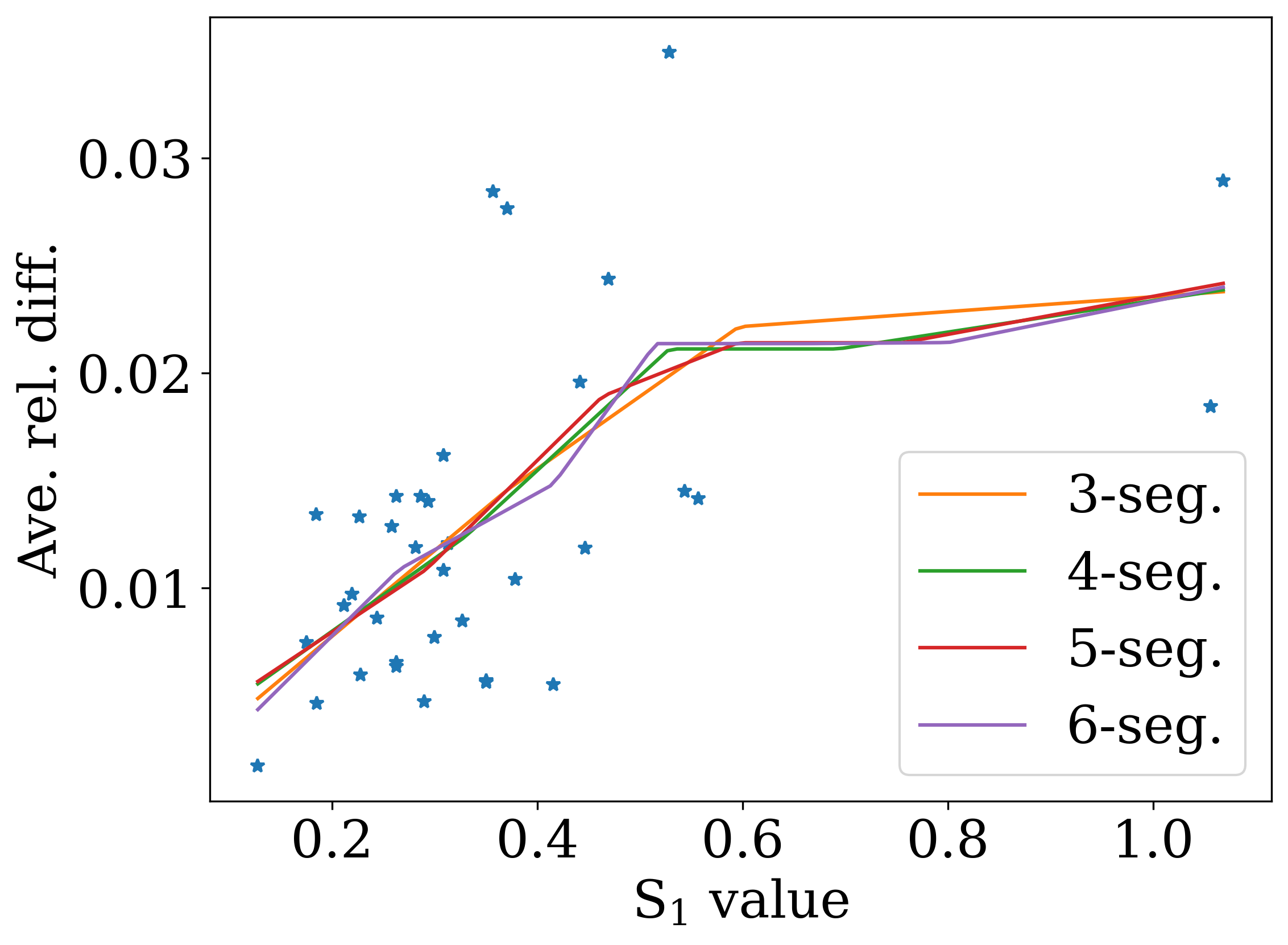}
    \caption{}
    \label{fig:smp_functions_0_0}
\end{subfigure}
\caption{
The averaged relative error in geometry optimization as a function of the $S_1$ value.
(\subref{fig:smp_function_0_0}) 
The orange line corresponds to a piecewise linear fit to the data using four segments for the piecewise linear function.
(\subref{fig:smp_functions_0_0}) Piecewise linear fits to the data with a varying number of segments.
}
\end{figure}

\begin{figure}
\centering
\begin{subfigure}[b]{0.48\textwidth}
    \centering
    \includegraphics[width=\textwidth]{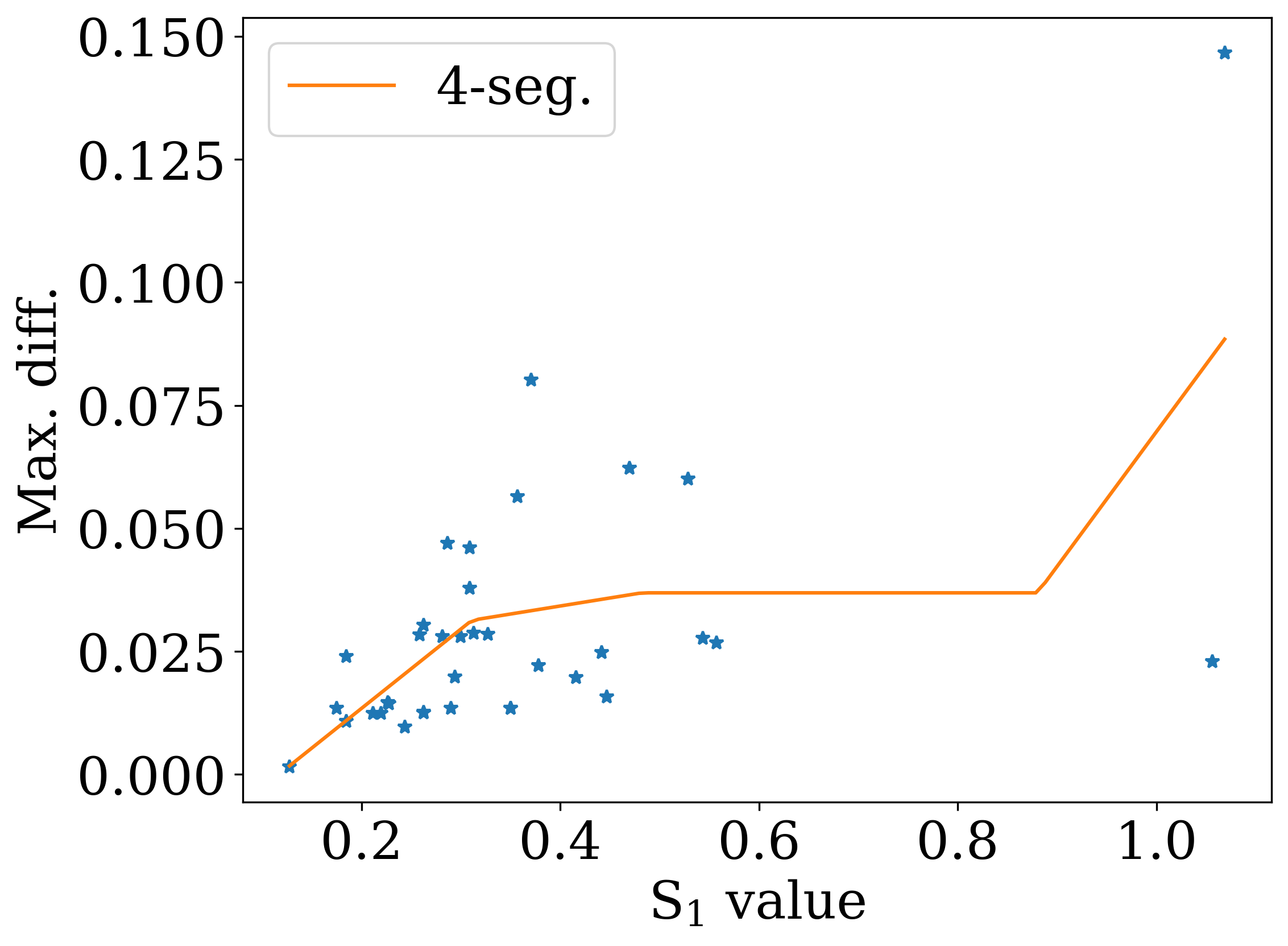}
    \caption{}
    \label{fig:smp_function_0_1}
\end{subfigure}
\hfill
\begin{subfigure}[b]{0.48\textwidth}
    \centering
    \includegraphics[width=\textwidth]{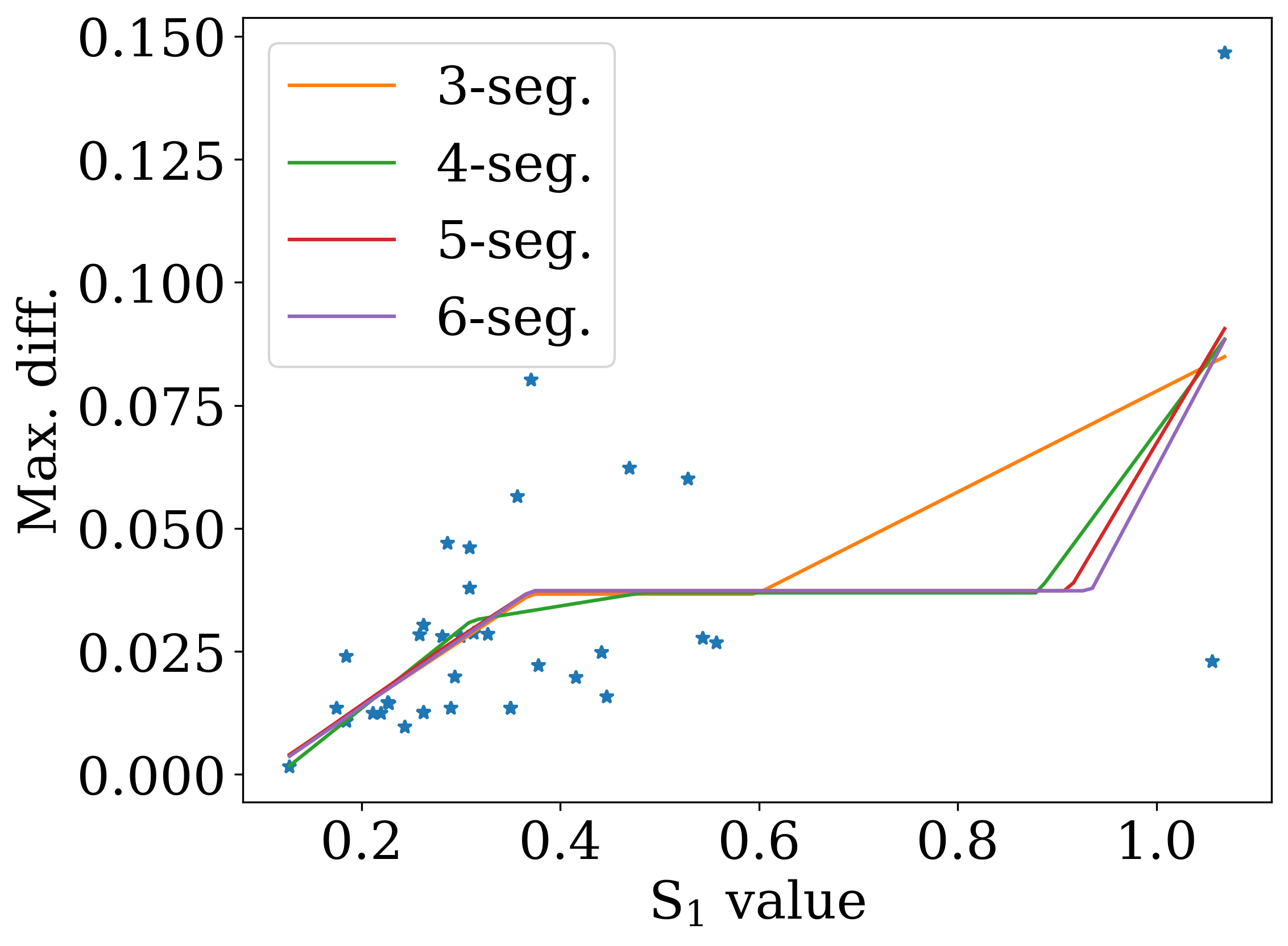}
    \caption{}
    \label{fig:smp_functions_0_1}
\end{subfigure}
\caption{
The maximal absolute error in geometry optimization as a function of the $S_1$ value.
(\subref{fig:smp_function_0_1}) 
The orange line corresponds to a piecewise linear fit to the data using four segments for the piecewise linear function.
(\subref{fig:smp_functions_0_1}) Piecewise linear fits to the data with a varying number of segments.
}
\end{figure}

\begin{figure}
\centering
\begin{subfigure}[b]{0.48\textwidth}
    \centering
    \includegraphics[width=\textwidth]{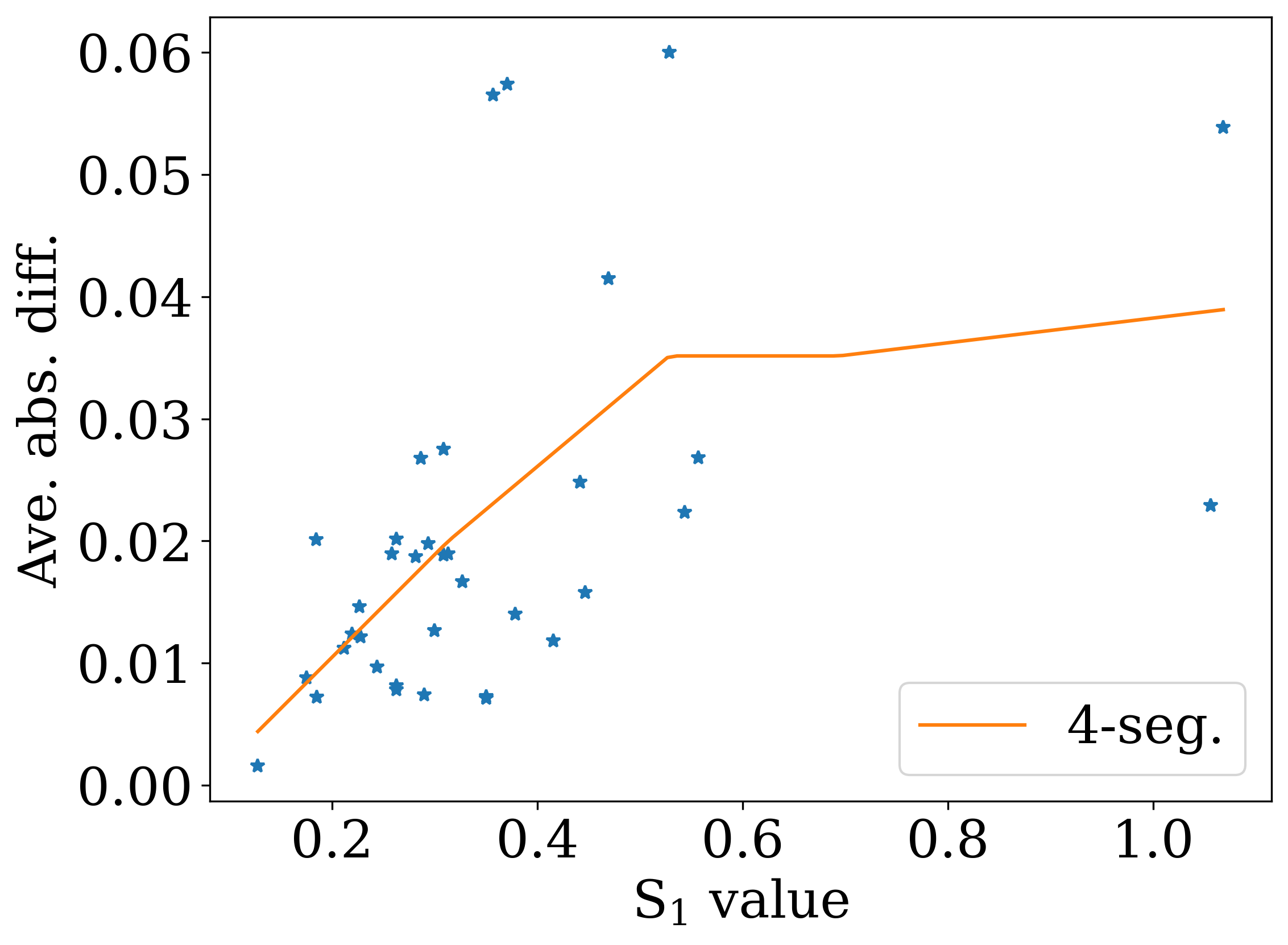}
    \caption{}
    \label{fig:smp_function_0_2}
\end{subfigure}
\hfill
\begin{subfigure}[b]{0.48\textwidth}
    \centering
    \includegraphics[width=\textwidth]{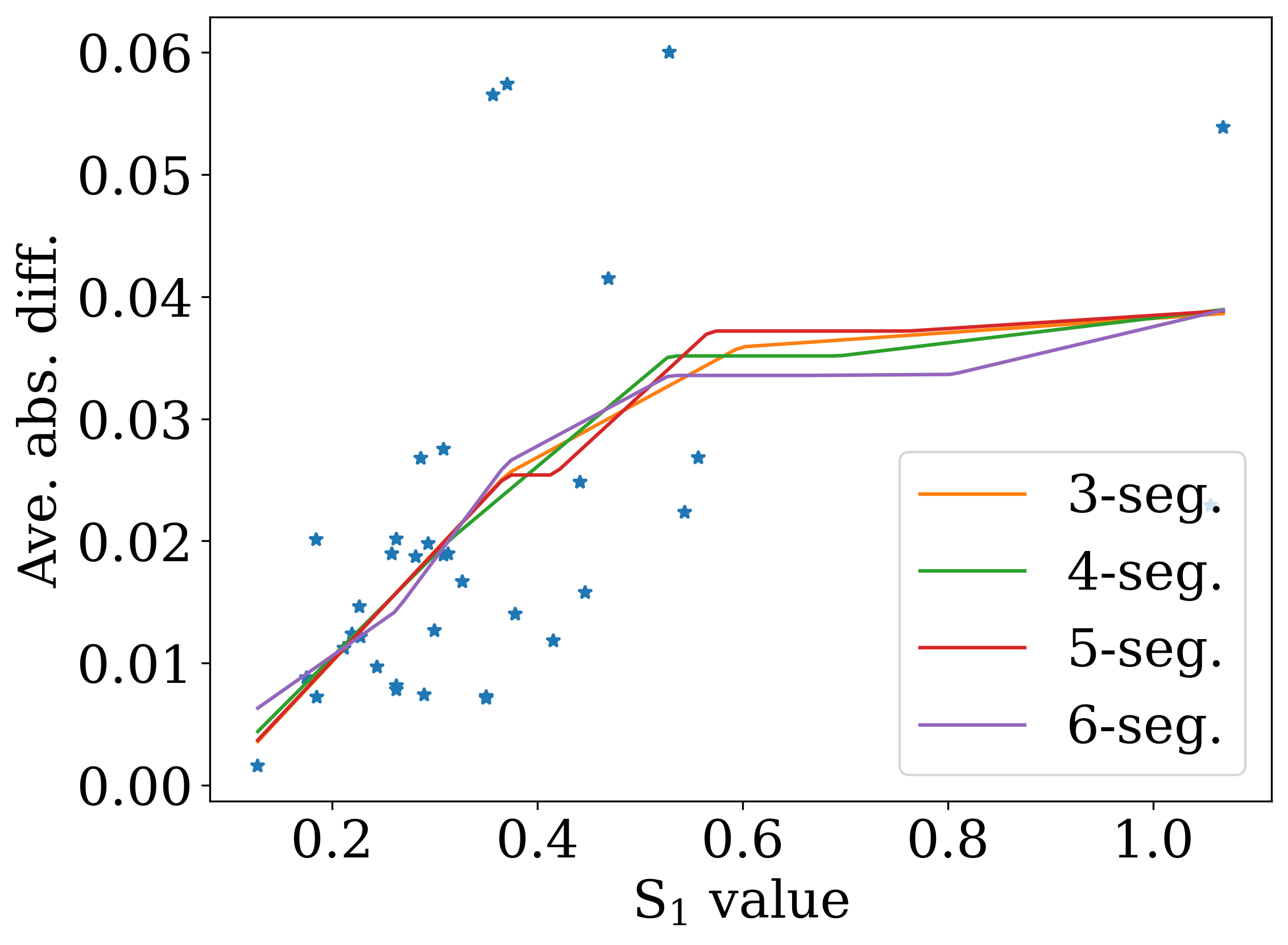}
    \caption{}
    \label{fig:smp_functions_0_2}
\end{subfigure}
\caption{
The averaged absolute error in geometry optimization as a function of the $S_1$ value.
(\subref{fig:smp_function_0_2}) 
The orange line corresponds to a piecewise linear fit to the data using four segments for the piecewise linear function.
(\subref{fig:smp_functions_0_2}) Piecewise linear fits to the data with a varying number of segments.
}
\end{figure}

\newpage
\subsubsection{S$_{2}$-diagnostic Correlations}

\begin{figure}
\centering
\begin{subfigure}[b]{0.48\textwidth}
    \centering
    \includegraphics[width=\textwidth]{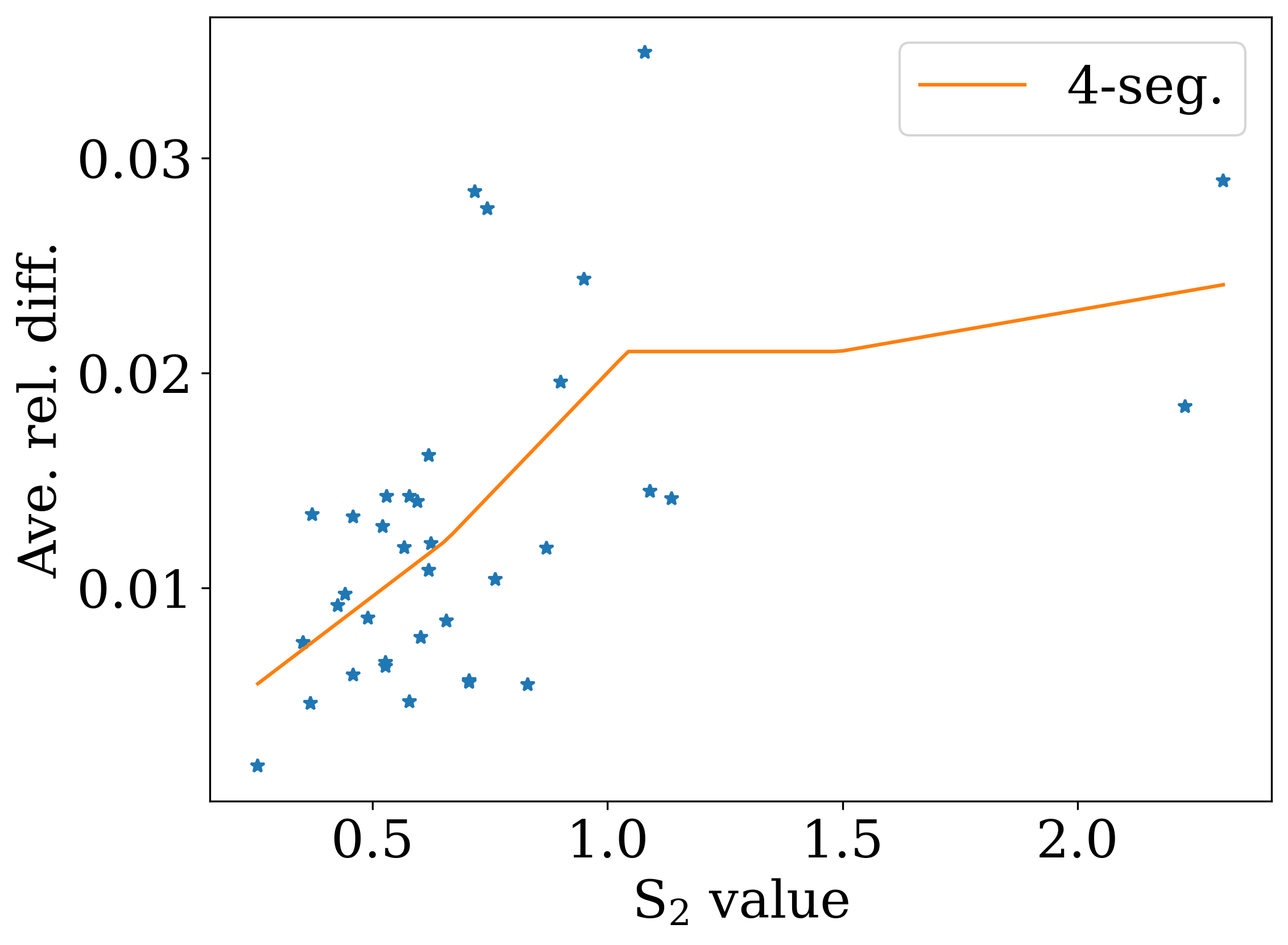}
    \caption{}
    \label{fig:smp_function_1_0}
\end{subfigure}
\hfill
\begin{subfigure}[b]{0.48\textwidth}
    \centering
    \includegraphics[width=\textwidth]{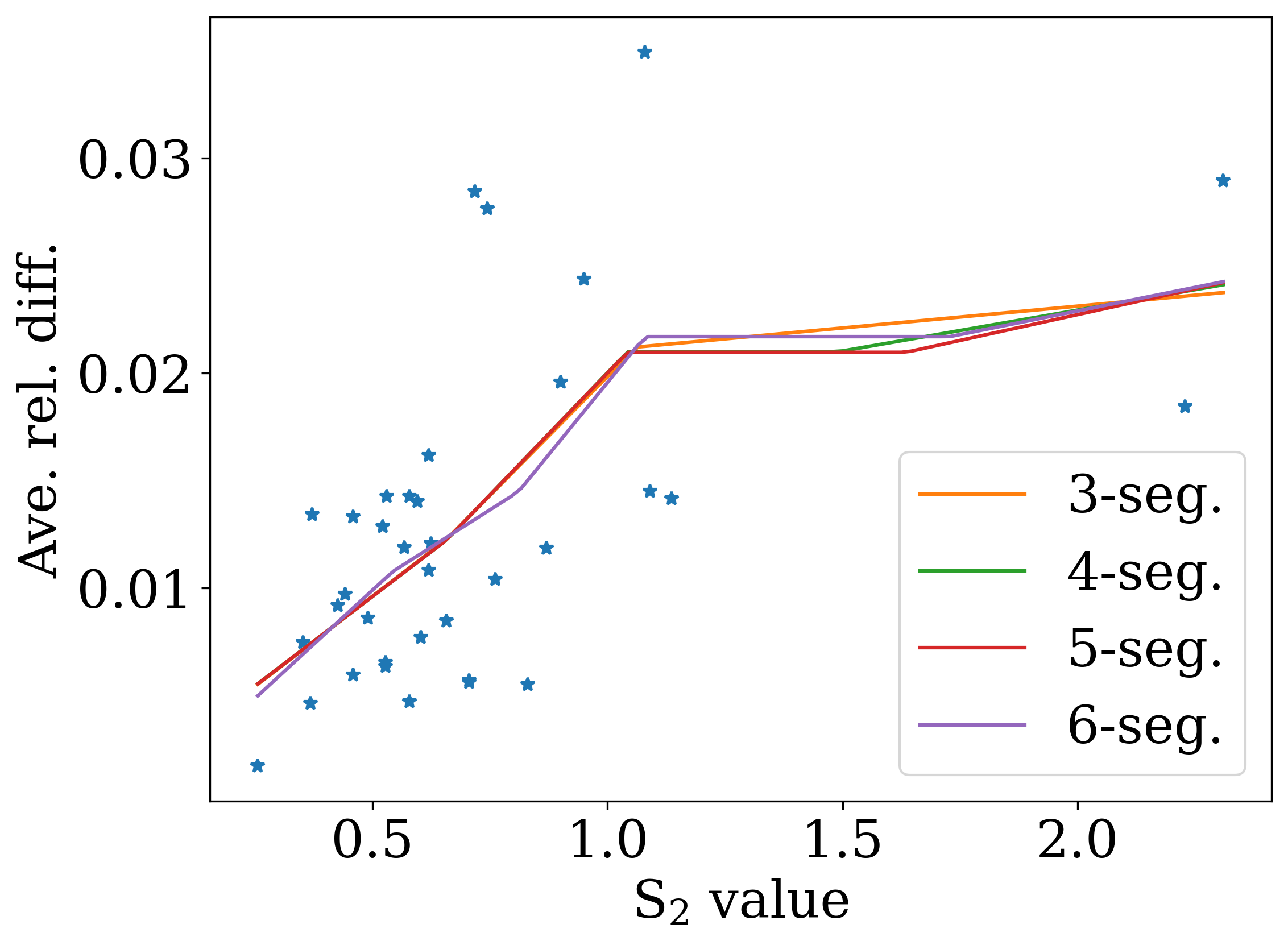}
    \caption{}
    \label{fig:smp_functions_1_0}
\end{subfigure}
\caption{
The averaged relative error in geometry optimization as a function of the $S_2$ value.
(\subref{fig:smp_function_1_0}) 
The orange line corresponds to a piecewise linear fit to the data using four segments for the piecewise linear function.
(\subref{fig:smp_functions_1_0}) Piecewise linear fits to the data with a varying number of segments.
}
\end{figure}

\begin{figure}
\centering
\begin{subfigure}[b]{0.48\textwidth}
    \centering
    \includegraphics[width=\textwidth]{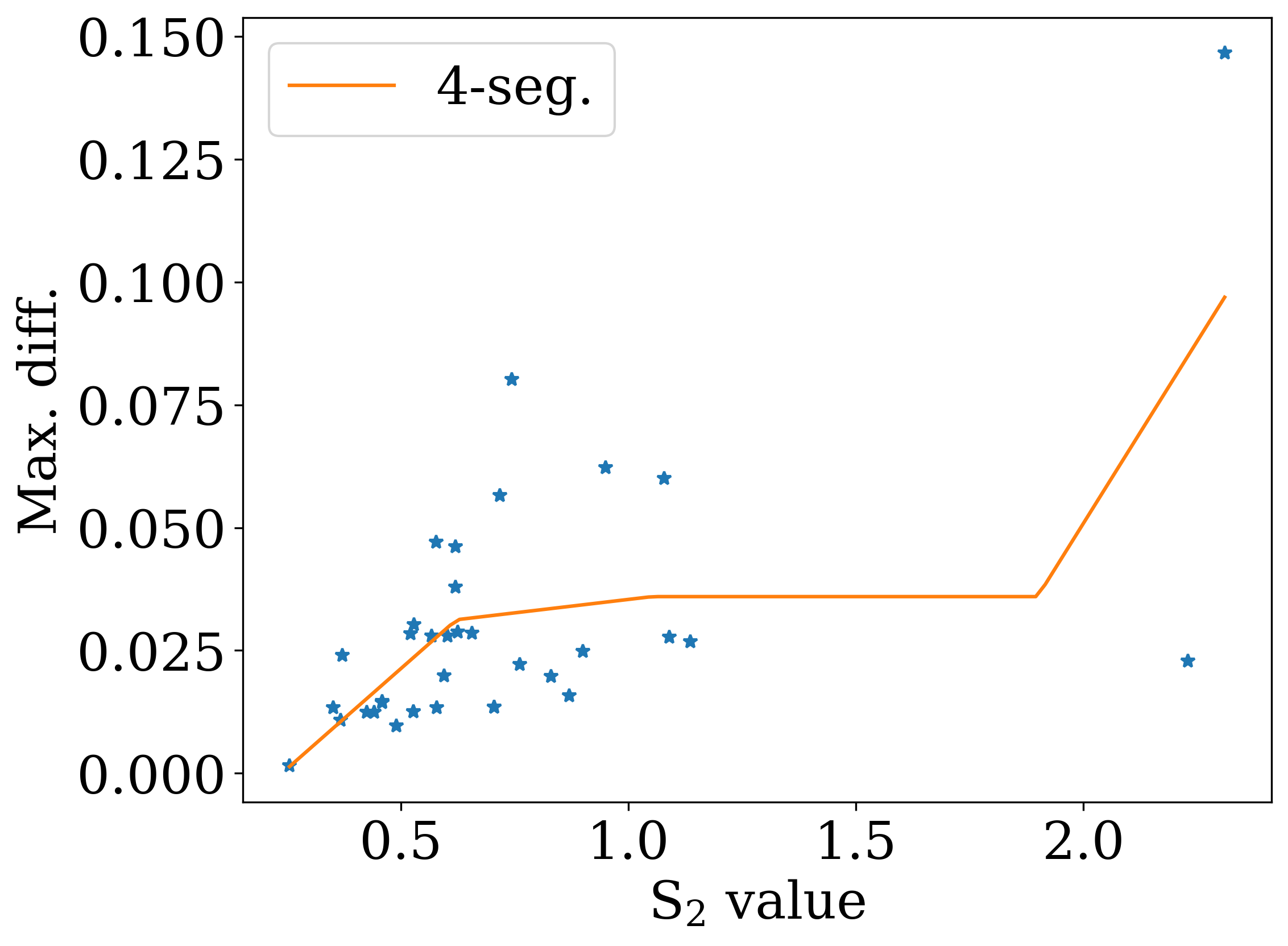}
    \caption{}
    \label{fig:smp_function_1_1}
\end{subfigure}
\hfill
\begin{subfigure}[b]{0.48\textwidth}
    \centering
    \includegraphics[width=\textwidth]{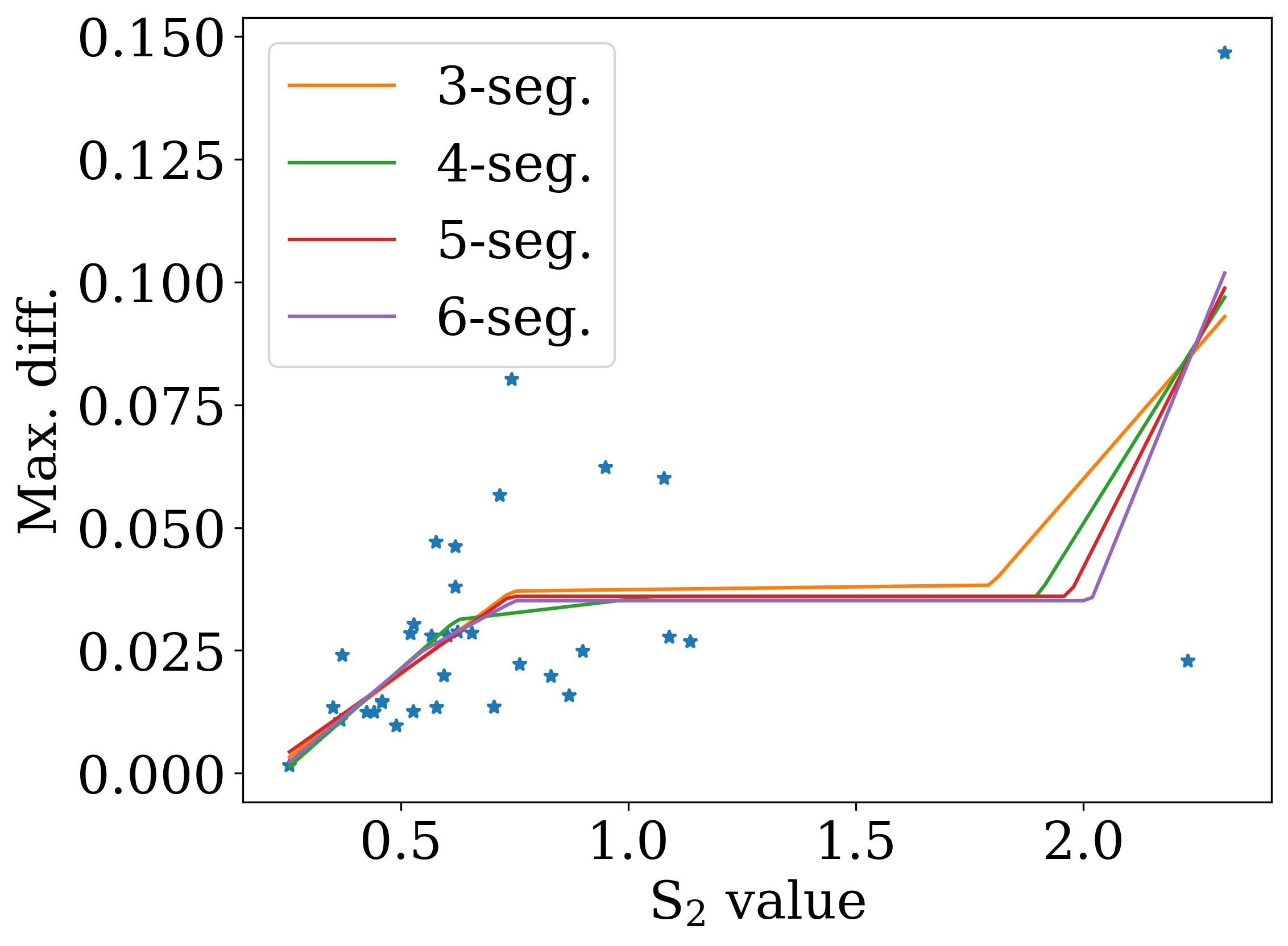}
    \caption{}
    \label{fig:smp_functions_1_1}
\end{subfigure}
\caption{
The maximal absolute error in geometry optimization as a function of the $S_2$ value.
(\subref{fig:smp_function_1_1}) 
The orange line corresponds to a piecewise linear fit to the data using four segments for the piecewise linear function.
(\subref{fig:smp_functions_1_1}) Piecewise linear fits to the data with a varying number of segments.
}
\end{figure}

\begin{figure}
\centering
\begin{subfigure}[b]{0.48\textwidth}
    \centering
    \includegraphics[width=\textwidth]{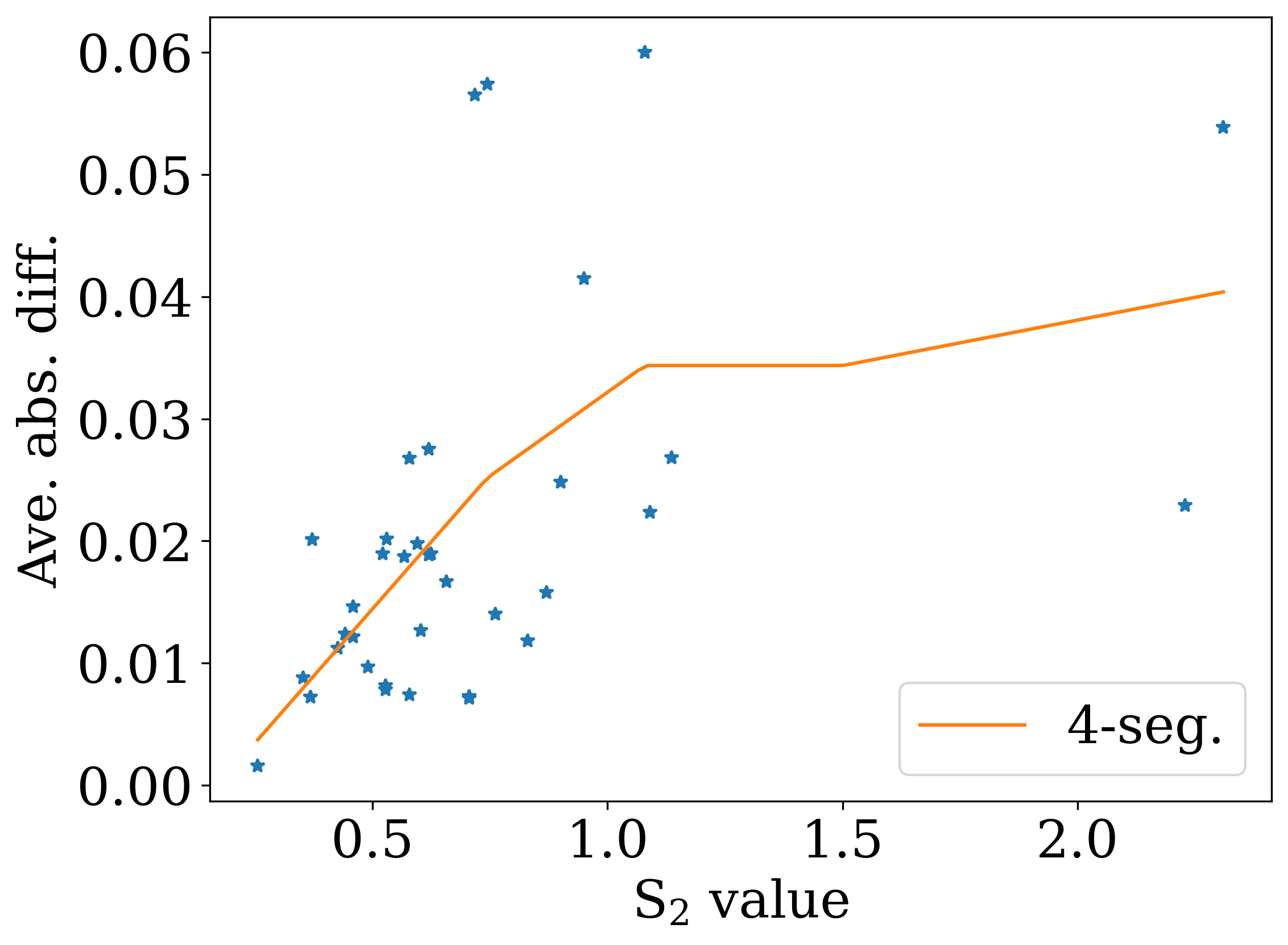}
    \caption{}
    \label{fig:smp_function_1_2}
\end{subfigure}
\hfill
\begin{subfigure}[b]{0.48\textwidth}
    \centering
    \includegraphics[width=\textwidth]{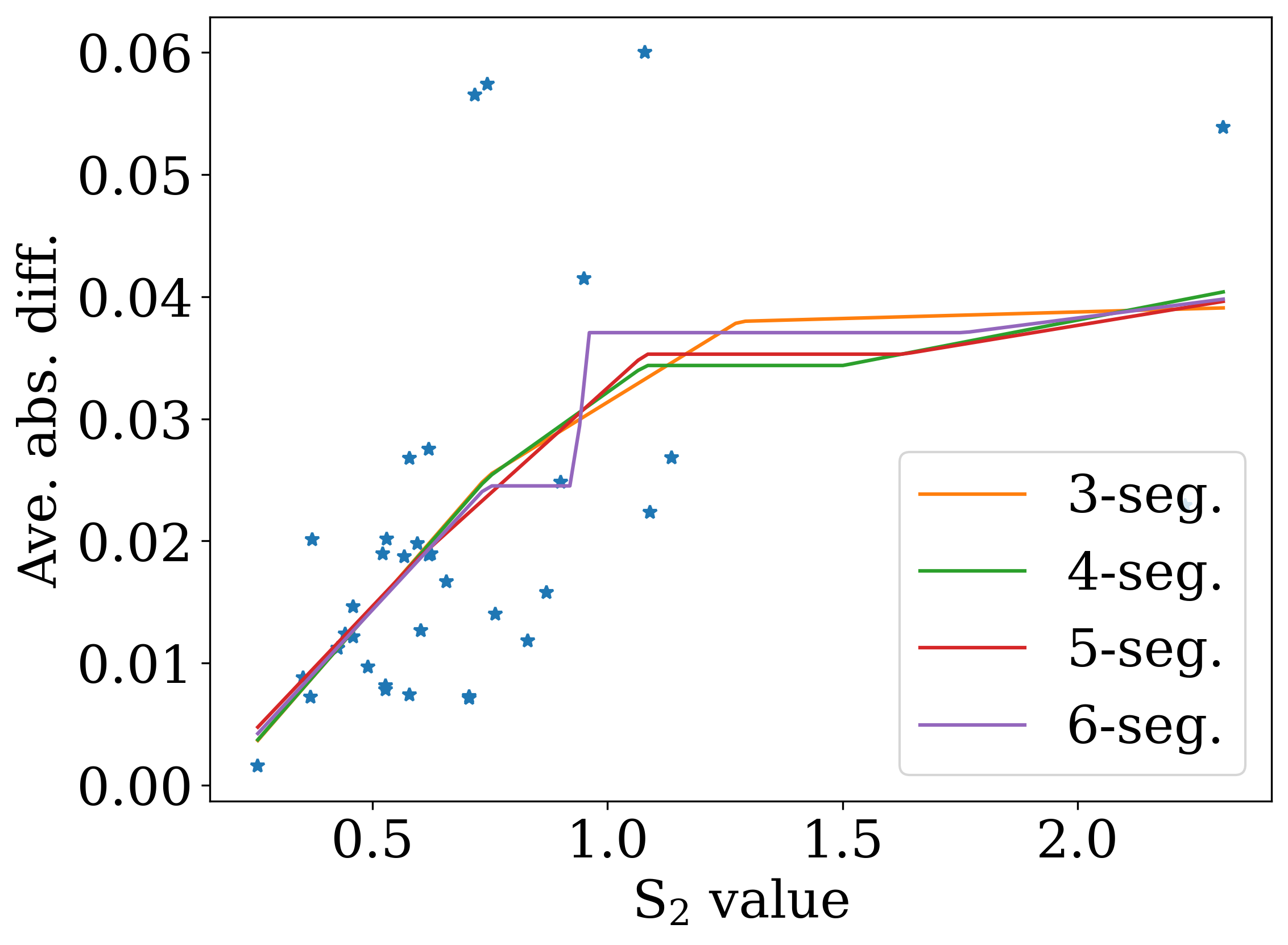}
    \caption{}
    \label{fig:smp_functions_1_2}
\end{subfigure}
\caption{
The averaged absolute error in geometry optimization as a function of the $S_2$ value.
(\subref{fig:smp_function_1_2}) 
The orange line corresponds to a piecewise linear fit to the data using four segments for the piecewise linear function.
(\subref{fig:smp_functions_1_2}) Piecewise linear fits to the data with a varying number of segments.
}
\end{figure}

\newpage
\subsubsection{S$_2$-diagnostic Correlations}

\begin{figure}
\centering
\begin{subfigure}[b]{0.48\textwidth}
    \centering
    \includegraphics[width=\textwidth]{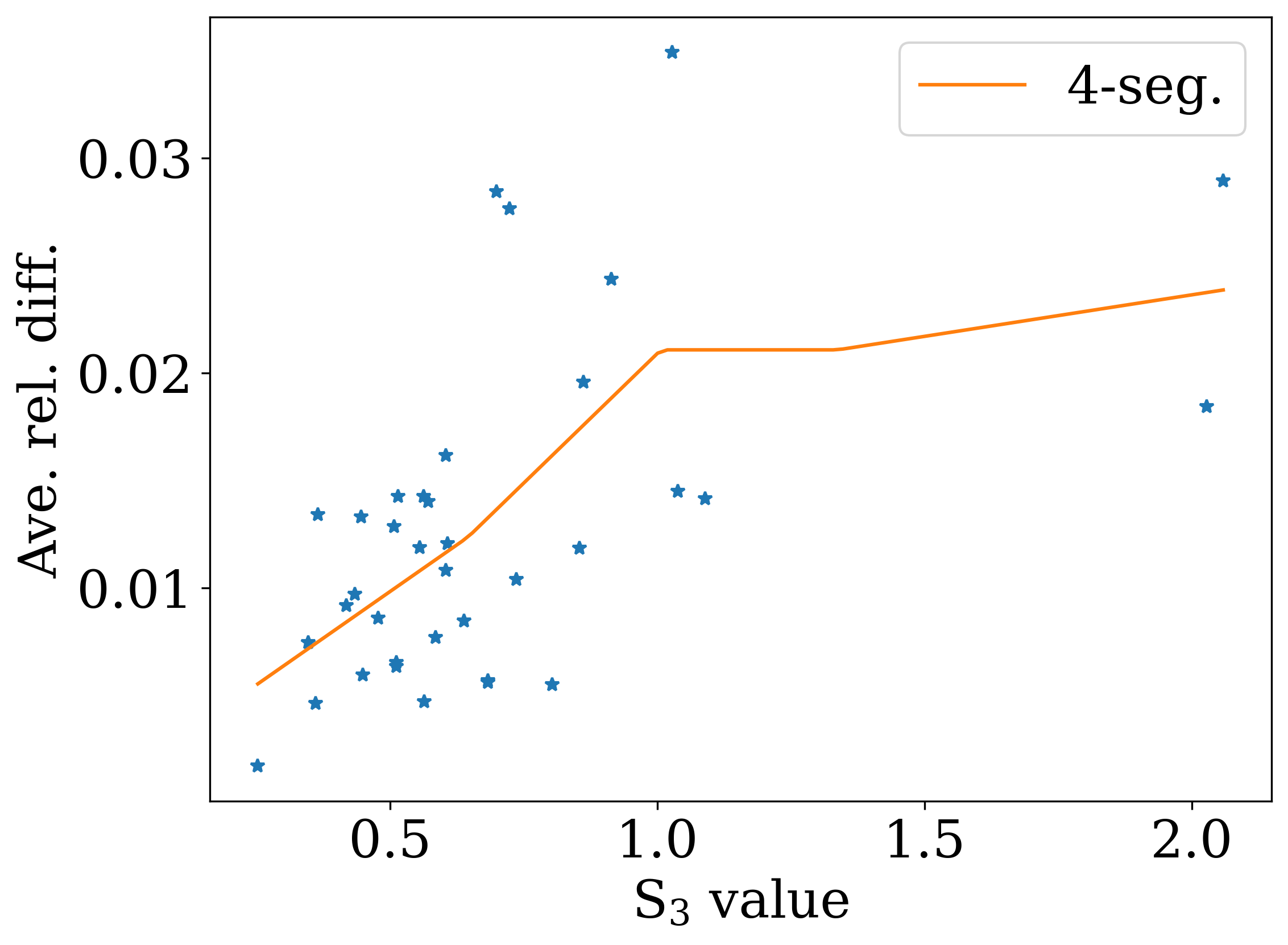}
    \caption{}
    \label{fig:smp_function_2_0}
\end{subfigure}
\hfill
\begin{subfigure}[b]{0.48\textwidth}
    \centering
    \includegraphics[width=\textwidth]{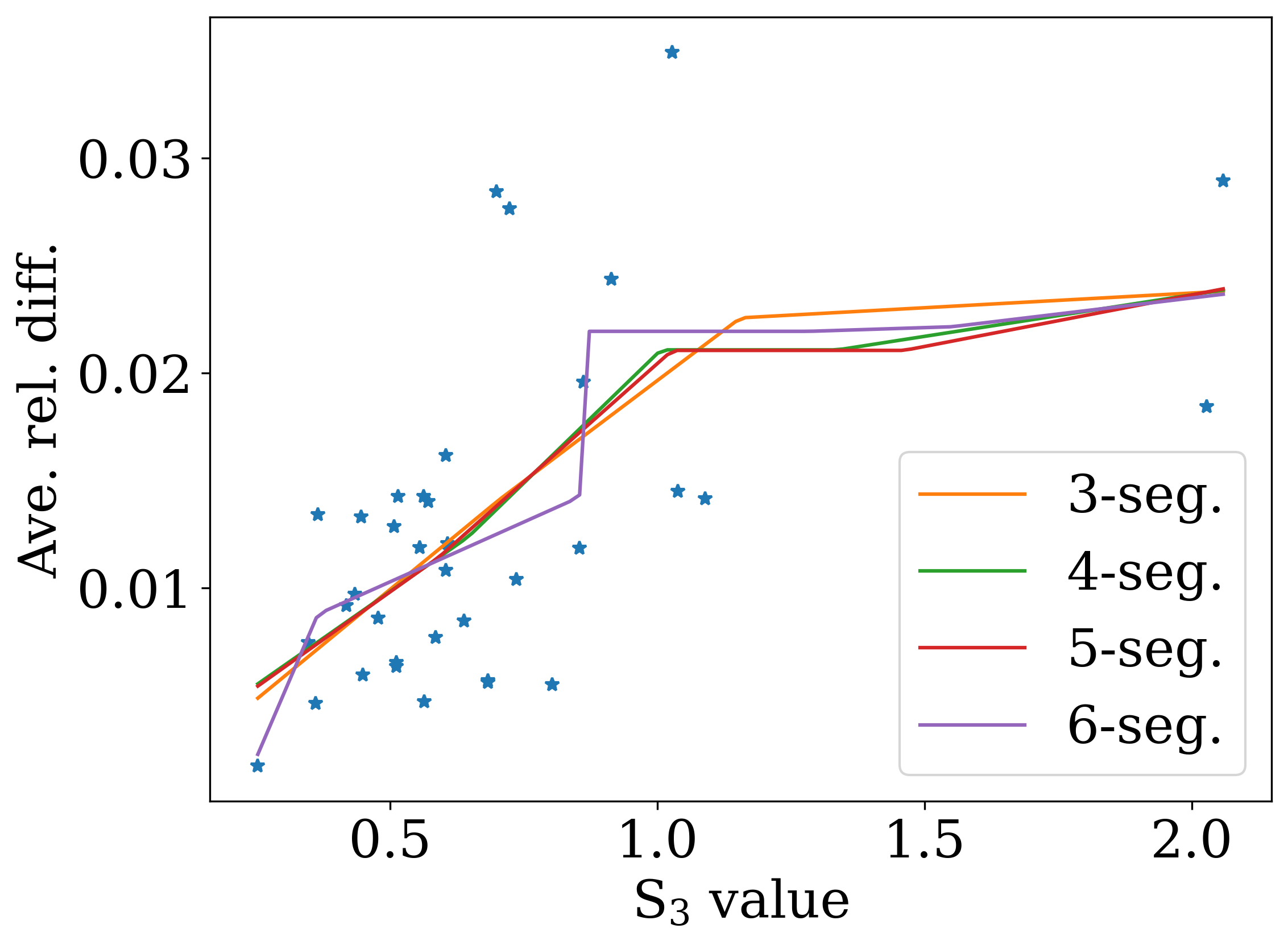}
    \caption{}
    \label{fig:smp_functions_2_0}
\end{subfigure}
\caption{
The averaged relative error in geometry optimization as a function of the $S_3$ value.
(\subref{fig:smp_function_2_0}) 
The orange line corresponds to a piecewise linear fit to the data using four segments for the piecewise linear function.
(\subref{fig:smp_functions_2_0}) Piecewise linear fits to the data with a varying number of segments.
}
\end{figure}

\begin{figure}
\centering
\begin{subfigure}[b]{0.48\textwidth}
    \centering
    \includegraphics[width=\textwidth]{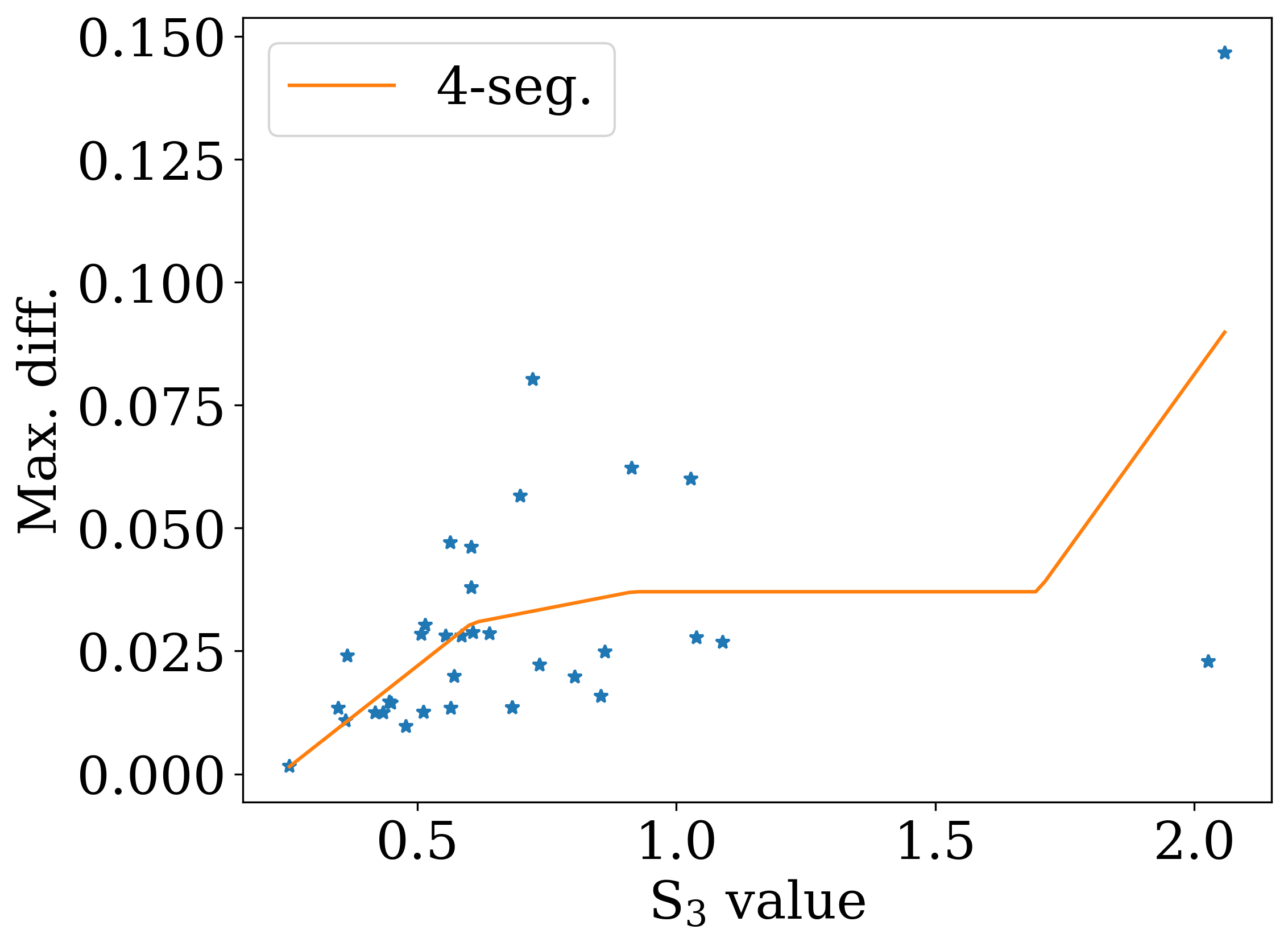}
    \caption{}
    \label{fig:smp_function_2_1}
\end{subfigure}
\hfill
\begin{subfigure}[b]{0.48\textwidth}
    \centering
    \includegraphics[width=\textwidth]{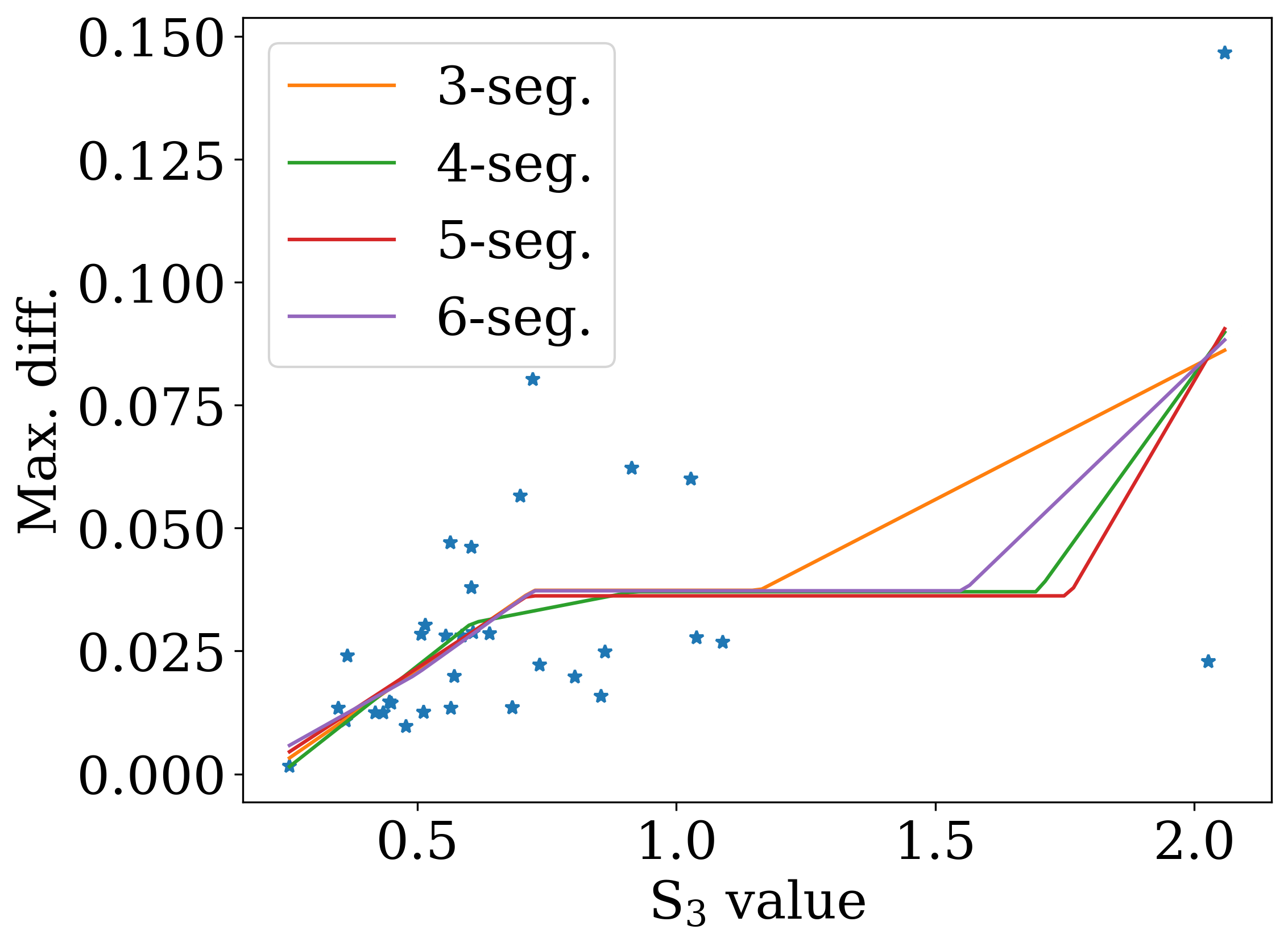}
    \caption{}
    \label{fig:smp_functions_2_1}
\end{subfigure}
\caption{
The maximal absolute error in geometry optimization as a function of the $S_3$ value.
(\subref{fig:smp_function_2_1}) 
The orange line corresponds to a piecewise linear fit to the data using four segments for the piecewise linear function.
(\subref{fig:smp_functions_2_1}) Piecewise linear fits to the data with a varying number of segments.
}
\end{figure}

\begin{figure}
\centering
\begin{subfigure}[b]{0.48\textwidth}
    \centering
    \includegraphics[width=\textwidth]{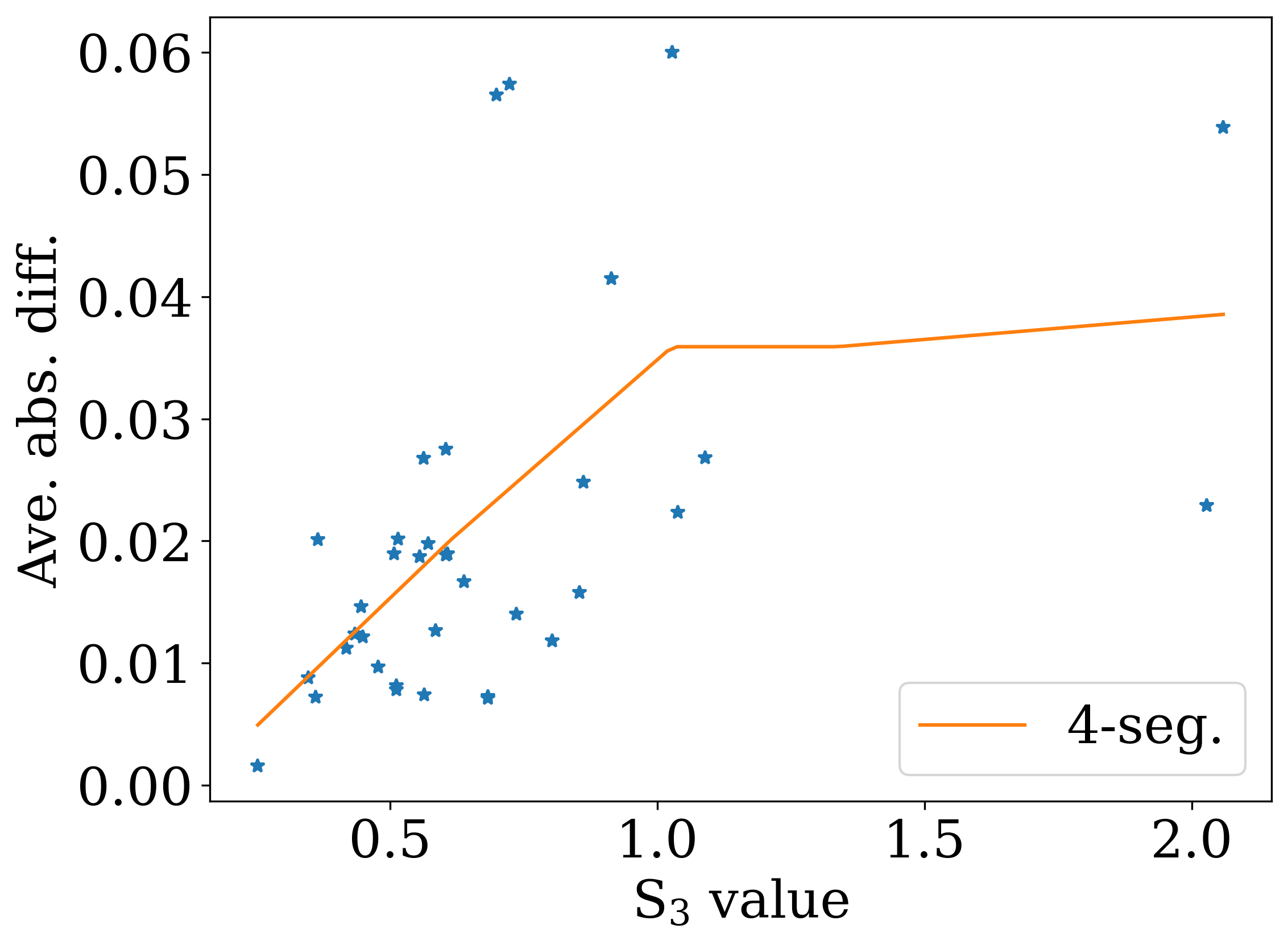}
    \caption{}
    \label{fig:smp_function_2_2}
\end{subfigure}
\hfill
\begin{subfigure}[b]{0.48\textwidth}
    \centering
    \includegraphics[width=\textwidth]{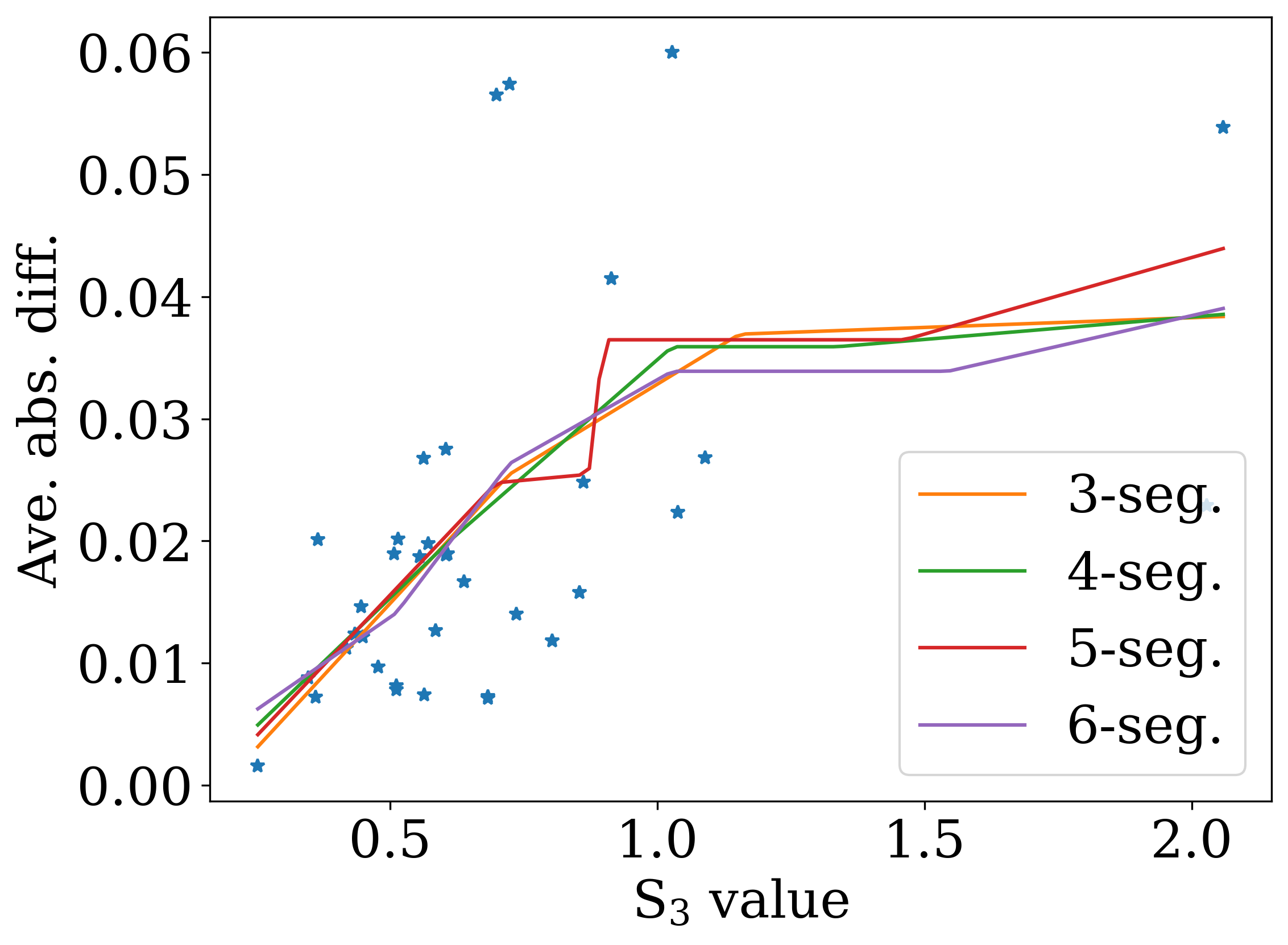}
    \caption{}
    \label{fig:smp_functions_2_2}
\end{subfigure}
\caption{
The averaged absolute error in geometry optimization as a function of the $S_3$ value.
(\subref{fig:smp_function_2_2}) 
The orange line corresponds to a piecewise linear fit to the data using four segments for the piecewise linear function.
(\subref{fig:smp_functions_2_2}) Piecewise linear fits to the data with a varying number of segments.
}
\end{figure}

\newpage
\subsection{Transition State Models}
Here we shall compare the performance of the $S$-diagnostics and the previously used $T_1$, $D_1$, and $D_2$ diagnostics. 

\begin{figure}
\centering
\begin{subfigure}[b]{0.48\textwidth}
    \centering
    \includegraphics[width=\textwidth]{Graphics/smp_daig_c2h4.png}
    \caption{}
    \label{fig:smp_c2h4_2}
\end{subfigure}
\hfill
\begin{subfigure}[b]{0.48\textwidth}
    \centering
    \includegraphics[width=\textwidth]{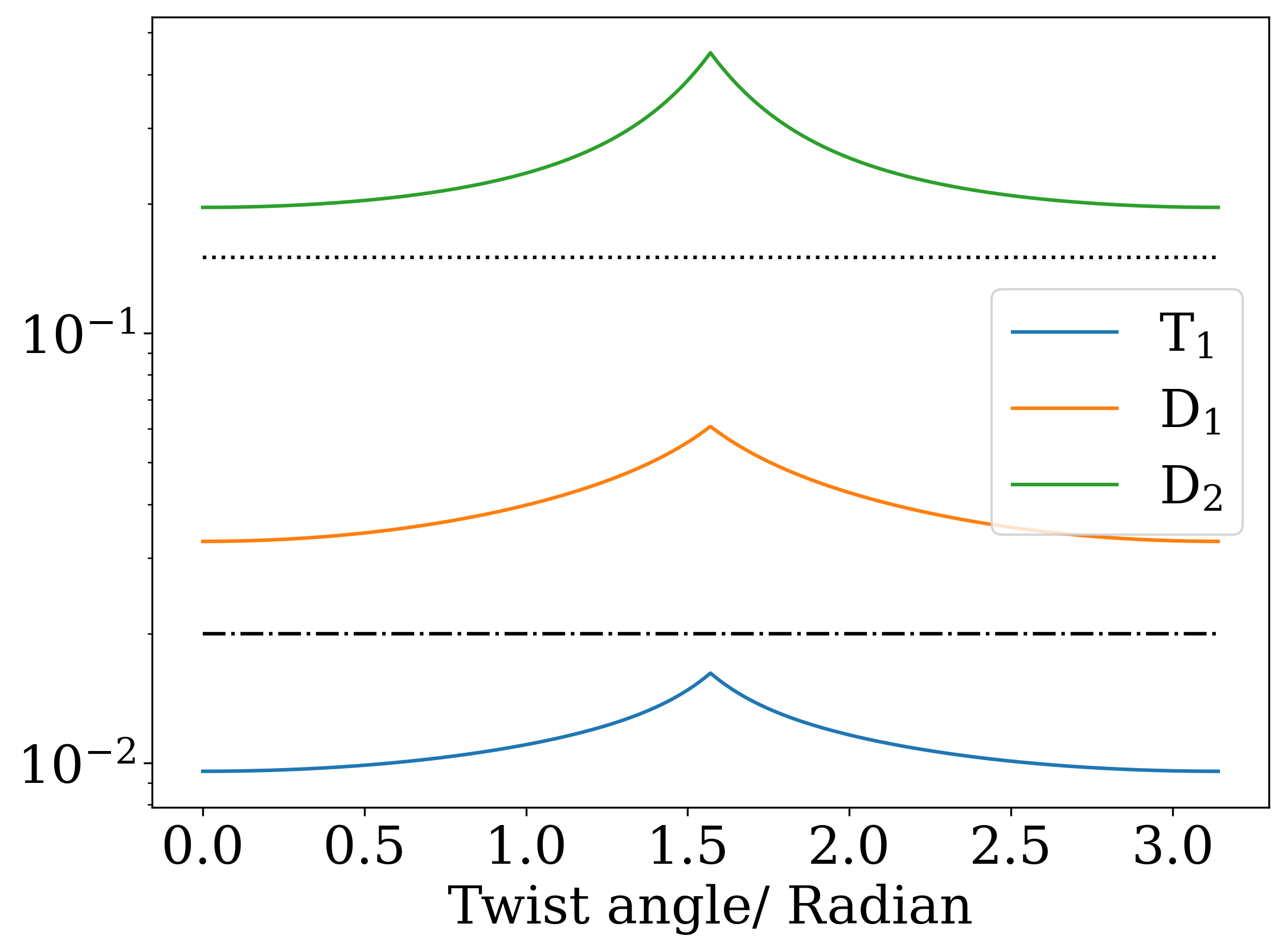}
    \caption{}
    \label{fig:prev_c2h4}
\end{subfigure}
\caption{\label{fig:prev_diag_c2h4}
(\subref{fig:smp_c2h4_2}) 
shows the $S$-diagnostics
(\subref{fig:prev_c2h4}) shows the
previously suggested $T_1$, $D_1$ and $D_2$ diagnostics
}
\end{figure}

\begin{figure}
\centering
\begin{subfigure}[b]{0.48\textwidth}
    \centering
    \includegraphics[width=\textwidth]{Graphics/SMP_BeH2.png}
    \caption{}
    \label{fig:smp_beh2_2}
\end{subfigure}
\hfill
\begin{subfigure}[b]{0.48\textwidth}
    \centering
    \includegraphics[width=\textwidth]{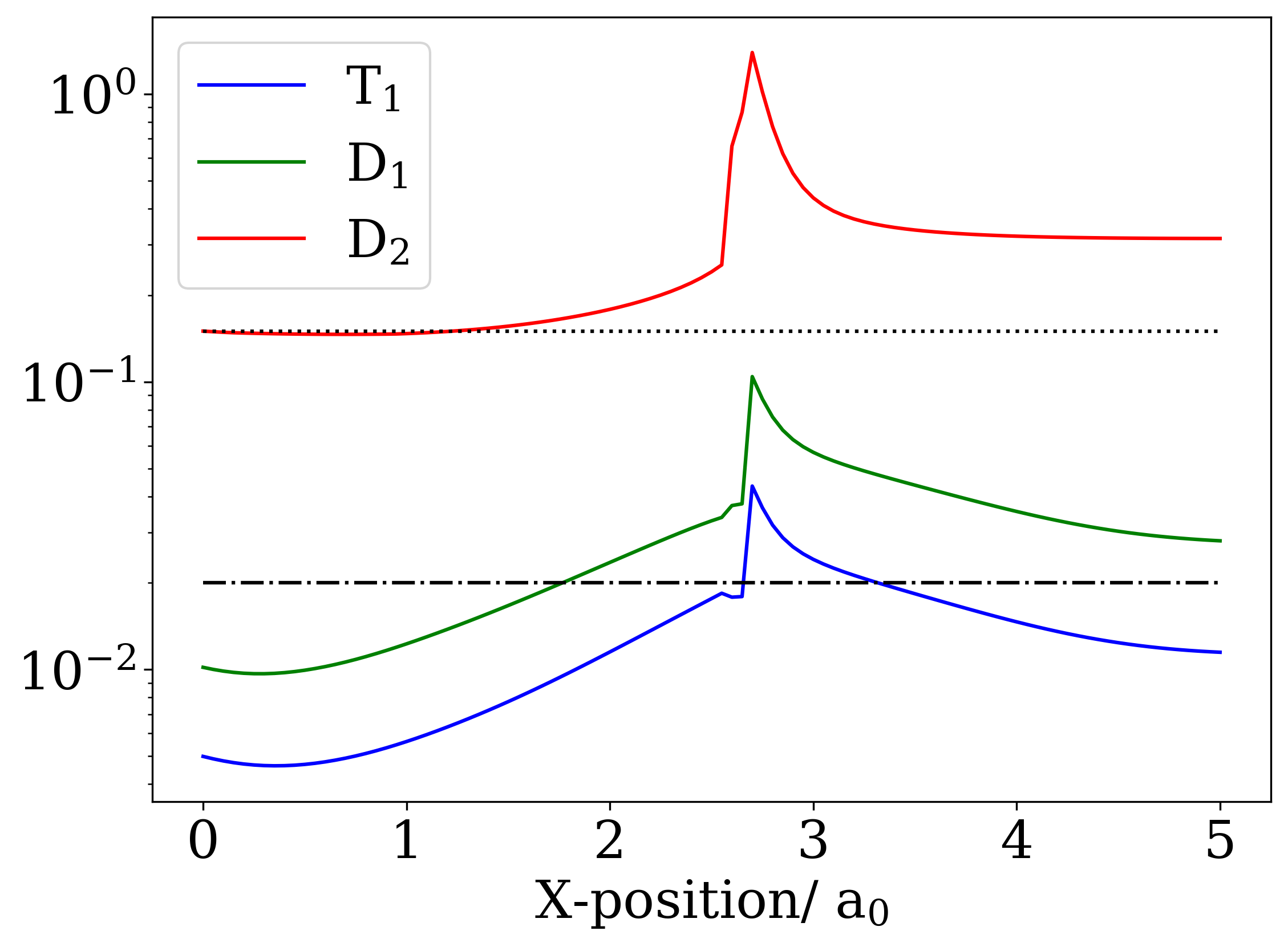}
    \caption{}
    \label{fig:prev_beh2}
\end{subfigure}
\caption{\label{fig:prev_diag_beh2}
(\subref{fig:smp_beh2_2}) 
shows the $S$-diagnostics
(\subref{fig:prev_beh2}) shows the
previously suggested $T_1$, $D_1$ and $D_2$ diagnostics
}
\end{figure}

\begin{figure}
\centering
\begin{subfigure}[b]{0.48\textwidth}
    \centering
    \includegraphics[width=\textwidth]{Graphics/SMP_H4.png}
    \caption{}
    \label{fig:smp_h4_2}
\end{subfigure}
\hfill
\begin{subfigure}[b]{0.48\textwidth}
    \centering
    \includegraphics[width=\textwidth]{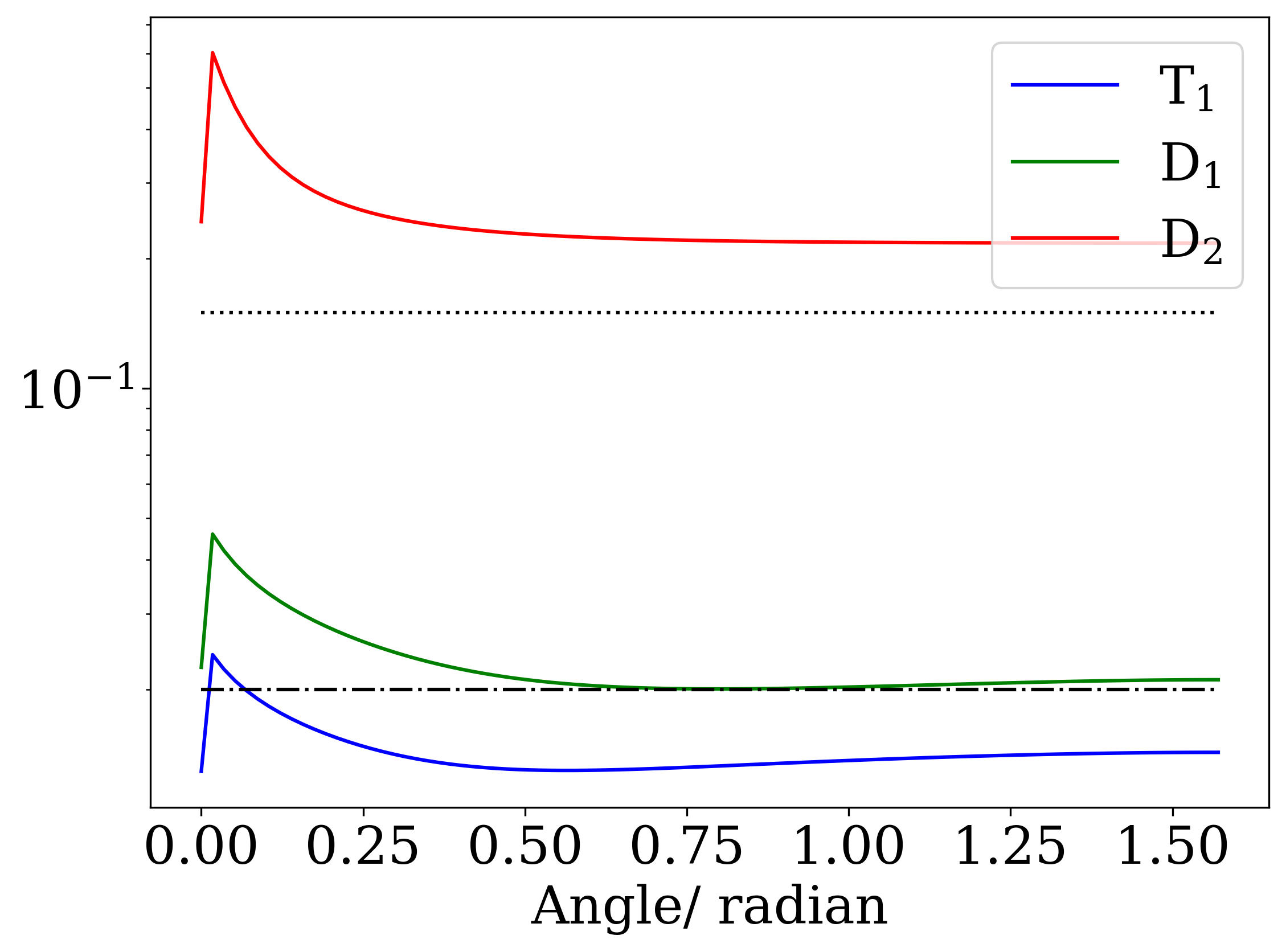}
    \caption{}
    \label{fig:prev_h4}
\end{subfigure}
\caption{\label{fig:prev_diag_h4}
(\subref{fig:smp_h4_2}) 
shows the $S$-diagnostics
(\subref{fig:prev_h4}) shows the
previously suggested $T_1$, $D_1$ and $D_2$ diagnostics
}
\end{figure}

\begin{figure}
\centering
\begin{subfigure}[b]{0.48\textwidth}
    \centering
    \includegraphics[width=\textwidth]{Graphics/smp_daig_h2_scus.png}
    \caption{}
    \label{fig:smp_h2_scus_2}
\end{subfigure}
\hfill
\begin{subfigure}[b]{0.48\textwidth}
    \centering
    \includegraphics[width=\textwidth]{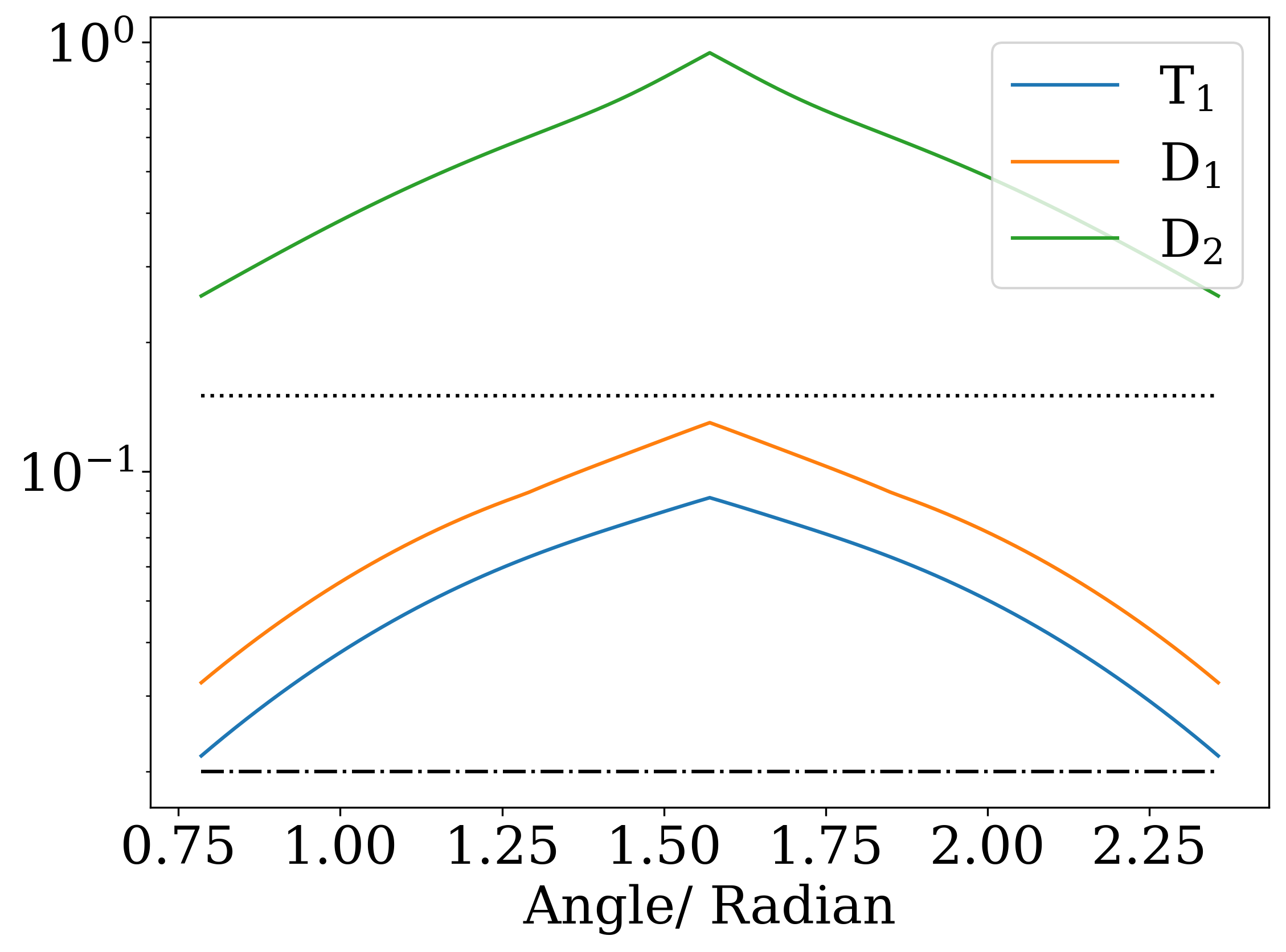}
    \caption{}
    \label{fig:prev_h2_scus}
\end{subfigure}
\caption{\label{fig:prev_diag_h2_scus}
(\subref{fig:smp_h2_scus_2}) 
shows the $S$-diagnostics
(\subref{fig:prev_h2_scus}) shows the
previously suggested $T_1$, $D_1$ and $D_2$ diagnostics
}
\end{figure}

\end{document}